\documentstyle[12pt,psfig]{article}
\newcommand{\be}{\begin{equation}}
\newcommand{\ee}{\end{equation}}
\newcommand{\ea}{\end{array}}

\def\Frac#1#2{\frac{\displaystyle{#1}}{\displaystyle{#2}}}

\begin{document}

\begin{flushright}
UG-DFM-2/97

FTUV/97-18 

IFIC/97-18
\end{flushright}

\vspace*{2cm}

\begin{center}
{\Large {\bf Many Body approach to the inclusive $(e,e^\prime)$ reaction from 
the quasielastic to the  $\Delta$ excitation region}}
\end{center}

\vspace{0.5cm}

\begin{center}
{\large {A. Gil$^1$, J. Nieves$^2$ and E. Oset$^1$  }}
\end{center}

\vspace{0.3cm}

{\small {\it

$^1$Departamento de F\'{\i}sica Te\'orica and IFIC, Centro Mixto Universidad
de Valencia - CSIC, 46100 Burjassot (Valencia) Spain.

$^2$ Departamento de F\'{\i}sica Moderna, Universidad de Granada, 
18071 Granada, Spain.}} 

\vspace{0.7cm}
\begin{abstract}

     We have performed a many body calculation of the inclusive 
 $(e,e^{\prime})$ cross section which runs over the three traditional
 regions at intermediate energies: the quasielastic peak, the dip region
 and the delta region. The longitudinal and transverse response functions
 in the quasielastic peak have also been evaluated. Traditional effects
 like polarization, meson exchange currents, final state interaction
 and delta renormalization in the nuclear medium have been included.
 Meson exchange currents are generated from a model of pion 
 electroproduction on the nucleon which reproduces accurately the
 experimental data.
 
       The inclusive cross section accounts for $1N,2N,3N$ mechanisms
       of virtual photon absorption and one pion production. Meson
       exchange currents associated to the $(\gamma^*,2\pi)$ 
       reaction are also accounted for.
       
We obtain good results for the $(e,e^{\prime})$ cross sections
in the whole energy range and for different nuclei.
The response functions are also in good agreement
with the latest experimental analysis. On the other hand, the method
provides the separation of the contribution to the inclusive cross
section from different physical channels which is a necessary
input to evaluate cross sections like $(e,e^{\prime}N)$,
$(e,e^{\prime}NN)$, $(e,e^{\prime} \pi)$ etc. 

\end{abstract}

\section{Introduction}

Inclusive electron scattering, particularly around the quasielastic
peak is probably one of the problems that has attracted more attention in
nuclear physics. One of the reasons for the proliferation of theoretical
work was the persistent difficulty of simple pictures based on the shell 
model of the nucleus, or the Fermi gas approach, to reproduce simultaneously
the longitudinal and the transverse response functions around the quasielastic
 peak. Particularly, the longitudinal response appeared systematically 
to be much larger than experiment \cite{MEZ,ALT,BAR}.

Even more worrisome was the fact that the integrated strength of the 
experimental longitudinal response function showed large discrepancies with
the expected result according to the Coulomb sum rule \cite{GIU}. At large
 values of the momentum transfer $q$ this integral should be the 
charge of the nucleus and there was some apparent missing strength.

Those experimental results have been generally accepted, even when some 
other experimental results seemed to challenge them. Indeed the results of 
the experiment of \cite{BLA} in $^{238}$U did not show the expected
suppression of the longitudinal response. More recently an experiment at Bates
on $^{40}$Ca \cite{KAR} showed a longitudinal response larger than in 
ref. \cite{MEZ} with only 20$\%$ reduction over the simple shell 
model expectations.

The growing experimental discrepancies stimulated the thorough and 
thoughtful work of  \cite{JOU}.
In this work the author analysed the world set of data and made
further improvements in the analysis, coming with new results for the 
response functions which show a smaller reduction of the longitudinal
response than previously assumed. At the heart of the issue was the fact
that, in a modified Rosenbluth plot in terms of the variable $\epsilon$ 
(which runs
from 0 to 1), the Saclay data~\cite{MEZ,ALT,BAR} concentrated their points
in the region of $\epsilon < 0.5$
which induces large errors in the slope of the straight
line which  correlates the points of the 
plot. Those results, complemented by others at SLAC and Bates, which 
fill up the region of $\epsilon \sim 1$
, lead to a more accurate determination of
the slope and thus the longitudinal response. The problems with the 
Coulomb sum rule are then automatically solved \cite{JOU} to the relief
of all \cite{DEN}.

As usually happens, some theoretical calculations were ``successful'' in
explaining the abnormal reductions of the longitudinal response, through
the use of small effective masses, swelling of nucleons, abnormal nucleon
form factors, correlations, etc. But in the benefit of the theoretical 
community it should be said that whenever this happened, the agreement
with the transverse response was spoiled, although excuses and good wishes
beyond the ideas used for the longitudinal response were invoked as 
possible solutions. Hence, a convincing simultaneous explanation of both the
longitudinal and transverse response functions was never obtained.

The interesting thing is that the existence of a ``problem'' induced so much
work that practically all resources of nuclear physics and many body theory
 have been used in this topic and a lot of things have been learned.
In the present work we shall benefit from all this previous work and by
using a selfcontained many body formalism we will incorporate all these 
effects which convincingly proved to be relevant in previous works. We also
incorporate some new ingredients which come out naturally within our many
body expansion and furthermore we include Delta--hole ($\Delta h$) 
 excitations, additional
to the particle--hole ($ph$) excitations, which allows us to simultaneously study the 
quasielastic peak, the $\Delta$ peak and the ``dip'' region between the
two peaks in the inclusive $(e,e^\prime)$ cross section.

\section{Brief review of different approaches and introduction to our
approach.}

The amount of ideas which have been studied in connection with the
inclusive $(e, e^\prime)$ scattering is quite large. We shall discuss them
briefly:

\subsection{Modification of the nucleon form factor (swelling of the nucleon)}

These ideas were soon invoked and explored within different frameworks
~\cite{NOB,SHA}. They appear naturally in some microscopic models of the
nucleon form factor when the intermediate baryon propagators are replaced
by those in the nuclear medium. This is the case, for instance in ref.~\cite{GOE}, where the underlying elementary model is the 
Nambu-Jona-Lasinio model
of the nucleon. Obviously there are many other medium effects not 
taken into account in these schemes, as we shall see. Furthermore,
although in a different language and in terms of a few relevant physical 
magnitudes, such medium modifications appear in a systematic many body
expansion, where they can be classified as vertex corrections.

\subsection{Relativistic effects}

Theories using relativistic scalar and vector potentials like in the Walecka
model \cite{WAL} have been popular. The scalar and vector potentials are
 about one 
order of magnitude larger than the ordinary non relativistic potential
which is roughly the sum of the two. This cancellation is missed in many
applications of these relativistic potentials leading to unrealistic 
predictions. The appealing thing of the relativistic approach is the small 
effective mass, $M_N^* \simeq M_N/2$
 of the nucleon and the fact that the nucleon response
for $ph$ excitations is roughly proportional to $M_N^*$. This reduces 
the longitudinal response but also the transverse one.

A clarifying view of these problems is exposed in \cite{HOR} where the
necessity to go beyond the relativistic mean field approach is shown in
order to avoid the pathological predictions tied to the small effective 
masses, like in the
computation of the nuclear response functions or the large relativistic
enhancements that drive magnetic moments outside their Schmidt lines.
In \cite{HOR} a relativistic RPA calculation is used. Similar conclusions are
found in \cite{PRI}, stating that ``selfconsistent calculations'' 
show cancellations 
between large relativistic effects on the single nucleon current and on
the many nucleon wave functions. In \cite{GIM} it is also shown, by    
solving numerically the Dirac equation with the relativistic potential, that
the genuine relativistic effects in the enhancement of the axial charge amount
to 20-30$\%$, while perturbative calculations give as much as 70-80$\%$
enhancement \cite{RIS,GRA}.

Even when improved with some selfconsistent steps, relativistic calculations
still rely on the concept of the effective mass~\cite{HOR,HOD,SUZ},
 which is only an
 approximation to the richer content of the nucleon self-energy. 
 The nucleon self-energy is a function of the energy and momentum, as
independent variables, and
leads to important dynamical properties of the nucleus 
~\cite{MAH} not contained
in static pictures like the mean field theories. One of the consequences is
that the nucleon effective mass is a strongly dependent function of the 
energy, with a peak around the Fermi surface \cite{MAH}. Furthermore, as shown
in \cite{PAL}, RPA correlations tied to the pionic degrees of freedom are
 essential 
in order to provide the energy dependence of the nucleon self-energy and hence 
the dynamical properties of the nucleus. Actually it is quite interesting to
see that recent sophisticated calculations in light nuclei using path integral
Monte Carlo methods \cite{CARL} find that ``pion degrees of freedom in both 
nuclear interaction and currents play a crucial role in reproducing the
experimental data''.

From this discussion it looks clear that improvements along the relativistic
line for the present problem should include ``selfconsistency'' in the
sense of ref.~\cite{PRI}, in order to
exploit the large cancellations between the scalar and vector potentials.

\subsection{Pionic effects}

As mentioned above \cite{CARL}, the pionic degrees of freedom play an important
role in quasielastic electron scattering. This has also been emphasized 
in \cite{WAN} where meson exchange currents driven by pion exchange are
evaluated, putting special emphasis in fulfilling the continuity equation 
and preserving gauge invariance in the many body system. This imposition
has as a consequence some changes in the results with respect to former
works along similar lines \cite{MAG,ORD,MIS}

In ref. \cite{RIN} similar ideas, but using the formalism of path integrals is
followed. RPA correlations are automatically generated in that scheme leading
to some quenching of the longitudinal response from a reduction in the 
isoscalar channel.

Pions are also explicitly used in approaches which include meson exchange 
currents, as we shall see below. They are usually taken static, as in    
~\cite{WAN,RIN}, meaning that the energy carried by the pion is neglected.
While this is a fair approximation for the exchange currents at energies
below pion production threshold, at higher energies the need to work will the
full pion propagator becomes apparent. This is particularly true if one wishes
to account for real pion production in the same many body scheme, as we
shall do.

In the resonance region primary pion production accounts for the largest
part of the response function (although some of the pions are absorbed in
their way out of the nucleus and show up in $2N$ or $3N$ emission channels).
Hence, the explicit treatment of pionic degrees of freedom allowing pions
to be produced, both as virtual as well as real states, becomes a necessity
in this region. Our scheme puts a special emphasis on pions. In fact it 
follows a different path to other schemes, beginning with real pion 
production and ensuring that a proper hand on the $(e, e' \pi)$ reaction is
held. Then exchange currents and further corrections in the many body 
system are generated from the model for the $e N \rightarrow
e' N \pi$ reaction.

\subsection{Meson exchange currents}

A large fraction of work has been devoted to the role played by meson 
exchange currents (MEC) in this reaction~\cite{WAN,MAG,ORD,MIS,GCO,AMA,LAN}.
The standard seagull, pion in flight and $\Delta$ terms are included mediated
by pions. The indirect 
effect of short range correlations in these terms is, however, neglected in those
works. The work of \cite{TAI} incorporates terms in the scheme which account
for virtual photon absorption on correlated pairs, hence accounting for
ground state correlations. Although the same concepts are shared in the
 previous
approaches, differences in the input and the way to implement them lead
to different results.

Our approach differs from the quoted works although conceptually it is quite
similar. First we realize that the two body currents appear in $(e, e')$ as
corrections to the main one body contribution. This is because virtual
photons can be absorbed by one nucleon. This is opposite to the case of 
real photons which require at least two nucleons to be absorbed (we are
thinking in terms of infinite nuclear matter). Thus it is clear that the
laboratory to test the effect of two nucleon currents is real photons not
virtual ones. This is the reason why prior to the present work we devoted 
energies to the problem of real photon absorption \cite{CAR}. Second, 
in order to
minimize sources of uncertainties, the MEC were generated from the model
$\gamma N \rightarrow \pi N$
 by allowing the pion to be produced in a virtual state and be 
absorbed by a second nucleon. After this is done, long range correlations 
from polarization phenomena, as well as short range correlations, are taken
into account. The model for the $\gamma N \rightarrow \pi N$
 reactions was tested against experimental
 data and was found to be good. This gives one some confidence in 
the strength that one generates for the MEC. The reliability of the
method gets extra support from recent measurements of two body photon
absorption \cite{CRO,GRO,helh} where the agreement with the
$(\gamma, np)$ emission channels
is rather 
good. Some discrepancies remain in the $(\gamma, pp)$
channel, but this channel has
 an experimental cross section nearly one order of magnitude smaller than
the $(\gamma, np)$ one, thus for the purpose of the total 
$(\gamma, NN)$ emission or the two 
body MEC in $(e,e')$ such discrepancies will not play an important role.

In view of the success in the real photon case, we adopt here the same
scheme and follow the same steps simply substituting the real photon by
the virtual one. This means that we begin with a model for the 
$e N \rightarrow e' N \pi$ reaction
and construct the MEC from it following identical steps to those of ref.
~\cite{CAR}. For this reason we begin in next section by showing our model for
the $e N \rightarrow e' N\pi$
 reaction and contrasting it with the experimental results.

\subsection{RPA correlations}

Several works have emphasized the role played by RPA correlations allowing
for a $ph$ excitation which propagates in the nuclear medium mediated by 
some residual $ph$ interaction \cite{HOR,RIN,DRE,WAB}. These works share the
feature that
a reduction is produced in the longitudinal channel. We shall also incorporate
these long range correlations or polarization effects. In addition 
we shall also include $\Delta h$
 excitations, as a source for polarization which 
will be relevant in the transverse channel as shown in \cite{WAG}.

\subsection{Final state interaction (FSI)}

This is another topic which has received some attention and a thorough work
devoted to this issue can be seen in \cite{CHI}. It is clear that once a
$ph$ excitation is produced by the virtual photon, the outgoing nucleon can
collide many times, thus inducing the emission of other nucleons. A distorted
wave approximation with an optical (complex) nucleon nucleus potential
would remove all these events. However, if we want to evaluate the inclusive
$(e, e')$ cross section these events should be kept and one must sum over all 
open final state channels \cite{NIMAI,CHI}. This is done explicitly in 
~\cite{CHI}
and the result of it is a certain quenching of the quasielastic peak of the
simple $ph$ excitation calculation and a spreading of the strength to the
sides of it, or widening of the peak. The integrated strength over energies 
is not much affected though.

The use of correlated wave functions, evaluated from realistic $NN$ forces and
incorporating the effects of the nucleon force in the nucleon pairs  has
also been advocated in connection with the effects from final state
interaction \cite{FAN}. If one incorporates
two particle--two hole ($2p 2h$) components in the final excited 
states one gets the spreading of the peaks as found in~\cite{CHI}. For the 
purpose of the response function it is like exciting $ph$ components which
have a decay width into the $2p 2h$  channel. This gives rise to the
quenching of the peak and spreading of the strength. Another source of these
effects is the momentum dependence of the nucleon self-energy which is also
accounted for in the scheme and which sometimes is taken into account 
approximately in terms of an effective mass (although our position on 
evaluations which  use this variable  has been already
made clear). The approach of \cite{FAN} using an orthogonal correlated basis with
functions obtained with variational methods also incorporates RPA 
correlations discussed in point 5) but does not account for the MEC discussed
in point 4), which rely upon the coupling of the photon to the pion or the
$\Delta$ excitations.

In our many body scheme we will account for this FSI by using nucleon 
propagators properly dressed with a realistic self-energy in the medium,
which depends explicitly on the energy and the momentum \cite{PED}. This 
self-energy leads to nucleon spectral functions in good agreement with other
accurate more microscopic approaches like the ones  
in \cite{ANG,ART}. The self-energy
of \cite{PED}
has the proper energy--momentum dependence plus an imaginary part from
the coupling to the
$2p 2h$ components, hence it has the ingredients to account for the FSI 
effects discussed in \cite{FAN} although using a different calculational
scheme.

Nuclear spectral functions with a language closer to the one we shall 
follow
have also been used in \cite{CIO}. They include the interesting result,
which we will employ here, that keeping  the width of the particle
states is important but one can disregard the width of the hole states.

\subsection{$\Delta$ excitation}

While many efforts have been devoted to the quasielastic peak very little
attention has been given to the $\Delta$ region and the dip region between
the quasielastic and $\Delta$ peaks. One exception is the work of 
\cite{TJON}
which looks at the effects of MEC in the dip region. 

A recent work \cite{ANGH} presents some experimental results and a theoretical
analysis of the $\Delta$ region based on the 
$e N \rightarrow e' N \pi $ model of \cite{NOZ}. One interesting
conclusion of the work is that the data have a broader
energy spectrum than the theoretical one based on the properties of a 
free $\Delta$ width. This suggest a larger $\Delta$ width
 from $\Delta$ coupling to many body channels, 
additional to the natural decay width with effects of Fermi motion included.
This is actually well known from pion physics \cite{WOL,THI}.

Our aim in this section is also to evaluate as accurately as possible the
response function in this region, for which we shall use results for the
$\Delta$ in a nuclear medium \cite{LOR} which have been tested thoroughly in a 
variety of pionic reactions: elastic~\cite{CARMEN}, quasielastic, 
charge exchange, absorption, etc... \cite{MAN}.

The discussion in this sections has served to expose relevant works done
in the literature and to present our approach in connection to the ideas
exposed in these works. We shall follow a microscopic many body description
of the inclusive $(e, e')$ process and will incorporate in our approach the
important effects discussed in this section. We are thus aiming
at an evaluation as
accurate as possible at the present moment. A small sacrifice is made
in order to allow one to treat microscopically in a tractable way all
the effects
discussed above: we use an infinite nuclear medium and obtain results for
finite nuclei using the local density approximation (LDA). This was also
used for the evaluation of the total cross section of real photons with 
nuclei with good results \cite{CAR} and a mathematical derivation was made there
to justify the accuracy of the approximation. A direct comparison of results
with the LDA with finite nucleus results, using the same input was done
in \cite{EUG} for deep inelastic lepton scattering, showing that indeed the LDA
is an excellent approximation to deal with volume processes (i.e no 
screening or absorption effects). In \cite{ANT}
a comparison of the results for finite nuclei with those of the Fermi gas
around the quasielastic 
$ (e,e')$ peak is done, 
showing that for some average Fermi momentum, the 
results of the Fermi gas and those of finite nuclei are nearly identical,
proposing that choice as an even better prescription than the LDA. We
shall follow the LDA, already tested for real photons, and sufficiently 
good for our purposes.

Another feature which is novel in our approach and rather important
in practical terms is the  following: we shall use a method which
relates the $(e,e')$
cross section to the imaginary part of the virtual photon self-energy. This
is actually not new and has been used before~\cite{HOR,ALB}. The
novelty is that
by using properly Cutkosky rules one can relate the different sources of
imaginary part to different channels which contribute to the inclusive
$(e,e')$ cross section. In this way, we lay the grounds to evaluate from
the present input the cross sections for exclusive processes $(e, e'N)$,
$(e, e'NN)$, $(e, e' \pi) \,  (e, e' N \pi)$
 etc., which will be the subject of a forthcoming 
paper \cite{JUA}. The results of the present paper and those of \cite{JUA} 
are part
of a PhD thesis \cite{AMP} where additional details to those given here
can be found if desired.

\section{The $e N \rightarrow e' N \pi$ reaction}

\subsection{Formalism}

 We shall follow the model used in ref.~\cite{CAR} for 
$\gamma N \rightarrow \pi N$ at intermediate
energies generalizing it to virtual photons. In fact this model is 
essentially the same as used for the 
$e N \rightarrow e N \pi$  reaction in \cite{NOZ}. However, having
in mind the application to nuclei we make a reduction of the relativistic
amplitudes keeping terms up to 
$O (\frac{P}{M_N})$, with,
$P$,$M_N$ the nucleon momentum and mass. 
The neglect of the $O (\frac{P}{M_N})^2$ terms is justified 
numerically as we shall see in the results, so we construct this non
relativistic amplitude for the $e N \rightarrow
e N \pi$ process ready to be used with ordinary
non relativistic nucleon wave functions.

\vspace*{0.2cm}
\vskip 0.2cm
\centerline{\protect\hbox{\psfig{file=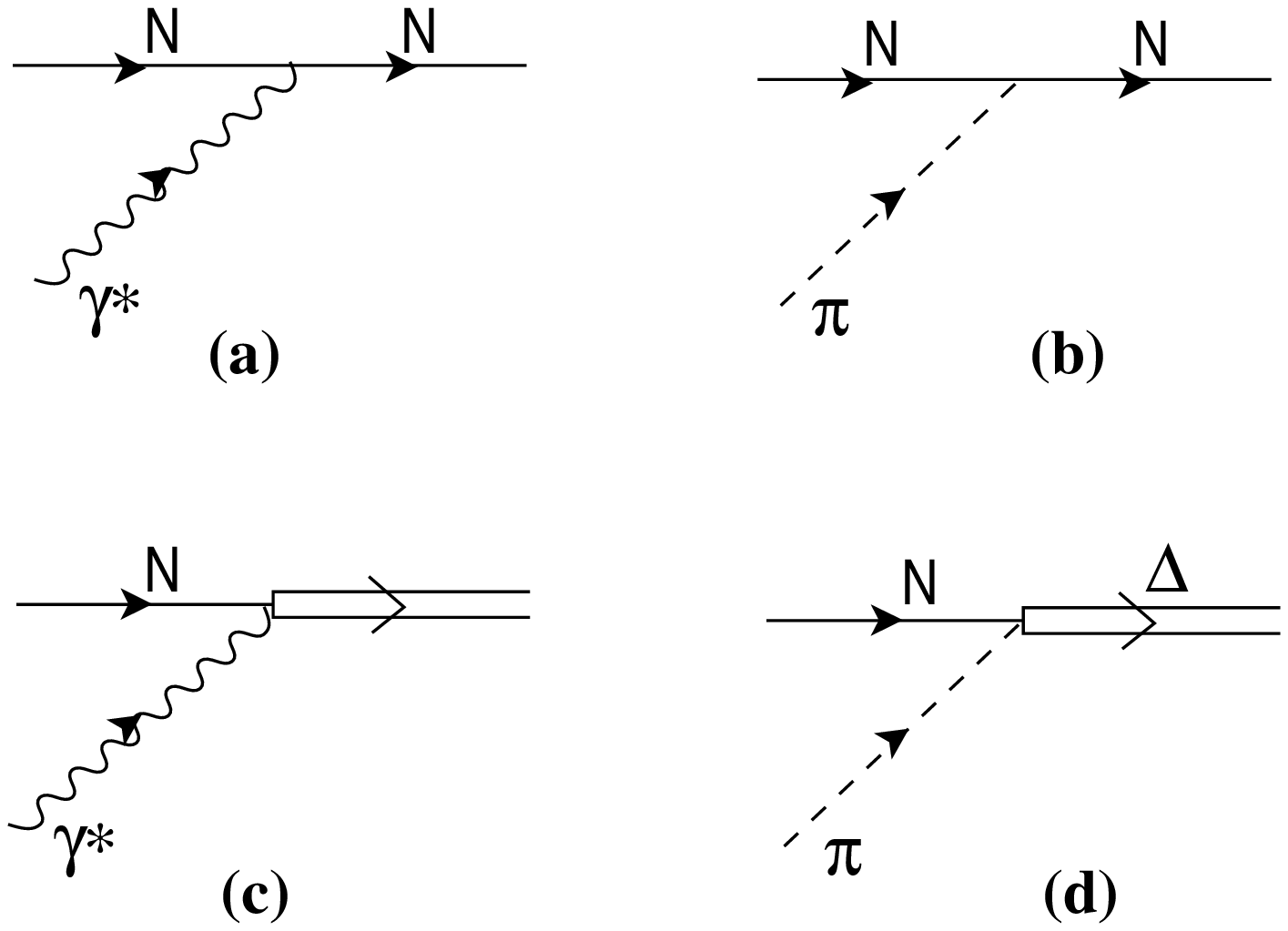,width=6.9cm}}}
\vskip 0.1cm
\noindent
{\small {\bf Fig.3.1} Basic couplings of the virtual photon and the pion
to the nucleon and to the $N \Delta$ transition.}
\vspace*{0.3cm}

The basic couplings which we need are those depicted in fig.3.1 which 
account for the coupling of the photon and the pion to the nucleon and 
to the  $N \Delta$ transition, plus the Kroll Ruderman term (KR), and the
coupling of the photon to the pion. The Kroll Ruderman term appears as a
gauge invariant term through minimal substitution when a pseudovector 
$\pi N N$
coupling is used, as we do. The analytical expressions for these vertices
are given in appendix A. For convenience, from the KR term of the appendix
coming from gauge invariance, and which we call there seagull term, we 
construct the KR term displayed in this chapter such that it contains all the 
non vanishing pieces of the amplitude when $p_\pi \rightarrow
0$. We will come back to this 
point later on.

\vspace*{1cm}

\subsection{The amplitudes for the $e N \rightarrow e N \pi$ model}

The Feynman diagrams which are considered in the model for 
$\gamma N \rightarrow \pi N$ of \cite{CAR}
or in $\gamma^*
N \rightarrow \pi N$ of \cite{NOZ} 
($\gamma^*$ will stand from now on for the virtual photon) are
depicted in fig. 3.2. 

\vskip 0.2cm
\centerline{\protect\hbox{\psfig{file=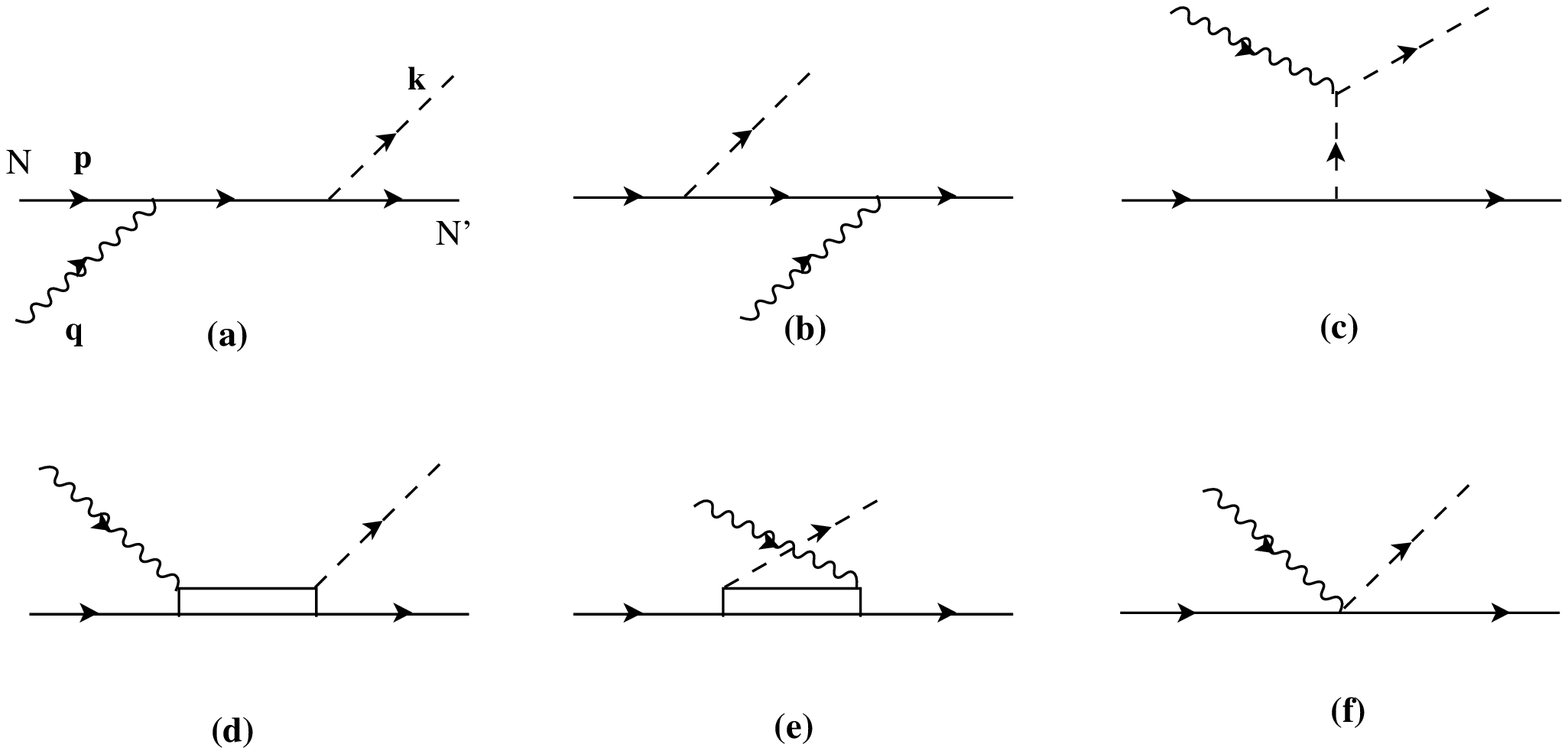,width=11.5cm}}}
\vskip 0.2cm
\noindent
{\small {\bf Fig.3.2} Feynman diagrams considered for the 
$\gamma^* N \rightarrow \pi N$ reaction. }
\vspace*{0.1cm}

They are the nucleon pole direct (NP) term (a), the nucleon
pole crossed (NPC) term (b), the pion pole (PP) term (c), 
the delta pole direct (DP) term (d),
delta pole crossed (DPC) term (e)  and Kroll Ruderman (KR) term (f). 
The expressions for
these amplitudes are obtained by doing the nonrelativistic reduction of
the relativistic amplitudes of \cite{NOZ}. There is some small 
contribution from 
the negative energy intermediate nucleon states which is kept in 
our expressions,
which as we mentioned neglect only terms of order 
$O (\frac{P}{M_{N}})^2$. The corresponding
expressions are:

\be
\begin{array}{rcl}
{\cal{M}}^{\mu}_{NP} & = &-e\frac{\displaystyle{f_{\pi N N}}}
{\displaystyle{m_{\pi}}}
B(N,N^{\prime} \pi)\frac{\displaystyle{1}}{\displaystyle{\sqrt{s}-M_{N}}}
F_{\pi}((q-k)^2)\times\\
& & \\
 & &\!\!\!\!\!\!\!\!\!\!\times\left(
\begin{array}{c}
F_{1}^{N}(q^{2})\vec{\sigma}\vec{k}\\
\\
F_{1}^{N}(q^{2})\vec{\sigma}\vec{k}\left[ \frac{\displaystyle{
2\vec{p}+\vec{q}}}{\displaystyle{2M_{N}}}\right]
+i\frac{\displaystyle{\vec{\sigma}\vec{k}}}{\displaystyle{2M_{N}}}(
\vec{\sigma}\times\vec{q}\,)G^{N}_{M}(q^{2})
\end{array}
\right)
\end{array}
\ee

\be\label{eq:KR}
{\cal{M}}^{\mu}_{KR} =e\frac{\displaystyle{f_{\pi N N}}}
{\displaystyle{m_{\pi}}}
B(N,N^{\prime} \pi)F_{A}(q^{2})C^{\mu}(\pi,N^{\prime})F_{\pi}((q-k)^2)
\ee

\noindent
where
$$
\begin{array}{lll}
\!\!\!\!\!\!\!\!\!\!\!\!\!\!\!\!\!\!\!\!\!\!\!C^{\mu}(\pi^{-}p)=
\left(
\begin{array}{c}
\frac{\displaystyle{\vec{\sigma}\vec{q}}}
{\displaystyle{2M_{N}}}\\
\\
\left(
1+\frac{\displaystyle{q^{0}}}{\displaystyle{2M_{N}}}\right)\vec{\sigma}
\end{array}
\right)
&\,\,\,\, ; &
C^{\mu}(\pi^{0} n)=0
\end{array}
$$
$$
\begin{array}{lll}
C^{\mu}(\pi^{+}n)=
\left(
\begin{array}{c}
\frac{\displaystyle{\vec{\sigma}\vec{q}}}
{\displaystyle{2M_{N}}}\\
\\
-\left(
1-\frac{\displaystyle{q^{0}}}{\displaystyle{2M_{N}}}\right)\vec{\sigma}
\end{array}
\right)
& ; &
C^{\mu}(\pi^{0}p)=
\left(
\begin{array}{c}
\frac{\displaystyle{\vec{\sigma}\vec{q}}}
{\displaystyle{M_{N}}}\\
\frac{\displaystyle{q^{0}}}{\displaystyle{M_{N}}}\vec{\sigma}
\end{array}
\right) 
\end{array}
$$

\be
{\cal{M}}^{\mu}_{PP} =-e_{\pi}\frac{\displaystyle{f_{\pi N N}}}
{\displaystyle{m_{\pi}}}
B(N,N^{\prime} \pi)\vec{\sigma}(\vec{k}-\vec{q}\,)F_{\gamma \pi \pi}(q^{2})
\frac{\displaystyle{(2k-q)^{\mu}}}{\displaystyle{(k-q)^{2}-m_{\pi}^{2}}}
F_{\pi}((q-k)^2)
\ee

\be
\begin{array}{lll}
{\cal{M}}^{\mu}_{NPC} = -e\frac{\displaystyle{f_{\pi N N}}}
{\displaystyle{m_{\pi}}}
B(N,N^{\prime} \pi)\frac{\displaystyle{1}}{\displaystyle{p^{0}-k^{0}-
E(\vec{p}-\vec{k})}}F_{\pi}((q-k)^2)
\times & &\\
& &
\\
 \times 
\left(
\begin{array}{c}
F_{1}^{N^{\prime}}(q^{2})\vec{\sigma}\vec{k}\\
F_{1}^{N^{\prime}}(q^{2})\left\{
\frac{\displaystyle{2\vec{p}+\vec{q}-2\vec{k}}}{\displaystyle{
2M_{N}}}\right\}\vec{\sigma}\vec{k}+G_{M}^{N^{\prime}}(q^{2})
i\frac{\displaystyle{
(\vec{\sigma}\times\vec{q})\vec{\sigma}\vec{k}}}{\displaystyle{2M_{N}}}
\end{array}\right)& &
\end{array}
\ee

\noindent
with
$$
\begin{array}{l}
B(n,n \pi^0)=-1\\
B(n,p \pi^-)=\sqrt{2}\\
B(p,p \pi^0)=1\\
B(p,n \pi^+)=\sqrt{2}
\end{array}
$$

\noindent
If we consider,

$$
\begin{array}{l}
I(\pi^{0})=I_{c}(\pi^{0})=2/3\\
I(\pi^{+})=-I_{c}(\pi^{+})=-\sqrt{2}/3\\
I(\pi^{-})=-I_{c}(\pi^{-})=\sqrt{2}/3
\end{array}
$$

\noindent
we get

\be
\begin{array}{rl}
{\cal{M}}^{\mu}_{DP}=&-if^{*}\frac{\displaystyle{f_{\gamma}(q^{2})}}
{\displaystyle{m_{\pi}^{2}}}\frac{\displaystyle{\vec{S}\left[
\vec{k}-\frac{k^{0}}{\sqrt{s}}\vec{p}_{\Delta}\right]}}
{\displaystyle{\sqrt{s}-M_{\Delta}+i\frac{\Gamma (s)}{2}}}
\Frac{\sqrt{s}}{M_{\Delta}}I(\pi)\times
\\
& \\
& \times
\left(
\begin{array}{c}
\Frac{\vec{p}_{\Delta}}{\sqrt{s}}\left(\vec{S}^{\dagger}
\times\vec{q}\right)\\
\\
\Frac{p^{0}_{\Delta}}{\sqrt{s}}\left[
\vec{S}^{\dagger}\times\left(\vec{q}-\Frac{q^{0}}{p^{0}_{\Delta}}\vec{p}_{\Delta}
\right)\right]
\end{array}\right) 
\end{array}
\ee

\be
\begin{array}{rl}
{\cal{M}}^{\mu}_{DPC}=&-if^{*}\frac{\displaystyle{f_{\gamma}(q^{2})}}
{\displaystyle{m_{\pi}^{2}}}\left(
\frac{\displaystyle{M_{N}}}{\displaystyle{M_{\Delta}}}\right)
\frac{\displaystyle{(E_{\Delta}+M_{\Delta})}}{\displaystyle{
(p^{2}_{\Delta}-M^{2}_{\Delta})}}I_{c}(\pi)\times
\\
& \\
& \times
\left(
\begin{array}{c}
0\\
(\vec{S}\times\vec{q}\,)(\vec{S}^{\dagger}\vec{k}\,)-\vec{B}
\end{array}
\right)
\end{array}
\ee

\noindent
with

$$
\vec{B}=\frac{1}{3}\left\{ i \frac{\displaystyle{(k^{0}+a)}}{\displaystyle{
E_{\Delta}+M_{\Delta}}}(\vec{q}^{\,2}\vec{\sigma}-({\vec{q}}\,\vec{\sigma})
\vec{q}\,)+\frac{\displaystyle{(k^{0}-a)}}{\displaystyle{2M_N}}
[({\vec{q}\,}\times\vec{p}\,)-i(\vec{q}\,\vec{p}\,)\vec{\sigma}
+i\vec{p}\,(\vec{\sigma}\vec{q}\,)]\right\}
$$

$$
a=(p_{\Delta}k)\frac{1}{M_{\Delta}}
$$

\newpage

\noindent
where $\sqrt{s}$ is the invariant energy of the system virtual photon-initial
nucleon, $e (e > 0)$ is the electron charge, $m_\pi=139.5$ MeV is the
pion mass,
$M_\Delta = 1238$ MeV, the $\Delta$ mass \cite{CAR}
and the free decay width of the $\Delta$ is given by

\be
\Gamma ({s})=
\frac{\displaystyle{1}}{\displaystyle{6\pi}}
\left(\frac{\displaystyle{f^*}}{\displaystyle{m_{\pi}}}\right)^2
\frac{\displaystyle{M_N}}{\displaystyle{\sqrt{s}}}
|\vec{k}_{cm}|^3\Theta(\sqrt{s}-M_{N}-m_{\pi})
\ee

On the other hand, $f_{\pi NN},
f_\gamma, f^*, F_1, G_M, F_A, F_{\gamma \pi \pi}, F_{\pi}$,
 are coupling constants and form factors which
we show in the appendix, as well as $\vec{S}$, the spin transition operator from
spin 3/2 to 1/2.

The expressions given keep the Lorentz covariance up to terms $O (\frac{
P}{m_{N}})^2$. The 
small $\Delta$ crossed term of eq.(6) holds strictly in the frame where
the outgoing nucleon is at rest. 

As mentioned before, in the KR term of eq.(~\ref{eq:KR}) we have
included corrections
in the zero component which come from the  zero component  of the NP
and NCP terms when using the vertex of eq.(~\ref{eq:piNN})
 (this means from positive 
energy intermediate states). On the other hand the $q^0/2M_N$
 terms in the spatial
components of the KR term come from the 
intermediate
negative  energy 
components of the NP and NCP relativistic amplitudes. This trick serves
us to concentrate on the KR term all the contributions which do not vanish
when the pion momentum goes to zero.

\subsection{Gauge invariance and form factors}

Gauge invariance implies that $q_\mu
{\cal{M}}^\mu = 0 $ where ${\cal{M}}^\mu = \sum_{i} {\cal{M}}^{\mu}_{i}$. 
From the expressions given above we can 
see that the delta terms are gauge invariant by themselves. As for the rest of  
the amplitudes they form a block of gauge invariance terms in the absence of 
form factors. However, we must impose some restrictions in order to keep 
gauge invariance when the form factors are included. If we consider the 
$\gamma^* n \rightarrow p \pi^-$
amplitude we see

\be
\begin{array}{c}
q_{\mu}{\cal{M}}^{\mu}\Frac{1}{e\frac{f_{\pi NN}}{m_{\pi}}B(N,N^{\prime}\pi)}= \
\\
=-\Frac{1}{p^{0}-k^{0}-E(\vec{p}-\vec{k})}
\left\{\left(
q_{0}F_{1}^{p}-\frac{\displaystyle{(2\vec{p}\,\vec{q}-
2\vec{k}\vec{q}+{\vec{q}\,}^{2})}}
{\displaystyle{2M_{N}}}F_{1}^{p}\right)\vec{\sigma}\vec{k}\right\}\\
\\
-\vec{\sigma}\vec{q}F_{A}-F_{\gamma\pi\pi}\vec{\sigma}(\vec{k}-\vec{q})=0
\end{array}
\ee

\noindent
which together imply

\be
\Rightarrow \left[F_{1}^{p}(q^{2})=F_{\gamma\pi\pi}(q^{2})=
F_{A}(q^{2})\right]
\ee

Gauge invariance in other isospin channels does not require extra
relationships. In the Appendix we can see that $\Lambda
\simeq M_A \simeq \sqrt{2} p_\pi$ and hence eq. (9)
is fulfilled to a good degree of approximation. However, in order to keep
strict gauge invariance we take only one form factor for all, which we
choose to be $F_1^p$ of eq.(~\ref{eq:f1p}). 
We have checked that by taking any of the other form
factors the changes induced in the cross sections are much smaller than 
the experimental errors (see fig.3.6), 
so we are rather safe with any of these choices.

On the other hand, since in the pion pole term we have a form factor 
corresponding to the coupling $\pi NN$ with a virtual pion,
$F_{\pi} ((q - k)^{2})$, we have included 
this form factor in the amplitudes of this block (NP,NCP,PP,KR) in order to
preserve gauge invariance.

\subsection{Unitarity}

Another refinement introduced in the model is unitarity. Watson's theorem
implies that the phase of the $\pi N \to  \pi N$
 and $ \pi N \to \gamma N $ amplitudes in each term of the 
partial wave decomposition must be the same. Such as our model stands, we
have a $\Delta$ term which by itself satisfies the theorem if we assume
the $\pi N \rightarrow \pi N$ 
 amplitude dominated by the $\Delta$, as it is the case. However, in 
the $\gamma N \rightarrow \pi N$
 amplitude we have a sizeable background that in our model is real
and the sum of the terms does not satisfy Watson's theorem. Although this
violation of unitary does not result in important numerical changes in the
cross section, we nevertheless unitarize the model as was done for real
photons \cite{CAR}. We follow the procedure of \cite{OLS} introducing 
a small phase
$\phi (\sqrt{s}, q^2)$  which corrects the $\Delta$ term, where 
$\sqrt{s}$  is the invariant energy 
of the virtual photon-initial nucleon system and $q^2$ the four--momentum 
squared of the virtual photon. By means of an iterative method we find
$\phi (\sqrt{s}, q^2)$ such that

 \be
   Im\left[ (T_{\Delta}(q^2)e^{i\phi (\sqrt{s}\,,\,q^2)}
     + T_{B}(q^2))^{(3/2,3/2)}e^{-i\delta_{(3/2,3/2)}(q^2) }\right]=0
 \ee

\noindent
where $T_\Delta(q^2)$ represents the $\Delta$ pole direct term amplitude, 
 $T_B(q^2)$ is the 
contribution of the rest of the terms to the (3/2, 3/2)
channel (see details in 
~\cite{CAR} for the projection of these terms in the  (3/2, 3/2) channel)
and $\delta_{(3/2,3/2)}(q^2)$  are the $\pi N$ phase shifts in the 3/2,3/2
spin-isospin channel.

\vspace*{0.4cm}

\subsection{Cross sections for the $e N \rightarrow e N \pi$ process}

We follow here the steps and nomenclature of ref.~\cite{AML}. In fig.3.3 we
show diagrammatically the process with the different variables which we use,
$k_e, k'_e, q, p, p'$
and $k$ representing the momenta of the 
incoming electron, outgoing electron, virtual photon,
initial nucleon, final nucleon and
pion respectively.

The unpolarized cross section is given by \cite{AML}:

\be
 \frac{\displaystyle{d^{2}\sigma}}{\displaystyle{d{\Omega}^{\prime}_{e}
  d{E}^{\prime}_{e}}}=\frac{\displaystyle{\alpha^{2}}}{\displaystyle{q^{4}}}
   \frac{\displaystyle{|\vec{k}_{e}^{\,\prime}|}}{\displaystyle{|\vec{k}_{e}}|}
    L_{\mu\nu}(e,e^{\prime}) \bar{W}^{\mu\nu}_{e.m.}
     \ee

 \vskip 0.2cm
   \centerline{\protect\hbox{\psfig{file=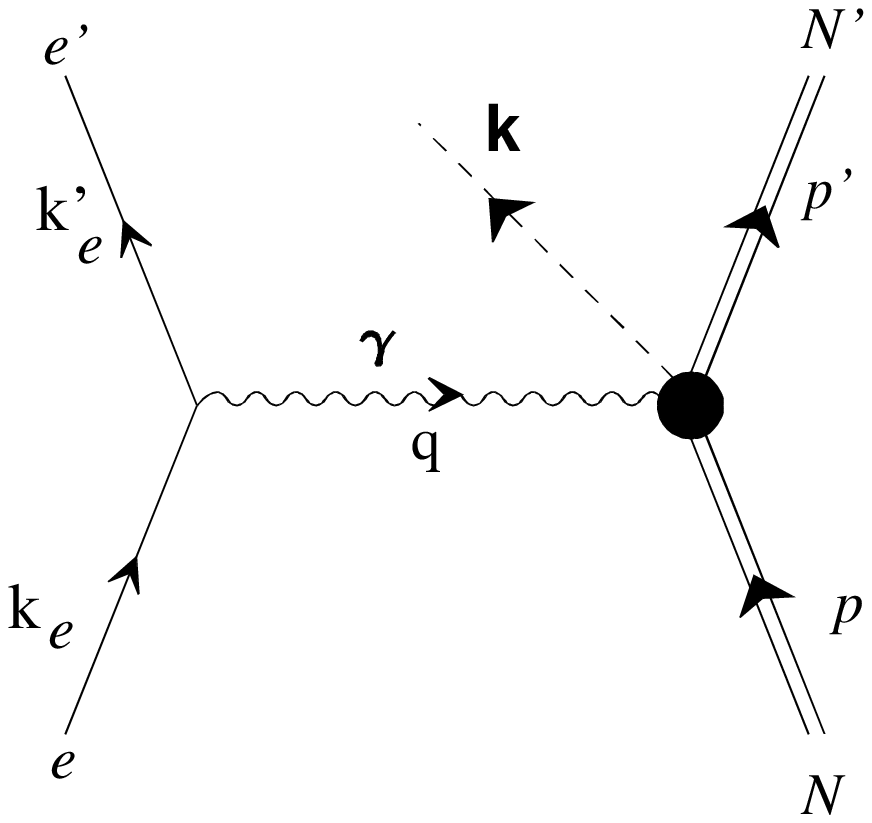,width=8cm}}}
      \vskip 0.2cm
\noindent      
{\small {\bf Fig.3.3} Feynman diagram for the 
$e N \rightarrow e ^{\prime} N \pi$ process. }
\vspace*{0.8cm}
          
\noindent
where $\alpha = 1/137.036$  is the fine structure constant $(e^2/4 \pi)$
and $L_{\mu \nu}$  the leptonic
tensor defined as

 \be
  L_{\mu\nu}(e,e^{\prime})=
   2({k_{e}}_{\mu}^{\prime}{k_{e}}_{\nu}+{k_{e}}_{\nu}^{\prime}{k_{e}}_{\mu}+
    \frac{\displaystyle{q^2}}{\displaystyle{2}}g_{\mu\nu})
     \ee

The hadronic tensor is given by

\be                           
\begin{array}{ll}
\bar{W}^{\mu \nu}_{em}=& \overline{\displaystyle \sum_{{}_{spin}}}
\displaystyle{\int}
\frac{\displaystyle{d^{3}p_{N}^{\prime}}}{\displaystyle{(2\pi)^{3}}}
\frac{\displaystyle{M_N}}{\displaystyle{E^{\prime}}}\int
\frac{\displaystyle{d^{3}k }}{\displaystyle{(2\pi)^{3}}}
\frac{\displaystyle{1}}{\displaystyle{2E_{\pi}}}(2\pi)^{3}\times
\\
&\\
&\!\!\!\!\!\!\!\!\!\!\!\!
\times \delta^{4}(p_{N}^{\prime}\!+\!k\!-\!p_{N}\!-\!q)
 <N^{\prime} \pi | j^{\mu}_{em}|N>^{*} <N^{\prime}
  \pi | j_{em}^{\nu}
   | N>
    \end{array}
\ee
     
\noindent
with $j^{\mu}_{em}$ the $\gamma^* N$ to $N^{\prime} \pi $ amplitude defined
 in sect. 3.2.

We can separate from there the angular dependence of the pion and get

\be
  \frac{\displaystyle{d^{3}\sigma}}{\displaystyle{d{\Omega}^{\prime}_{e}
   dE^{\prime}_{e}d\Omega_{\pi}}}=
    \frac{\displaystyle{\alpha^{2}}}{\displaystyle{q^{4}}}
     \frac{\displaystyle{|\vec{k}_{e}^{\,\prime}|}}{\displaystyle{|\vec{k}_{e}|}}
      L_{\mu\nu}(e,e^{\prime}) {W}^{\mu\nu}_{e.m.}(N)
       \ee

\noindent
where now
 
\be
\begin{array}{ll}
{W}^{\mu \nu}_{em}=&\overline{\displaystyle \sum_{{}_{spin}}}
\displaystyle{\int}
\frac{\displaystyle{d^{3}p_{N}^{\prime}}}{\displaystyle{(2\pi)^{3}}}
\frac{\displaystyle{M_N}}{\displaystyle{E^{\prime}}}\int
\frac{\displaystyle{d k\,{\vec{k}^{2}}}}
{\displaystyle{(2\pi)^{3}2E_{\pi}}}
(2\pi)^{3}\times\\
 &\\
   &\!\!\!\!\!\!\!\!\!\!\!\!\!\!\! \times \delta^{4}
     (p_{N}^{\prime}\!+\!k\!-\!p_{N}\!-q)
       <N^{\prime} \pi | j^{\mu}_{em}|N>^{*}
        <N^{\prime} \pi | j_{em}^{\nu}
         | N>
          \end{array}
           \ee

Now by taking $\vec{q}$ along the $z$ direction, using gauge invariance and the
explicit expressions for $L_{\mu \nu}$
 we can write, following exactly the same steps 
as in  \cite{AML,NOZ}

\newpage

\be
\begin{array}{ll}
\frac{\displaystyle{d^{5}\sigma}}{\displaystyle{d\Omega_{e}^{\prime}
dE_{e}^{\prime} d\Omega^{*}_{\pi}}}= & \Gamma \left\{
\frac{\displaystyle{d\sigma_{T}}}{\displaystyle{d\Omega^{*}_{\pi}}}
+{\cal \epsilon}
\frac{\displaystyle{d\sigma_{L}}}{\displaystyle{d\Omega^{*}_{\pi}}}
+{\cal \epsilon}
\frac{\displaystyle{d\sigma_{p}}}{\displaystyle{d\Omega^{*}_{\pi}}}
cos2\Phi^{*}_{\pi}+ \right.\\
&\\
& \left. +\sqrt{\displaystyle{2{\cal \epsilon}(1+{\cal \epsilon})}}
\frac{\displaystyle{d\sigma_{I}}}{\displaystyle{d\Omega^{*}_{\pi}}}
cos\Phi^{*}_{\pi}\right\}
\end{array}
\ee
\noindent

where

$$
\epsilon=\left[ 1- 2\frac{\displaystyle{|\vec{q}\,|^{2}}}{\displaystyle{
q^{2}}}tg^{2}\left(\frac{\displaystyle{\theta_e}}{\displaystyle{2}}
\right)\right]^{-1}
$$

\be
\begin{array}{l}
\Gamma=\frac{\displaystyle{\alpha}}{\displaystyle{2\pi^{2}}}
\frac{\displaystyle{|\vec{k}_{e}^{\,\prime} |}}
{\displaystyle{|\vec{k}_{e} |}}
\left[-\frac{\displaystyle{1}}{\displaystyle{q^{2}}}\right]
\frac{\displaystyle{k_{\gamma}}}{\displaystyle{1-{\cal \epsilon}}}\\
\\
k_{\gamma}=\frac{\displaystyle{s-M_{N}^{2}}}{\displaystyle{2M_{N}}}
\end{array}
\ee

\noindent
and
\be
\begin{array}{l}
\frac{\displaystyle{d\sigma_{T}}}{\displaystyle{d\Omega^{*}_{\pi}}}
=\frac{\displaystyle{e^{2}}}{\displaystyle{64\pi^{2}k_{\gamma}}}
\frac{\displaystyle{M_{N}}}{\displaystyle{\sqrt{s}}}
|\vec{k}_{CM}|\left[(J^{xx}+J^{yy})\right]^{CM}_{\Phi^{*}_{\pi}=0}\\
\\
\frac{\displaystyle{d\sigma_{L}}}{\displaystyle{d\Omega^{*}_{\pi}}}
=\frac{\displaystyle{-q^{2}}}{\displaystyle{(q^{0}_{cm})^{2}}}
\frac{\displaystyle{e^{2}}}{\displaystyle{32\pi^{2}k_{\gamma}}}
\frac{\displaystyle{M_{N}}}{\displaystyle{\sqrt{s}}}|\vec{k}_{CM}|
\left[ J^{zz}_{CM}\right]_{\Phi^{*}_{\pi}=0}
\\
\\
\frac{\displaystyle{d\sigma_{p}}}{\displaystyle{d\Omega^{*}_{\pi}}}
=\frac{\displaystyle{e^{2}}}{\displaystyle{64\pi^{2}k_{\gamma}}}
\frac{\displaystyle{M_{N}}}{\displaystyle{\sqrt{s}}}|\vec{k}_{CM}|
\left[(J^{xx}-J^{yy})\right]^{CM}_{\Phi^{*}_{\pi}=0}
\\
\\
\frac{\displaystyle{d\sigma_{I}}}{\displaystyle{d\Omega^{*}_{\pi}}}=
-\sqrt{\displaystyle{\frac{\displaystyle{-q^{2}}}
{\displaystyle{(q^{0}_{cm})^{2}}}}}
\frac{\displaystyle{e^{2}}}{\displaystyle{64\pi^{2}k_{\gamma}}}
\frac{\displaystyle{M_{N}}}{\displaystyle{\sqrt{s}}}
|\vec{k}_{CM}|\left[(J^{zx}+J^{xz})\right]^{CM}_{\Phi^{*}_{\pi}=0}
\end{array}
\ee

\noindent
where the variables of the electron are in the lab frame while those of
the pion are in the $\gamma^* N$ CM frame. $J^{\mu \nu}$
in the former expressions is given
by

\be
J^{\mu\nu}=Tr(j^{\dagger\mu}_
{em}j^{\nu}_{em})
\ee

\noindent

The angular variables $\theta_e, \theta^{*}_{\pi}, \Phi^{*}_\pi$
 are depicted in fig. 3.4. The variables $k_e, k'_e$ determine the
$(e, e')$ reaction plane and $\vec{q}$, the virtual photon
momentum, determines the $z$ direction. The $\vec{q}$ direction and the pion
momentum $\vec{k}$ determine the $\pi N$ reaction plane, and the angle
between this plane and the ($e, e')$ plane is the
angle $\Phi^{*}_\pi $.

The normalization of $d \sigma_i/ d \Omega^*_\pi$
 is chosen in such a way that in the limit
of real photons $d \sigma_i/ d \Omega^*_\pi$  coincides with the unpolarized 
cross section of $\gamma N \rightarrow \pi N$ 
with real photons. In this case the variable $k_\gamma$
 becomes the lab momentum
of the real photon. The variables $\sigma_T, \sigma_L,
\sigma_p, \sigma_I$  are the so called transverse, 
longitudinal, polarization and interference cross sections, respectively.

\vskip 0.2cm
\centerline{\protect\hbox{\psfig{file=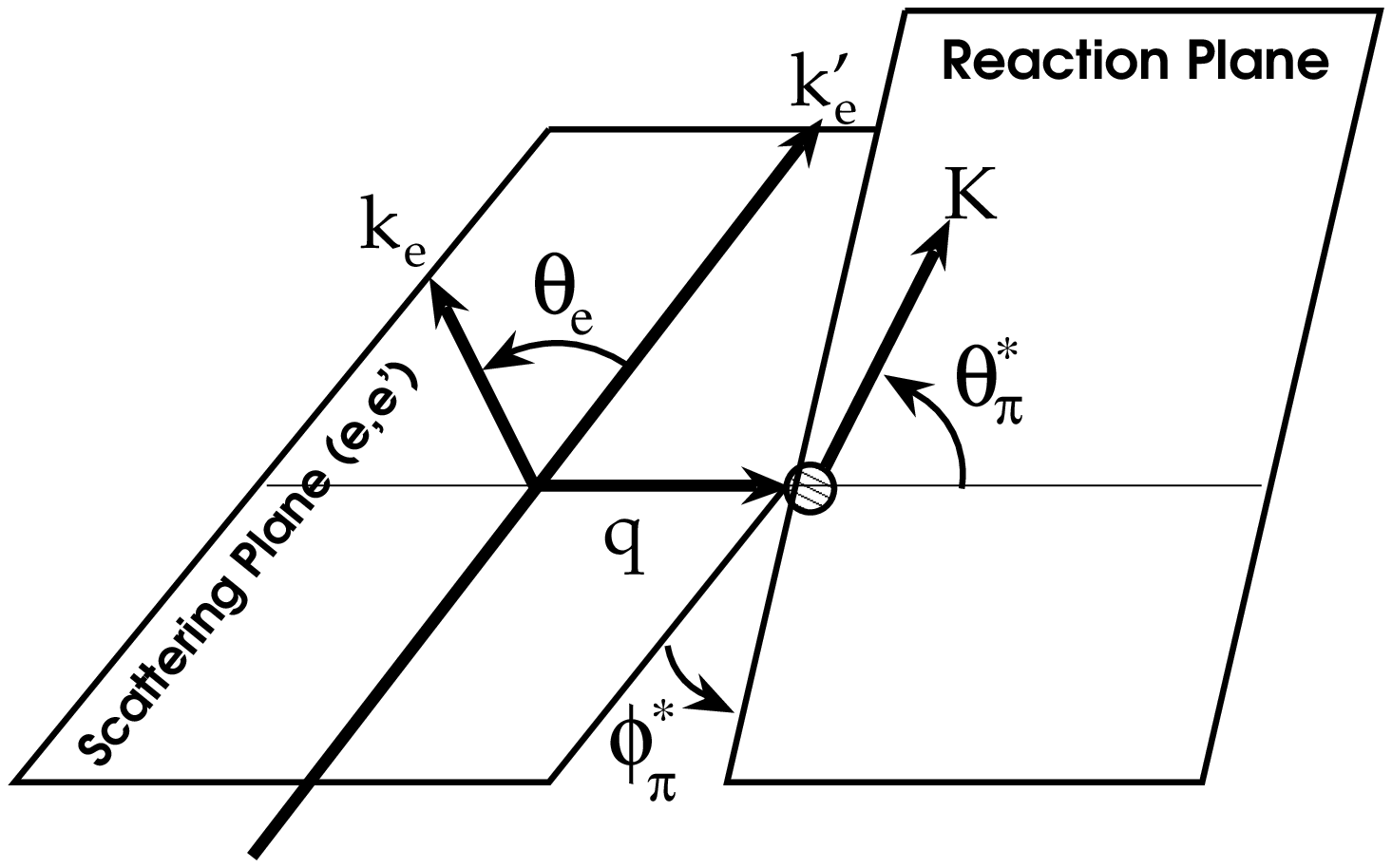,width=11cm}}}
\vskip 0.2cm
\noindent
{\small {\bf Fig.3.4} Angular variables
$\theta_e, \theta^{*}_{\pi}, \Phi^{*}_\pi$ in the
$e N \rightarrow e^{\prime} N \pi$ process.}
\vspace*{0.1cm}

\vspace*{1cm}

\subsection{Results for $e N \rightarrow e N \pi$}

In fig. 3.5 we show the $W=\sqrt{s}$ dependence of the longitudinal (lower line)
and 
transverse (upper line) cross 
sections summing over the final charge of the pion and integrating over
$\Omega_\pi^*$.
The peak position, the strength and the shape of our results are in good 
agreement with the data of  \cite{BAE}. In general terms our results are very 
close to those obtained in \cite{NOZ} with a similar good agreement with 
experiment as found there.

In fig. 3.6, and as an illustration of our discussion about gauge invariance
in sect. 3.3, 
we show the  differences between the results
  for the inclusive cross sections
$\sigma_T$ and $\sigma_L$,  obtained:
 by using  $F_A$ and taking 
(as gauge invariance forces)
$F_{1}^{p}=F_{\gamma \pi \pi}=F_A$ (dotted lines);
 by using $F_{\gamma \pi \pi}$ and taking 
 $F_{1}^{p}=F_A=F_{\gamma \pi \pi}$ (dash lines);
 by using $F_{1}^{p}$ and taking
  $F_A=F_{\gamma \pi \pi}=F_{1}^{p}$ (full lines) and,
  finally, by using $F_A$, $F_{\gamma \pi \pi}$ and $F_{1}^{p}$ 
  (dash-dotted lines). As one can see, the differences are negligible,
  relative to present experimental errors.
  
In fig. 3.7 we show the results for 
$d \sigma_I / d \Omega^{*}_{\pi} / (\sin \theta_\pi^{*} \sqrt{2})$
 in the channel $e p \rightarrow e n \pi^+$  and compare
them with the experimental data of \cite{BRE}. We can see that the agreement
is reasonably good and so is the case for 
$ d \sigma_P / d \Omega_{\pi}^{*} / \sin^2\theta_\pi^{*} $
shown in fig. 3.8, where the
data are again from \cite{BRE}.

In fig. 3.9 we show the $\Phi_{\pi}^{*}$ dependence of the cross section
in the channel
$e p \rightarrow e n \pi^+$
 for three 
different kinematics. The data are from~\cite{BRE} and we see again a
reasonable agreement with experiment.

Finally in fig. 3.10 we show the $\Phi_\pi^{*}$ dependence for another channel, the
$e p \rightarrow e p \pi^0$,
 in order to show a case where the agreement with the data, in this 
case from ref.~\cite{MIST}, is not as good as in general terms. Given the fact
that the angle $\Phi_\pi^{*}$ will be integrated in the $(e,e')$ 
reactions in nuclei,
such punctual discrepancies will not  matter in our study of the
nuclear processes.

\newpage
\vspace*{2.cm}
\centerline{\protect\hbox{\psfig{file=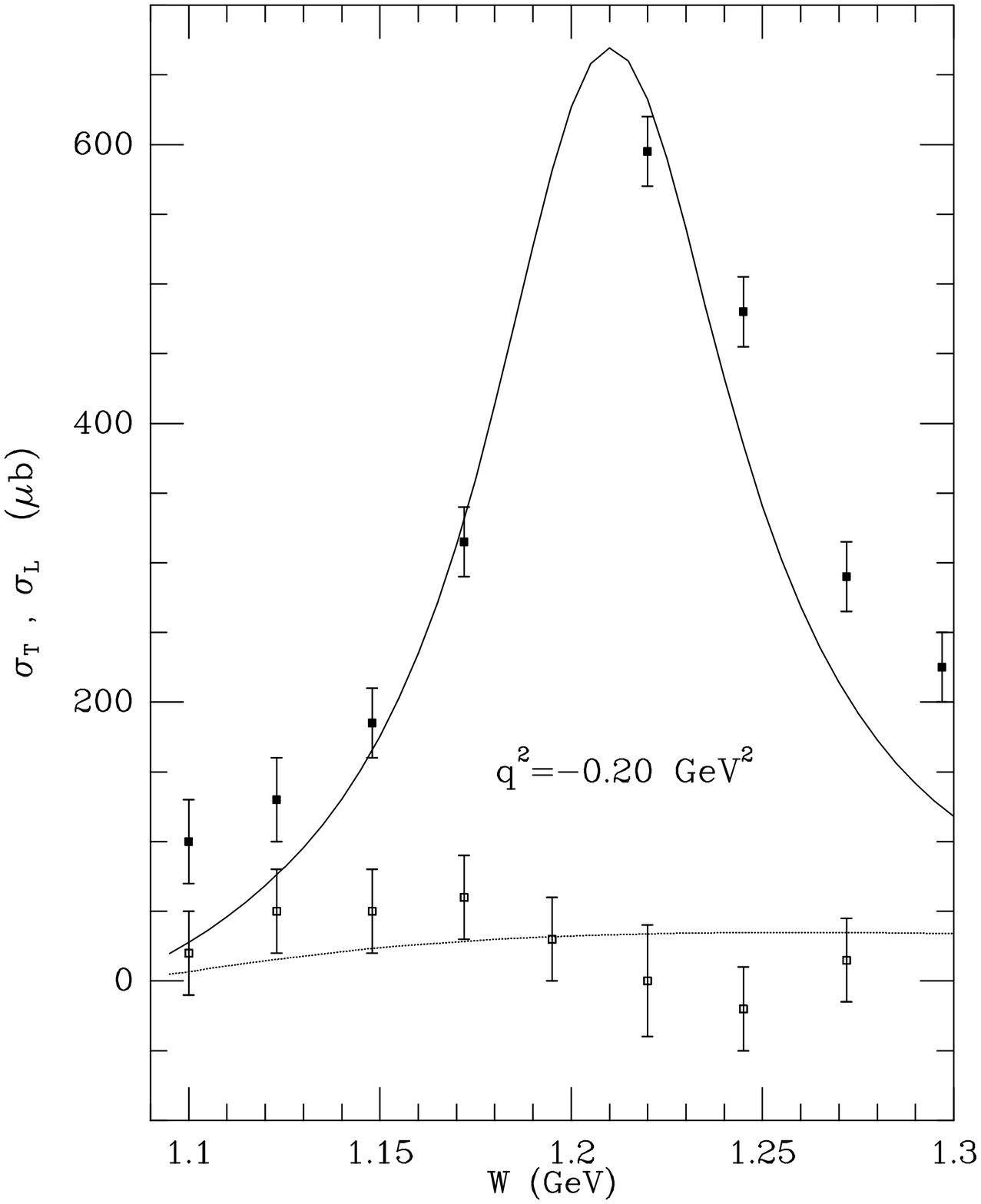,width=11.cm}}}
\vskip 0.2cm
\noindent
{\small {\bf Fig.3.5} 
$W=\sqrt{s}$ dependence of the longitudinal (lower line)
and
transverse (upper line) cross
sections summing over the final charge of the pion and integrating over
$\Omega_\pi^*$. Experimental data from \cite{BAE}.
  }

\newpage
\vspace*{2.cm}
\centerline{\protect\hbox{\psfig{file=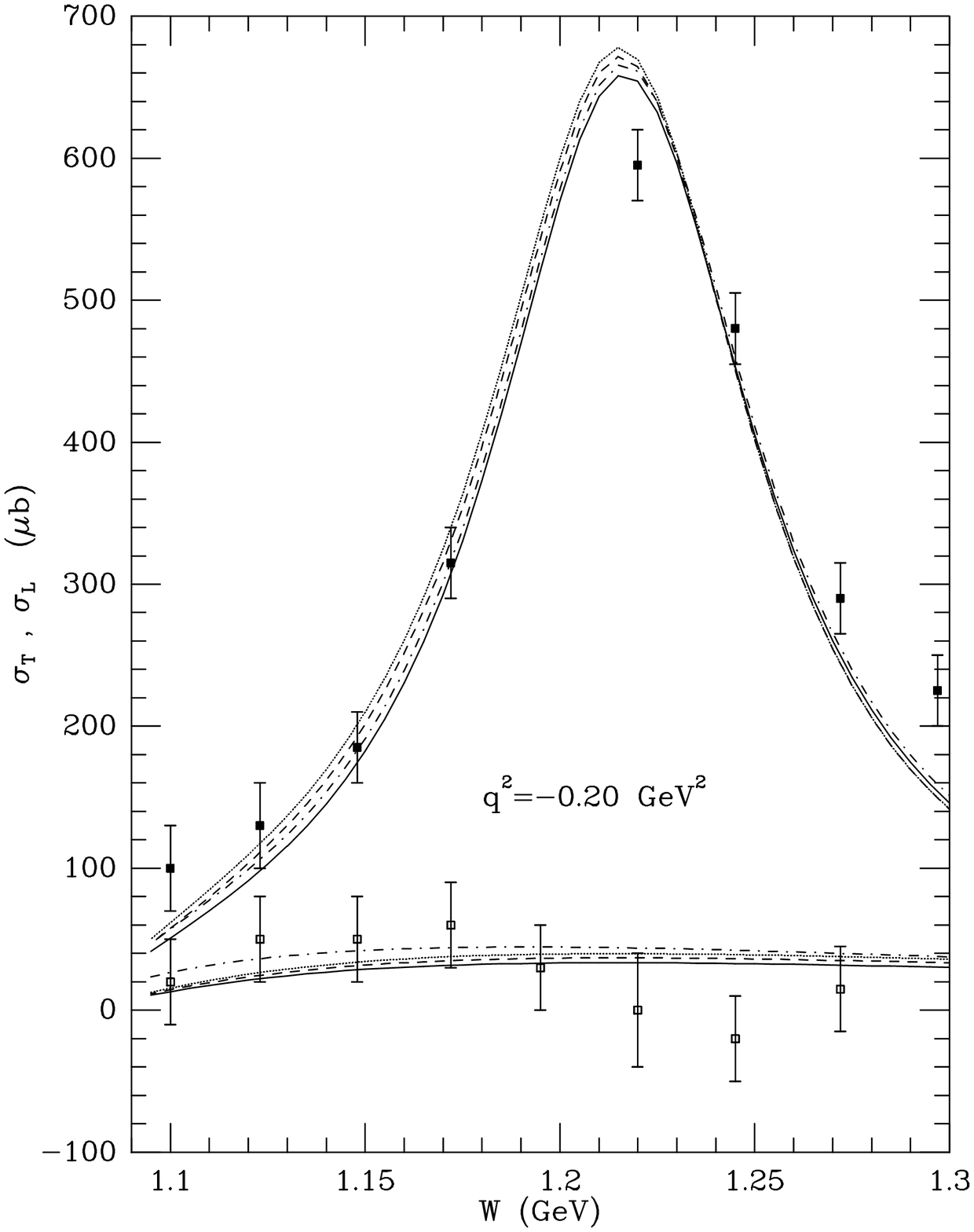,width=11.cm}}}
\vskip 0.2cm
\noindent
{\small {\bf Fig.3.6}
 Inclusive cross sections
   $\sigma_T$ and $\sigma_L$  obtained:
    by using  $F_A$ and taking
    $F_{1}^{p}=F_{\gamma \pi \pi}=F_A$ (dotted lines);
     by using $F_{\gamma \pi \pi}$ and taking
      $F_{1}^{p}=F_A=F_{\gamma \pi \pi}$ (dash lines);
       by using $F_{1}^{p}$ and taking
         $F_A=F_{\gamma \pi \pi}=F_{1}^{p}$ (full lines) and,
           finally, by using $F_A$, $F_{\gamma \pi \pi}$ and $F_{1}^{p}$
             (dash-dotted lines). Experimental data from \cite{BAE}.}

\newpage
%\vskip 1cm
\vspace*{0.8cm}
\centerline{\protect\hbox{\psfig{file=sigmai.ps,width=12.cm}}}
\vskip 0.2cm
\noindent
{\small {\bf Fig.3.7} 
Calculation of 
$d \sigma_I / d \Omega^{*}_{\pi} / (\sin \theta_\pi^{*} \sqrt{2})$
 in the $e p \rightarrow e n \pi^+$ channel . Experimental data from 
 \cite{BRE}.}

%\newpage
\vspace*{0.8cm}
   \centerline{\protect\hbox{\psfig{file=sigmap.ps,width=10.5cm}}}
      \vskip 0.2cm
         \noindent
{\small {\bf Fig.3.8} 
Calculation of $ d \sigma_P / d \Omega_{\pi}^{*} / \sin^2\theta_\pi^{*} $
in the  $e p \rightarrow e n \pi^+$ channel. Experimental data from
 \cite{BRE}.}

\begin{minipage}[c]{14cm}
\begin{minipage}[c]{6.8cm}
\centerline{\protect\hbox{\psfig{file=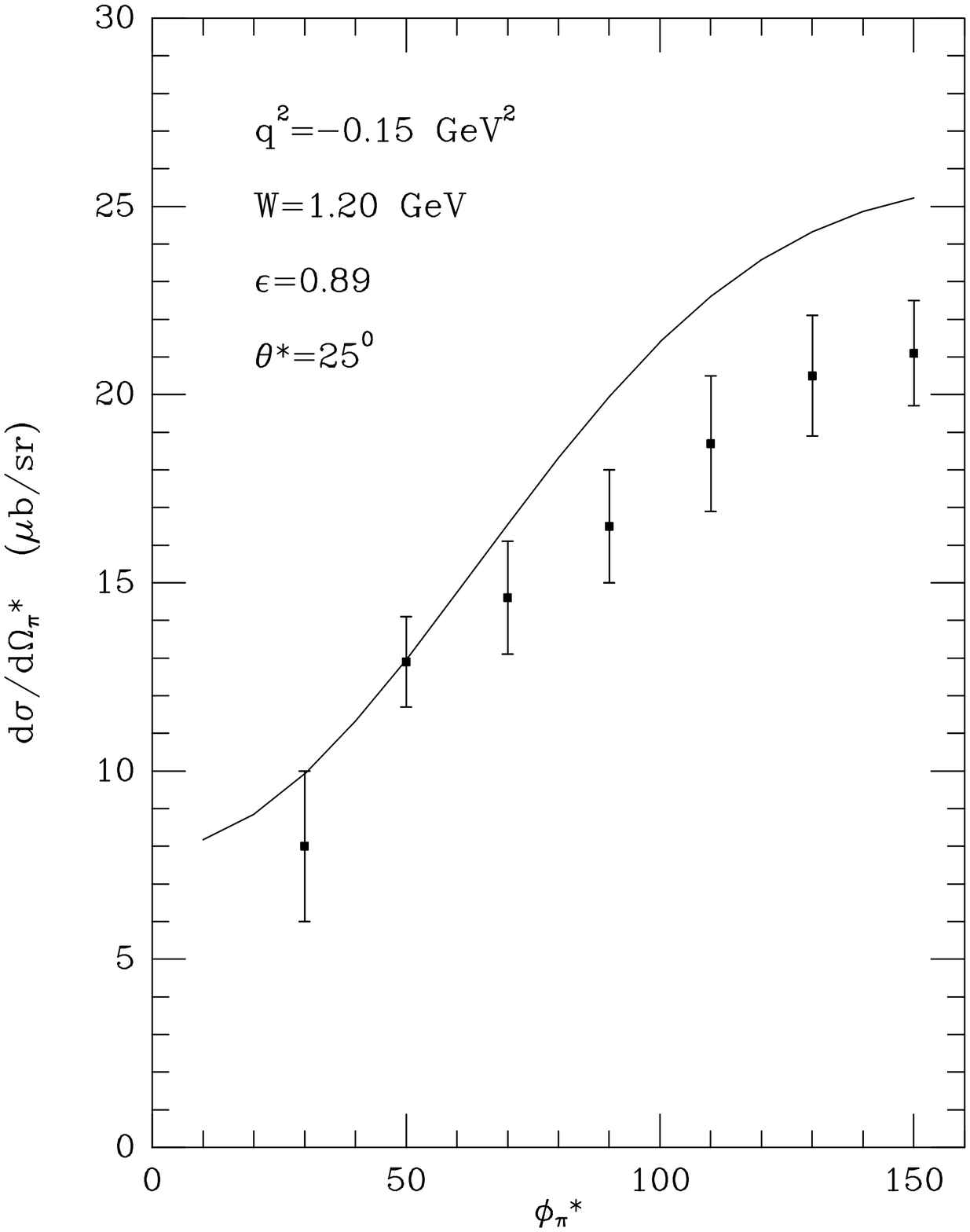,width=6.2cm}}}
\end{minipage}
\begin{minipage}[c]{6.8cm}
\centerline{\protect\hbox{\psfig{file=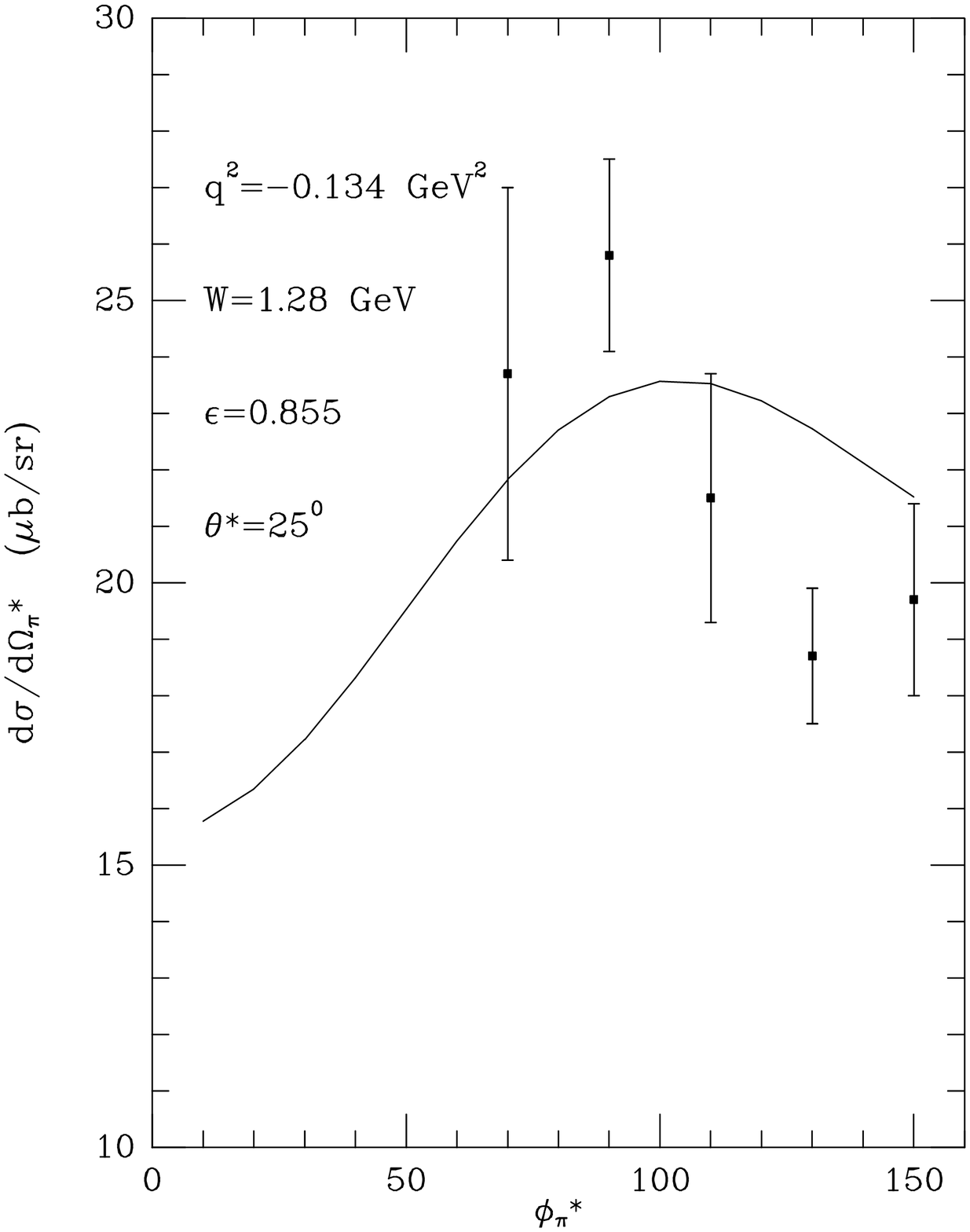,width=6.2cm}}}
\end{minipage}
\end{minipage}

\vspace*{0.6cm}
\centerline{\protect\hbox{\psfig{file=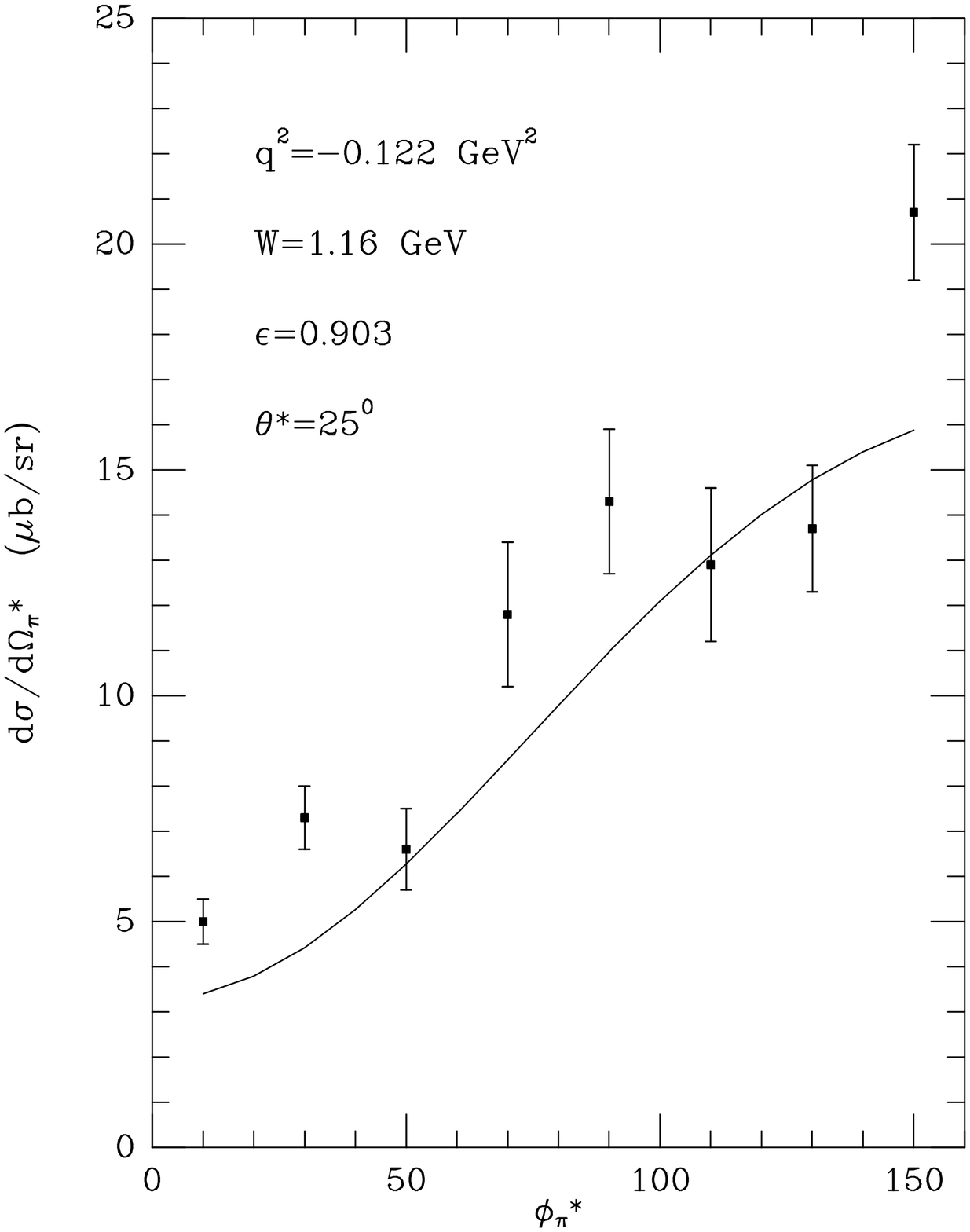,width=6.5cm}}}
\vspace*{0.3cm}
\noindent
{\small {\bf Fig.3.9} 
Calculation of the $\Phi^{*}_{\pi}$ dependence of the
cross section in the $e p \rightarrow e n \pi^+$ channel. 
Experimental data from \cite{BRE}.}

\newpage
\vskip 0.5cm
\centerline{\protect\hbox{\psfig{file=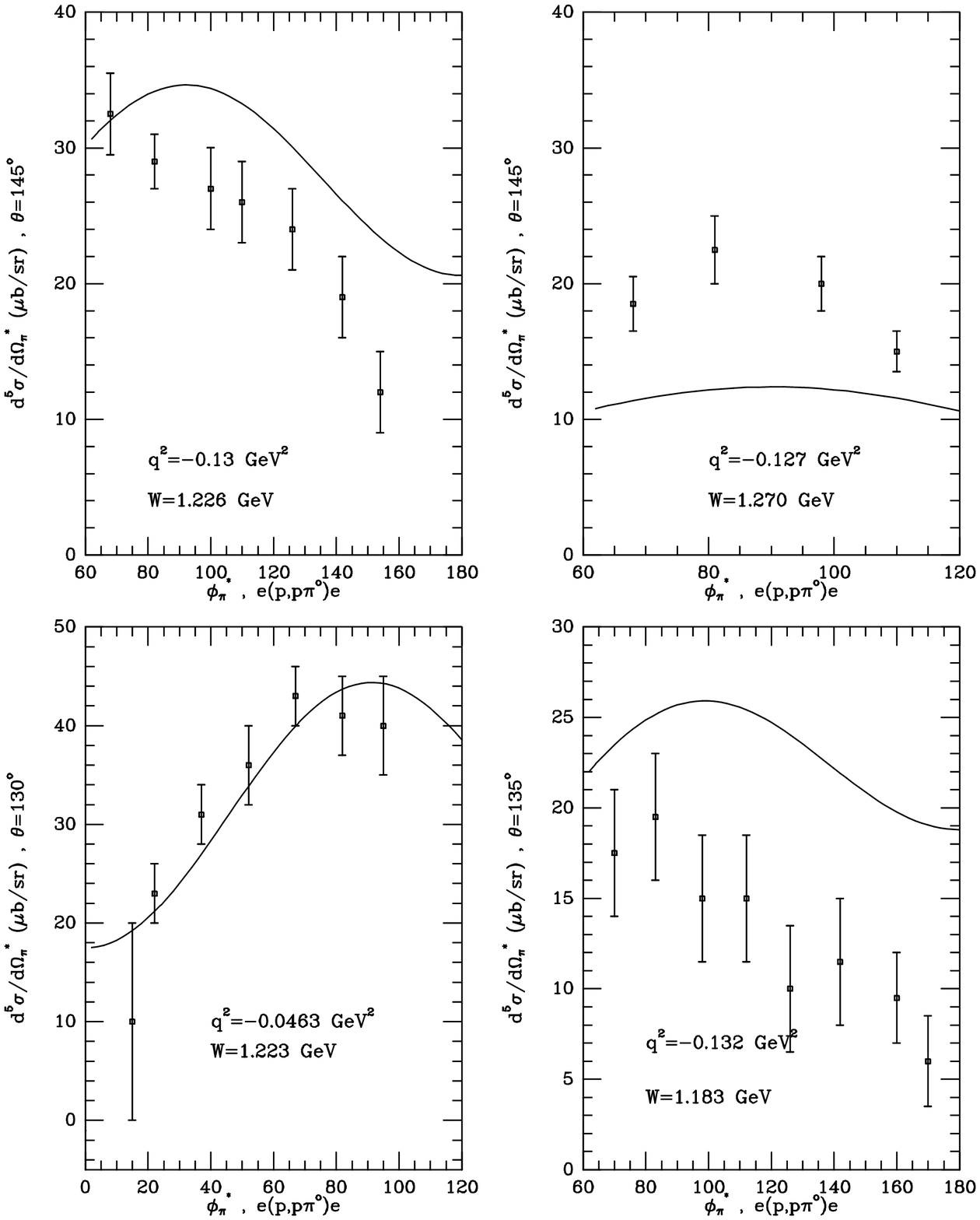,width=12.5cm}}}
\noindent
{\small {\bf Fig.3.10} Calculation of the $\Phi^{*}_{\pi}$ dependence of the
cross section in the 
$e p \rightarrow e n \pi^0$ channel.
Experimental data from \cite{MIST}. }
\vspace*{0.1cm}

In the next sections we shall use this model to evaluate the pion 
production contribution to the $(e,e')$ cross section, as well as the exchange
currents which contribute to the  $2 N$ emission channel.

\section{The $(e, e')$ reaction in nuclei.}

\subsection{Formalism}

We want to use a covariant many body formalism to evaluate
the $(e, e')$ cross section. For this purpose we evaluate the electron
self-energy for an electron moving in infinite nuclear matter. 
Diagrammatically this is depicted in fig. 4.1 

\vskip 0.2cm
\centerline{\protect\hbox{\psfig{file=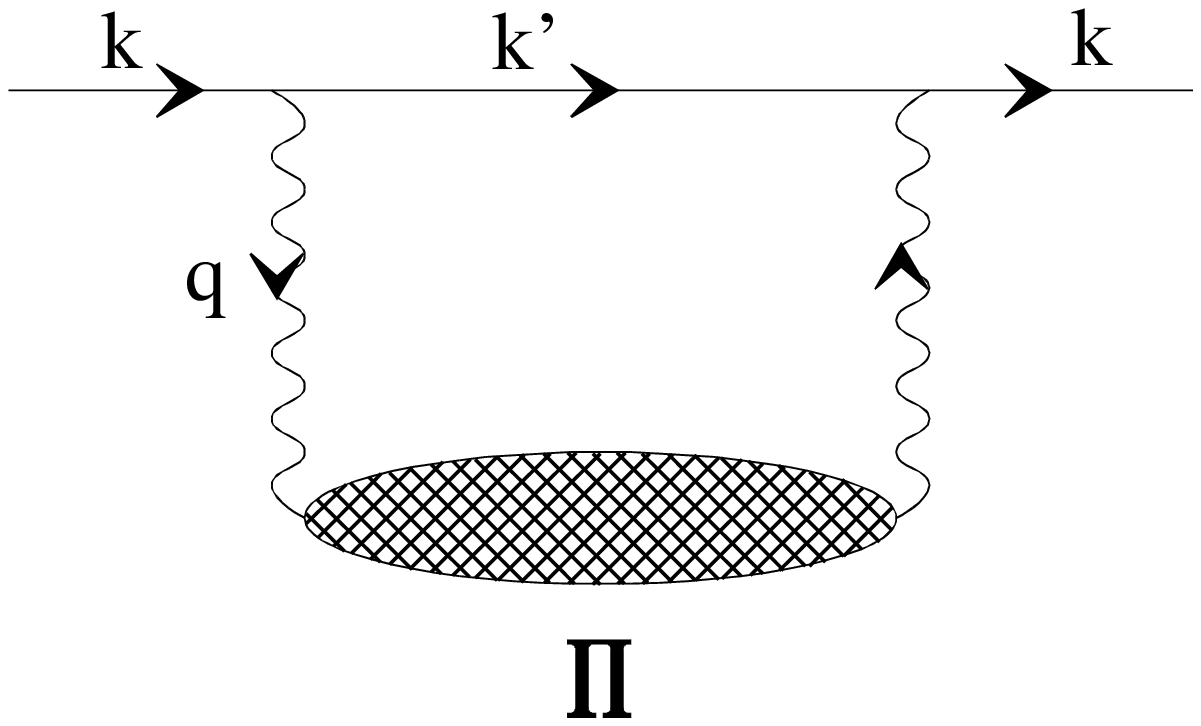,width=6cm}}}
\vskip 0.2cm
{\small {\bf Fig.4.1} Diagrammatic representation of the electron self-energy 
in nuclear matter.}
\vspace*{0.1cm}

The electron disappears
from the elastic flux, by inducing $1p 1h, 2p 2h...$ excitations or creating
pions, etc., at a rate given by

\be
\Gamma(k)=-2\Frac{m_e}{E_e}Im\Sigma.
\ee

\noindent
where $Im \Sigma$
 is the imaginary part of the electron self-energy. This latter
magnitude can be readily evaluated from the diagram of fig. 4.1 and we 
find:

\be
\Sigma_{r}(k)=ie^{2}{\displaystyle \int \frac{\displaystyle{d^{4}q}}
{\displaystyle{(2\pi)^{4}}}\bar{u}_{r}(k)\gamma_{\mu}\frac{
\displaystyle{(\slash \!\!\! k^{\prime}+{m}_{e})}}
{\displaystyle{k^{\prime\, 2}
-{m}_{e}^{2}+
i\epsilon}}\gamma_{\nu}u_{r}(k)\frac{\displaystyle{\Pi^{\mu\nu}_{\gamma}(q)}}
{\displaystyle{(q^{2}+i\epsilon)^2}}}
\ee

\noindent
where $\Pi^{\mu \nu}_\gamma$
 is the virtual photon self-energy. Eq.(21) 
displays explicitly the electron propagator (fraction after ${\gamma}_{\mu}$)
and the photon propagator $(q^2+i \epsilon)^{-1}$ which appears twice. 
 By averaging over the spin
of the electron, $r$, we find

\be
 \Sigma(k)=\frac{\displaystyle{ie^{2}}}{\displaystyle{2m_{e}}}\displaystyle{
  \int\frac{\displaystyle{d^{4}q}}
   {\displaystyle{(2\pi)^{4}}} \frac{\displaystyle{L_{\mu\nu}
    \Pi_{\gamma}^{\mu\nu}(q)}}{\displaystyle{q^{4}}}\frac{\displaystyle{
     1}}{\displaystyle{(k^{\prime\, 2}-{m}_{e}^{2}+i\epsilon)}}
      }
\ee

\noindent
and since we are interested in the imaginary part of $\Sigma$ we can obtain
it by following the prescription of the Cutkosky's rules. In this case we 
cut with a straight horizontal line the intermediate $e'$ state and those 
implied by the photon polarization (shaded region). Those states are then
placed on shell by taking the imaginary part of the propagator, 
self-energy, etc. Technically the rules to obtain $Im \Sigma$ reduce to making 
the substitutions:

     \be
      \begin{array}{l}
       \Sigma(k)\rightarrow 2iIm\Sigma(k)\Theta(k^{0})\\
        \\
         \Xi (k^{\prime})\rightarrow 2i Im\Xi(k^{\prime})\Theta({k^{\prime}}^{0})\\
          \\
           \Pi^{\mu\nu}(q)\rightarrow 2i Im\Pi^{\mu\nu}(q)\Theta(q^{0})
            \end{array}
             \ee

\noindent
where

 \be
  \Xi(k^{\prime})=\frac{\displaystyle{1}}
   {\displaystyle{k^{\prime\, 2}-{m}_{e}^{2}+i\epsilon}}
 \ee

\noindent
and $\Theta$ is the Heaviside, or step, function.
By proceeding according to these rules we obtain

\be
 Im\Sigma(k)=\frac{\displaystyle{2\pi\alpha}}{\displaystyle{m_{e}}}
  {\displaystyle \int\frac{\displaystyle{d^{3}q}}{\displaystyle{
   (2\pi)^{3}}}\left(Im\Pi_{\gamma}^{\mu\nu}L_{\mu\nu}(k,k^{\prime})\right)
    \frac{\displaystyle{1}}{\displaystyle{q^{4}}}\frac{\displaystyle{1}}{
     \displaystyle{2E_{e}(\vec{k}^{\prime})}}
      }\Theta(q^{0})
       \ee

The relationship of $Im \Sigma$ to the $(e, e')$ cross section is easy:
$\Gamma dt dS$  provides a probability times a differential of area, which is a 
contribution to a cross section. Hence we find

     \be
   d\sigma=\Gamma (k)dt dS=
    -\frac{\displaystyle{2m}}{\displaystyle{E_e}}
           Im\Sigma dl\,dS=
   -\frac{\displaystyle{2m}}{\displaystyle{|\vec{k}\,|}}
       Im\Sigma d^{3}r
   \ee

\noindent
and hence the nuclear cross section is given by
\be
 \sigma= -{\displaystyle \int d^{3}r \frac{\displaystyle{2m}}{\displaystyle{
  |\vec{k}\,|}}Im\Sigma(k, \rho(\vec{r}\,))
   }
 \ee

\noindent
where we have substituted $\Sigma$ as a function of the nuclear
density at each 
point of the nucleus and integrate over the whole nuclear
volume. Eq. (27) 
assumes the local density approximation, which, as shown in \cite{CAR},
is an excellent approximation for volume processes like here, hence we are 
neglecting the electron screening and using implicitly plane waves for the
electrons (corrections to account for the small distortion are usually
done in the experimental analysis of the data, see \cite{GUE} ).

Coming back to eq. (25) we find then

\be
  \frac{\displaystyle{d^{2}\sigma}}{\displaystyle{d\Omega^{\prime}_e 
dE^{\prime}_e}} = 
    -\frac{\displaystyle{\alpha}}{\displaystyle{q^{4}}}\frac{
      \displaystyle{|\vec{k}^{\,\prime}|}}{\displaystyle{|\vec{k}|}}\frac{
        \displaystyle{1}}{\displaystyle{(2\pi)^{2}}}{\displaystyle\int d^{3}r
          \left(Im\Pi_{\gamma}^{\mu\nu}L_{\mu\nu}\right)
            }
 \ee

\noindent
which gives us the $(e,e')$ differential cross section in terms of the
imaginary part of the photon self-energy.

If one compares eq. (28) with the general expression for the inclusive
$(e, e')$ cross section \cite{MUL,GIO} (see also eq. (11))

 \be
  \frac{\displaystyle{d^{2}\sigma}}{\displaystyle{d\Omega^{\prime}_e 
dE^{\prime}_e}}
   =\frac{\displaystyle{\alpha^{2}}}{\displaystyle{q^{4}}}\frac{\displaystyle{
    |\vec{k}^{\,\prime}|}}{\displaystyle{|\vec{k}|}}L^{\mu\nu}W_{\mu\nu}
     \ee

\noindent
we find
 
   \be
      W^{\mu\nu}=-\frac{\displaystyle{1}}{\displaystyle{\pi e^{2}}}
         {\displaystyle \int d^{3}r
            \frac{\displaystyle{1}}{\displaystyle{2}}
               \left(Im\Pi^{\mu\nu}+Im\Pi^{\nu\mu}\right)
                  }
    \ee
 
Once again, by choosing $\vec{q}$ in the $z$ direction and using gauge 
invariance one
can write the cross section in terms of the longitudinal and transverse
structure functions $W_L, W_T$ as

 \be
  \frac{\displaystyle{d^{2}\sigma}}{\displaystyle{d\Omega^{\prime}_e
    dE^{\prime}_e}}=
     \left(\frac{\displaystyle{d\sigma}}{\displaystyle{d\Omega}}\right)_{Mott}
      \left(-\frac{\displaystyle{q^{2}}}{\displaystyle{|\vec{q}\,|^{2}}}\right)
       \left\{W_{L}(\omega ,|\vec{q}\,|)+
        \frac{\displaystyle{W_{T}(\omega,|\vec{q}\,|)}}
         {\displaystyle{\epsilon}}\right\}
          \ee
\noindent
where

\be
 \begin{array}{c}
  q^{2}=\omega^{2}-|\vec{q}\,|^{2}\\
   \\
    \left.\frac{\displaystyle{d\sigma}}{\displaystyle{d\Omega}}\right|_{Mott}=
     \frac{\displaystyle{\alpha^{2}\cos_{e}^{2}(\theta /2)}}{\displaystyle{
      4E_{e}^{2}\sin_{e}^{4}(\theta /2)}}\\
       \\
\end{array}       
\ee

\noindent
and

\be
 \begin{array}{l}
  W_{L}=-\frac{\displaystyle{q^{2}}}{\displaystyle{\omega^{2}}}W^{zz}
   =-\frac{\displaystyle{q^{2}}}{\displaystyle{|\vec{q}\,|^{2}}}W^{00}\\
    \\
     W_{T}=W^{xx}
      \end{array}
        \ee

Hence using eq. (30) we can write $W_L$ and
$W_T$ in terms of the photon self-energy as

 \be
    \begin{array}{l}
       W_{L}=\frac{\displaystyle{q^{2}}}{\displaystyle{\pi e^{2} |\vec{q}\,|^{2}}}
          {\displaystyle \int d^{3}r Im \Pi^{00}(q,\rho(\vec{r}\,))
             }\\
                \\
       W_{T}=-\frac{\displaystyle{1}}{\displaystyle{\pi e^{2}}}{\displaystyle
        \int d^{3}r Im \Pi^{xx}(q,\rho(\vec{r}\,))}
         \end{array}
       \ee

\noindent
where we see that we only need the components $\Pi^{00}$ and $\Pi^{xx}$.

\subsection{The virtual photon self-energy in pion production}

We must construct a self-energy diagram for the photon which contains pion 
production in the intermediate states. This is readily accomplished by 
taking any generic diagram of the 
$\gamma^* N \rightarrow \pi N$  amplitude of fig. 3.2 and folding
it with itself. One gets then the diagram of fig. 4.2 where the circle 
stands for any of the 6 terms of the elementary model for 
$\gamma^* N \rightarrow \pi N$. 
The
lines going up and down in fig. 4.2. follow the standard many body
nomenclature and stand for particle and hole states respectively.
\vskip 0.2cm
  \centerline{\protect\hbox{\psfig{file=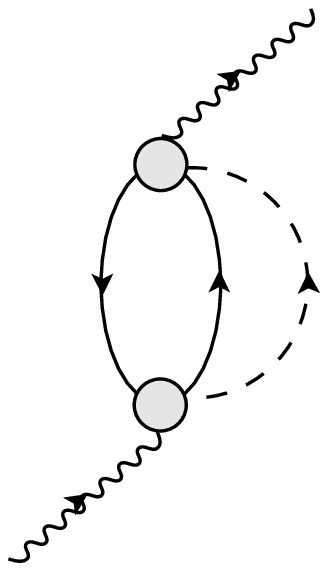,width=3.cm}}}
    \vskip 0.2cm
%      \vspace*{-4.7cm}
\noindent
 {\small {\bf Fig.4.2} Photon self-energy obtained by folding the 
 $\gamma^* N \rightarrow \pi N$ amplitude.}
 \vspace*{0.1cm}

 The photon
self-energy corresponding to this diagram (actually 36 diagrams) is 
readily evaluated and gives

\be
\begin{array}{ll}
\Pi_{NN^{\prime}}^{\mu\nu}(q)=i{\displaystyle \int \frac{\displaystyle{
d^{4}k}}{\displaystyle{(2\pi)^{4}}}} & 2 {\displaystyle \int\frac
{\displaystyle{d^{3}p}}{
\displaystyle{(2\pi)^{3}}}\frac{\displaystyle{n_{N}(p)[1-n_{N^{\prime}} (
p+q-k)]}}{\displaystyle{q^{0}-k^{0}+E(p)-E(p+q-k)+i\epsilon}}
 }\times\\
  &\\
   &\times D_{\pi}(k)\frac{1}{2}Tr^{Spin}(T^{\mu}{{T}}
    ^{\dagger \nu})_{NN^{\prime}}
     \end{array}
      \ee

\noindent
where $T^\mu$ is the amplitude for 
$\gamma^* N \rightarrow \pi N$. The indices $N, N'$ in eq. (35) stand for 
the hole and particle nucleon states respectively and 
$n_N (\vec{p})$ is the occupation number
in the Fermi local sea. $E (\vec{p})$ is the energy of the nucleon 
$\sqrt{\vec{p}\,^2 + M_N^2}$ and $D_\pi$ is the
pion propagator

\be
 D_{\pi}(k)=\frac{\displaystyle{1}}
  {\displaystyle{k^{2}-m_{\pi}^{2}+i\epsilon}}
   \ee

A further simplification can be done by evaluating the $T^\mu$
 amplitudes at 
an average Fermi momentum. Explicit integration over $\vec{p}$ and also this 
approximation were done is \cite{CAR} and the approximation was found to be
rather good. We take $< \vec{p} >= \sqrt{\frac{3}{5}} k_F$
with $k_F$ the local Fermi momentum ($3\pi^2 \rho (r)/2)^{1/3}$
and a direction
orthogonal to that of the virtual photon. The errors induced by 
this approximation are
smaller than 5$\%$. Then we can use the Lindhard function $\bar{U}_{N,N'}$
defined as 

 \be
  \bar{U}_{r,s}(q-k)=2{\displaystyle \int \frac{\displaystyle{d^{3}k}}
   {\displaystyle{(2\pi)^{3}}}\frac{\displaystyle{n_{r}(\vec{p}\,)[1-n_{s} (
   \vec{p}+\vec{q}-\vec{k}\,)]}}{\displaystyle{q^{0}-k^{0}+{E}(\vec{p}\,)
   -{E}(\vec{p}+\vec{q}-\vec{k}\,)+i\epsilon}}} 
   \ee

\noindent
where the indices, $r,s$ correspond to protons or neutrons.
 
For the evaluation of the imaginary part we need an extra Cutkosky rule

$$
U (p) \rightarrow 2 i Im U (p) \Theta (p^0)
$$

\noindent
which is the general rule considering that the Lindhard
function plays the role of a $ph$ propagator. 
 
 Hence we apply the Cutkosky
rules of eq. (23) and the former one and we find

\be
 \begin{array}{ll}
   Im\Pi^{\mu\nu}=& {\displaystyle \int\frac{\displaystyle{d^{3}k}}
     {\displaystyle{(2\pi)^{3}}}
       Im
         \bar{U}_{NN^{\prime}}(q-k)\frac{\displaystyle{1}}{\displaystyle{2
           \omega(\vec{k}\,)}}
             \theta(q^{0}-\omega(\vec{k}\,))
               }\times\\
                 &\\
       &\times \left.\frac{1}{2}Tr^{Spin}(T^{\mu}T^{\dagger\nu}_{NN^{\prime}})
                      \right|_{k^{0}=\omega(\vec{k}\,)}
                        \end{array}
                          \ee

Since there are analitycal expressions for $Im\bar{U}_{NN^{\prime}}$
(see Appendix B of \cite{CAR}), the 
approximation done saves us three integrals and a considerable amount of 
computational time.

There is an interesting test to eq.(38). Indeed, in
 the limit of small densities
the Pauli blocking factor $1-n$ becomes 1 and
$Im\bar{U}_{NN^\prime} (q) 
\simeq - \pi \rho_N \delta (q^0 - \frac{\vec{q}\,^2}{2 M_N})$. Substituting
this into eq. (38) and (29),(30) one easily obtains that
$\sigma_{eA} = \sigma_{ep} Z + \sigma_{en} N$, the
strict impulse approximation. By performing the integral in eq. (38) one 
accounts for Pauli blocking and in an approximate way for Fermi motion. Later
on we shall introduce other corrections due to medium polarization.

\subsection{The $\Delta$ excitation term}

One of the terms implicit in eq. (38) is the one where one picks up the
$\Delta$ excitation term both in  $T^\mu$ and is $T^{\dagger \nu}$.
 This term is depicted diagrammatically in fig. 4.3(a) and, like in
pion-nuclear and photo-nuclear reactions at intermediate energies,
plays a major role in this reaction.

In order to evaluate this piece one can
go back to eq. (35) and perform the $d^4 k$
integration to factorize the $\Delta$
width and on the other hand one will also have the modulus 
squared of the $\Delta$ propagator
present in the $\Delta$ term of eq. (5). This, however, can be obtained
more economically by reinterpreting the diagram 4.3 (a) as a $\Delta h$
excitation with a $\Delta$ width.
We can also divert a little from the
general formulation, and in order to gain some extra accuracy we can
implement Lorentz covariance exactly simply boosting the tensor
$ \Pi^{\mu \nu}$ from
a frame where the $\Delta$ is at rest
$(\vec{q} + \vec{p} = \vec{p}_\Delta = 0)$,
where the amplitude of eq.
(5) would be (by construction) equivalent to the relativistic amplitude.

\vskip 0.15cm
\centerline{\protect\hbox{\psfig{file=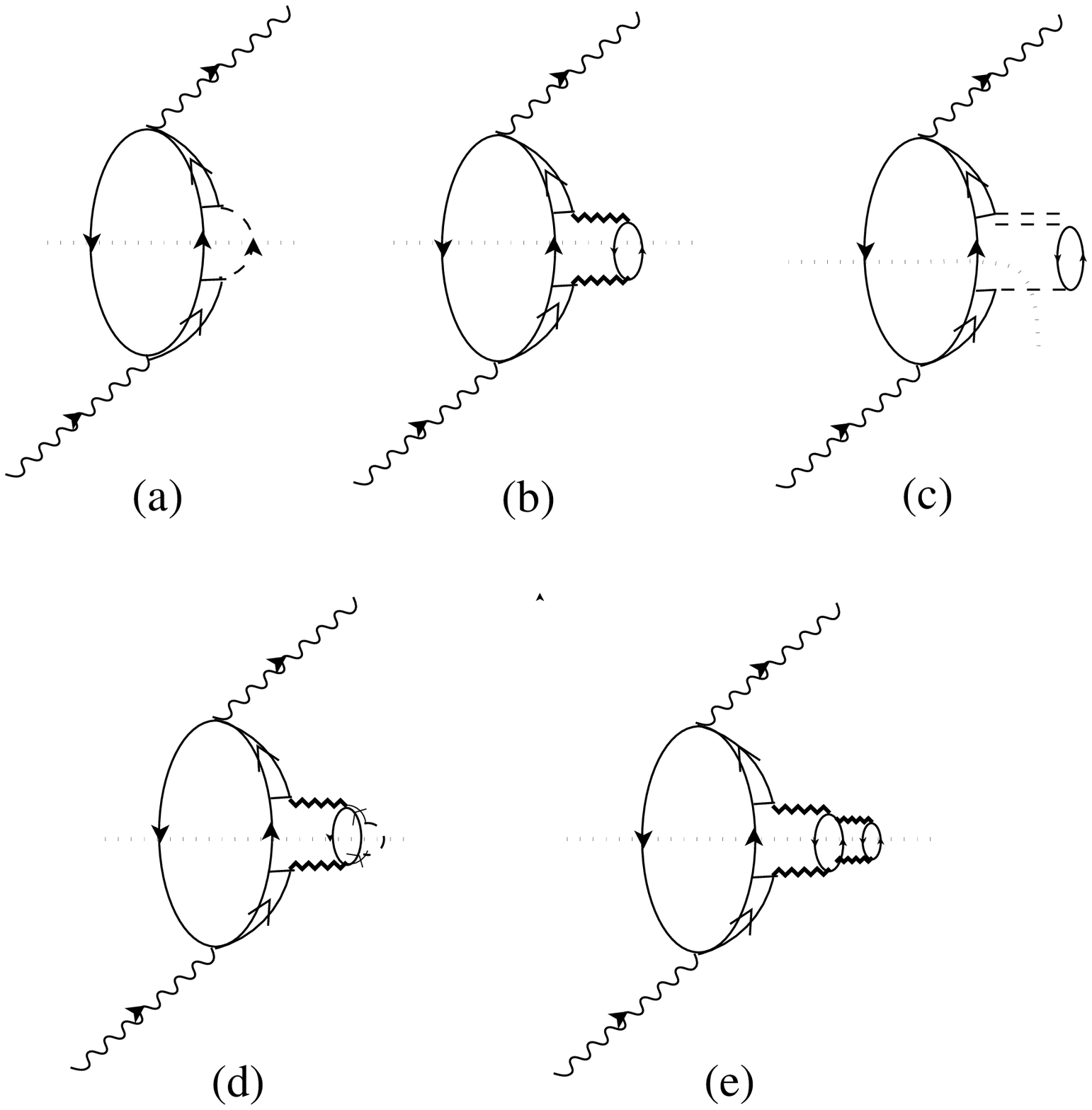,width=8cm}}}
\vskip 0.1cm
\noindent
{\small {\bf Fig.4.3} Diagrammatic representation of the $\Delta h$
photonuclear excitation piece.}

Hence we get

\vspace*{-0.2cm}
\be
\begin{array}{ll}
Im\Pi_{\Delta}^{\mu\nu}=&\sum_{ij}|c_{ij}|^{2}\frac{\displaystyle{f_{\gamma}^{2}
(q^{2})}}{\displaystyle{m_{\pi}^{2}}}{\displaystyle \int \frac{\displaystyle{
d^{3}p}}{\displaystyle{(2\pi)^{3}}}n_{i}(p)\Lambda^{\mu}_{m}(p,q)
\Lambda^{\nu}_{l}(p,q)
}\\
&\\
&\times Tr\left((\vec{S}^{\dagger}\times\vec{q}_{cm})^{m}
(\vec{S}\times\vec{q}_{cm})^{l}\right)\frac{\displaystyle{s}}{
\displaystyle{M_{\Delta}^{2}}}\times
\\
&\\
&\times \frac{\displaystyle{Im\Sigma^{j}_{\Delta}(p+q)-
\frac{\displaystyle{\bar{\Gamma}}}{\displaystyle{2}}(p+q)}}
{\displaystyle{\left|\sqrt{s}-M_{\Delta}+i\frac
{\displaystyle{\bar{\Gamma}(s)}}
{\displaystyle{2}}-\Sigma_{\Delta}(s)\right|^{2}}}
\end{array}
\ee

The Lorentz matrix $\Lambda$ is such that
${\Lambda_\gamma}^\mu_\nu q^\nu_{cm} = q^\mu$ and $s=(p+q)^{2}$. 
The coefficients $C_{ij}$, $i,j= 1,2$ stands 
for proton or neutron, account for isospin and are given by

\be 
|C_{ij}|=\left(\frac{2}{3}\right)^{2}\delta_{ij}+\frac{2}{9}
\delta_{i,j+1}+\frac{2}{9}\delta_{i,j-1}
\ee

Eq. (39), however, seems to 
neglect the Pauli blocking factor $1-n$ of eq. (35).
This factor, however, is taken into account implicitly and leads to the Pauli
blocked width  $\bar{\Gamma}$, which is 
evaluated in \cite{RAFAPI}. Furthermore, 
in a nuclear 
medium the $\Delta$ is renormalized and acquires a self-energy
$\Sigma_\Delta$, which
is also accounted for in eq.(39). The results of \cite{LOR}
 for $\Sigma_\Delta$ are used
in the calculation.
 This self-energy accounts for the diagrams 
depicted in fig. 4.3, where the double dashed line stands for the effective
spin -isospin interaction, while the serrated line accounts for the induced
interaction. The effective spin-isospin interaction is originated by a pion
exchange in the presence of short range correlations and includes 
$\rho$-exchange as well. It is obtained by substituting

\be
\hat{q}_{i}\hat{q}_{j} D_{\pi}(q)
 \rightarrow \hat{q}_{i}\hat{q}_{j}V_{l}(q) +({\delta}_{ij}-
 \hat{q}_{i}\hat{q}_{j})V_{t}(q)
\ee

\noindent
and expressions for $V_l, V_t$ are found in \cite{LOR}
( $(f_{\pi NN}/m_{\pi})^{2}V_{l,t}$ here is equivalent to $V_{l,t}$
of \cite{LOR}). The induced interaction
accounts for the series of diagrams depicted in fig. 4.4.

\vspace*{0.2cm}
\centerline{\protect\hbox{\psfig{file=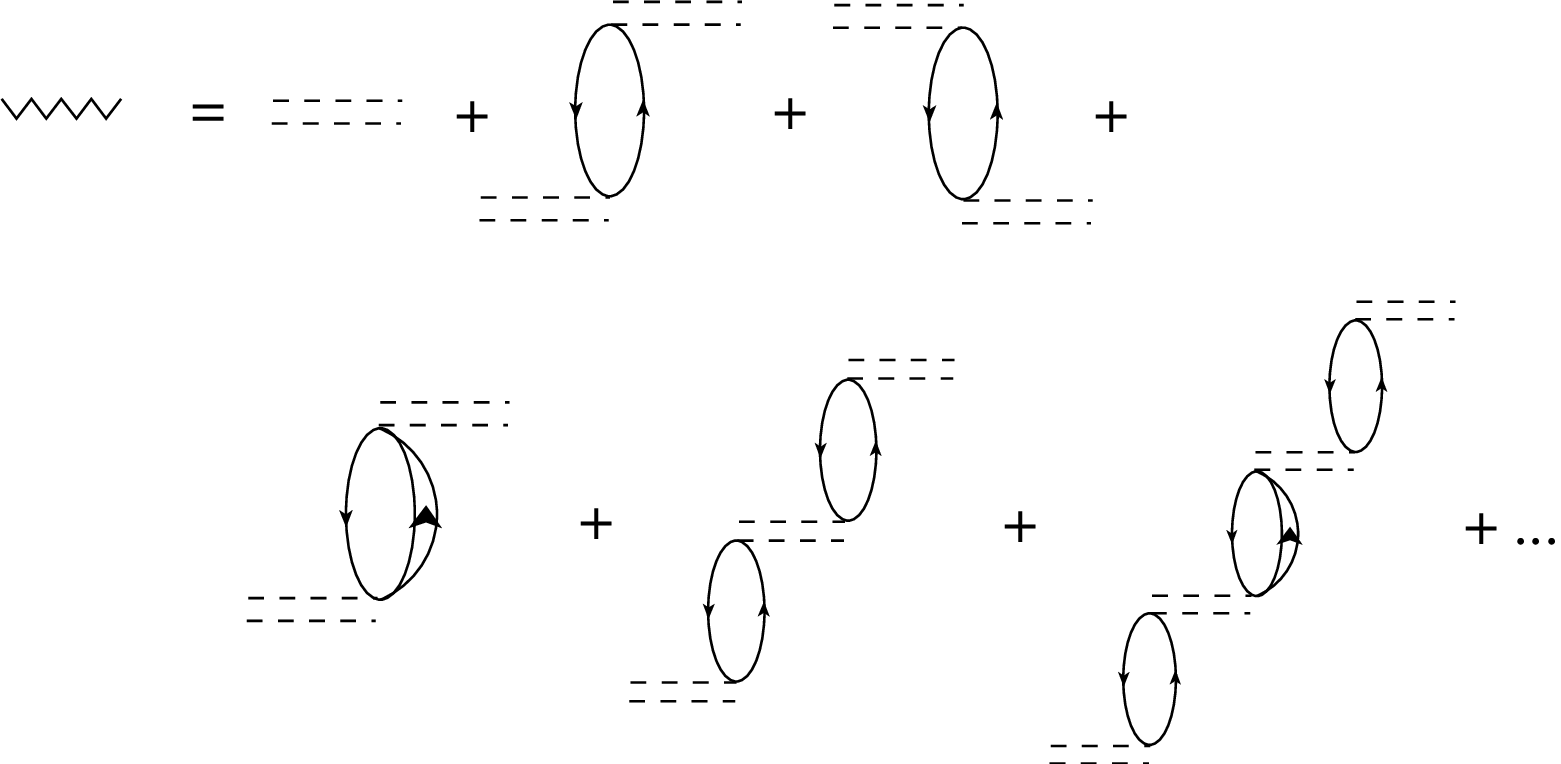,width=8cm}}}
\vskip 0.2cm
\noindent
{\small {\bf Fig.4.4} Feynman diagrams 
included in the evaluation of the $\Delta$
self-energy.}
\vspace*{0.1cm}

There is an RPA sum 
through $ph$ and $\Delta h$ excitation and is readily obtained as

\be
\begin{array}{lll}
V_{ind}&=&
\hat{q}_i\hat{q}_j\,
\Frac{V_l(q)}{1-U(q)V_l(q)\left(\Frac{f_{\pi NN}}{m_{\pi}}\right)^2}
+\\
&&\\
&&
(\delta_{ij}-\hat{q}_i\hat{q}_j)
\Frac{V_t(q)}{1-U(q)V_t(q)\left(\Frac{f_{\pi NN}}{m_{\pi}}\right)^2}
\end{array}
\ee

\noindent
where now $U (q) = U_N (q) + U_\Delta (q)$
is the Lindhard function for $ph\,+\,\Delta h$
 excitations including forward
going and backward going bubbles \cite{LOR} in contrast to $\bar{U}$
which only contains 
the forward going bubble of a $ph$ excitation (the only one which
contributes to
$Im U_N $ for
$q^0 > 0$).  $U_N$ in addition incorporates a factor two of isospin 
with respect to
$\bar{U}$, such that $Im U_N =
2 Im \bar{U}$ for symmetric nuclear matter.
However, all the work which goes into the evaluation of 
$\Sigma_\Delta$
is done in
ref. \cite{LOR}, where a useful analytical parameterization of the numerical
results is given that we use here. The imaginary part is parametrized 
as

\be
Im\Sigma_{\Delta}=-\left\{C_{Q}(\rho /\rho_{0})^{\alpha}+C_{A_{2}}
 (\rho /\rho_{0})^{\beta}+C_{A_{3}}(\rho /\rho_{0})^{\gamma}\right\}
  \ee

\noindent 
where the different coefficients are given in \cite{LOR} as a function of the
energy.

The separation of terms in eq. (43) is useful because the term $C_Q$ comes
from the diagrams (c) and (d)
of fig. 4.3
 when the lines cut by the dotted line are
placed on shell, and hence the term is related to the $(
\gamma^*, \pi)$ channel, while
$C_{A2}, C_{A3}$
come from the diagrams (b) and (e) and are related to two and three
body absorption. Hence, the separation in this formula allows us to separate
the final cross section into different channels.

\vskip 0.2cm
 \centerline{\protect\hbox{\psfig{file=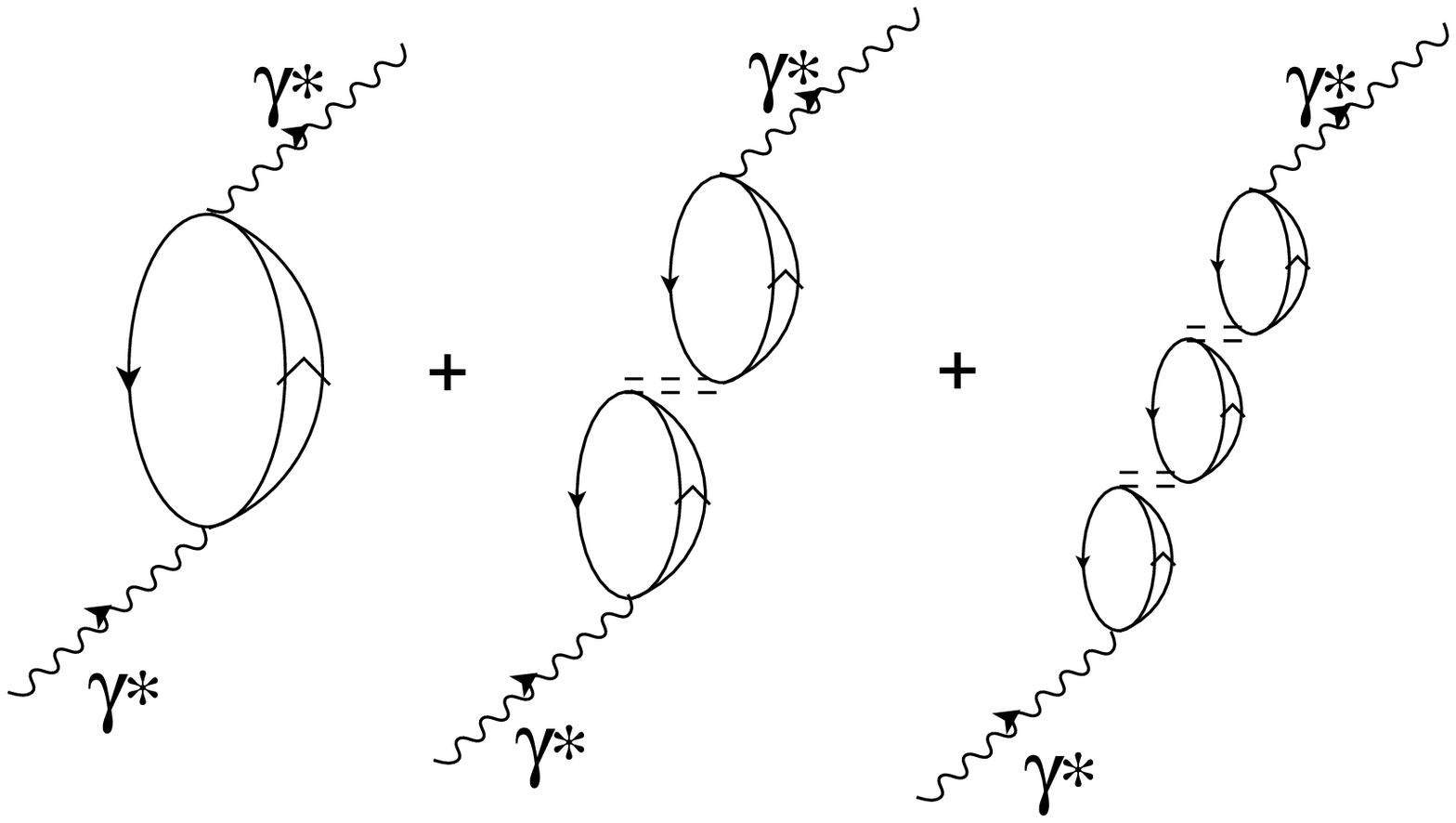,width=8.cm}}}
  \vskip 0.2cm
   \vspace*{-1.cm}
\noindent   
 {\small {\bf Fig.4.5} Irreducible pieces in the $\Delta h$ channel from the
 $\Delta h$ interaction.}
\vspace*{0.1cm}

It is also easy to realize that the RPA sum of $\Delta$h excitations, shown
in fig. 4.5, can be taken into account by substituting $Re \Sigma_\Delta$
by \cite{CAR}

\be
  Re\Sigma_{\Delta}\rightarrow Re\Sigma_{\Delta}+\frac{4}{9}
 \left(\frac{\displaystyle{f^{*}}}{\displaystyle{m_{\pi}}}\right)^{2}
 \rho V_{t}
      \ee

\vskip 0.2cm
  \centerline{\protect\hbox{\psfig{file=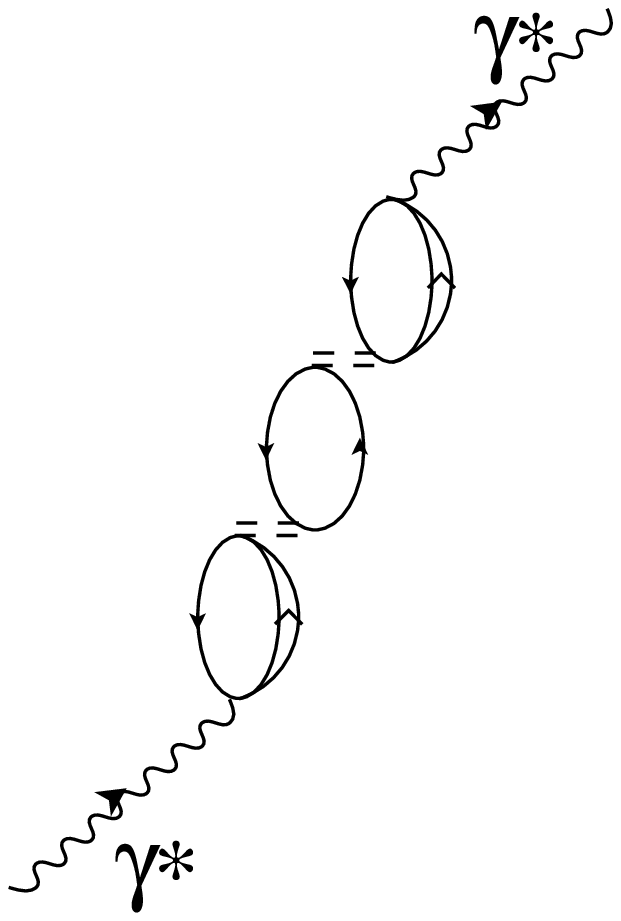,width=4.3cm}}}
    \vskip 0.2cm
      \vspace*{-0.5cm}
\noindent
 {\small {\bf Fig.4.6} Diagrammatic representation of the inclusion
 of a $ph$ excitation between $\Delta h$ excitations.}
\vspace*{0.1cm}

\noindent
and furthermore, if we wish to include some $ph$ excitation in between, 
see fig. 4.6, (which is actually not relevant numerically), this is done
easily by substituting  $Re \Sigma_\Delta$ by

 \be
    Re \Sigma_{\Delta}+\frac{4}{9}
        \left(\frac{f^{*}}{m_{\pi}}\right)^{2}\frac{\displaystyle{V_{t}}}{
   \displaystyle{\left(1-\frac{{f}_{\pi NN}^{2}}{{m}_{\pi}^{2}}{U_N}
   V_{t}\right)}}
                \rho
                    \ee
         
\subsection{Results for the $\Delta$ contribution}

In fig. 4.7 we show the results coming from the $\Delta$ term discussed
in the former section. The experimental data are coming from \cite{BAR}.
 We
have separated the contribution from the different channels. Besides
the upper solid line which stands for the total contribution, looking from
up to down at about $\omega = 350$ MeV
 the next line corresponds to pion production,
the following one  is two nucleon absorption and the lowest one three body absorption.
We can see that most of the experimental strength in the $\Delta$ region
is provided by this $\Delta$
excitation term, but there is still some strength missing.
In fig. 4.8 we show the contribution of the delta piece for the  $^{208}$ Pb
nucleus. The data are now
from \cite{ZGH}. The results are similar to those found in 
$^{12}$C and there is
still some strength  missing. The study of this missing strength will 
occupy the next sections.

%\newpage
\centerline{\protect\hbox{\psfig{file=deltacarbon.ps,width=10.cm}}}
\vskip 0.2cm
\noindent
{\small{\bf Fig.4.7} Contribution of the
$\Delta$ piece to the $(e,e^{\prime})$
cross section in $^{12}$C. Experimental data from \cite{BAR}. 
See text for different contributions.}  
\vskip 0.1cm
\centerline{\protect\hbox{\psfig{file=plomodelta.ps,width=10.cm}}}
\vskip 0.1cm
\noindent
{\small{\bf Fig.4.8}
 Contribution of the
 $\Delta$ piece to the $(e,e^{\prime})$
 cross section in $^{208}$Pb. Experimental data from \cite{ZGH}.
}

\subsection{Two body photoabsorption}

Let us go back to the generic diagram of pion electroproduction  of
fig. 4.2. Let us take the pion line and allow the pion to excite a
$ph$. This leads us to the diagram of fig. 4.9.

 \vskip 0.1cm
   \centerline{\protect\hbox{\psfig{file=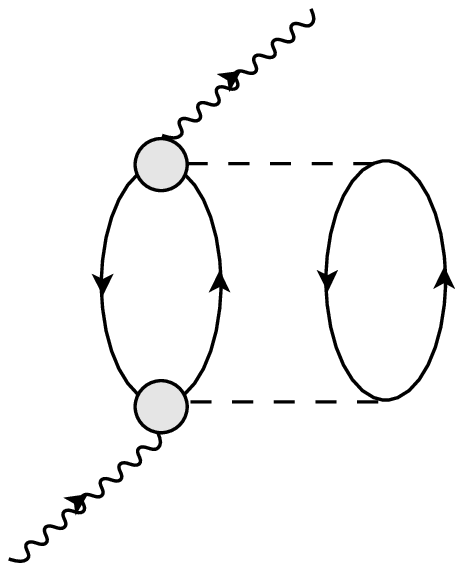,width=4.7cm}}}
        \vskip 0.1cm
\noindent        
{\small {\bf Fig.4.9} Photon self-energy obtained from the one in fig. 4.2
when the pion is allowed to excite a $ph$.}
\vspace*{0.1cm}

This is still a generic diagram which actually
contains 36 diagrams when in the shaded circle we put each one of the 
terms of the $\gamma^* N \rightarrow \pi N$ amplitude of fig. 3.2. One 
must avoid the temptation of factorizing these amplitudes in order
to evaluate these diagrams since some of them might be symmetric
and then have a symmetry factor 1/2. This is the case here with one diagram
implicit in fig. 4.9, which is the one corresponding to the pion pole
term in each one of the $\gamma^* N \rightarrow \pi N$ amplitudes.
This diagram is shown explicitly in fig. 4.10.

  \vskip 0.2cm
    \centerline{\protect\hbox{\psfig{file=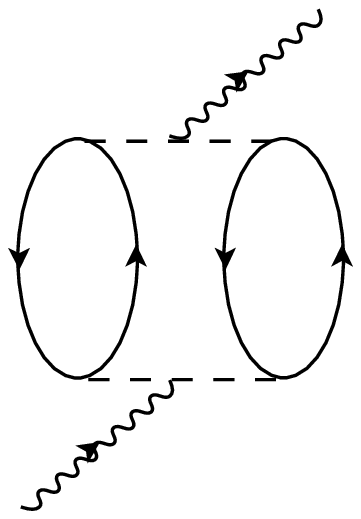,width=4.4cm}}}
      \vskip 0.2cm
\noindent
  {\small {\bf Fig.4.10} Pion pole term included in fig. 4.9.}
\vspace*{0.1cm}
        
One can see that the diagrams in fig. 4.9 contribute to $Im \Pi$ 
according to Cutkosky rules when we cut by
a horizontal line and the $2 p 2h$ are placed on shell.

The contribution of the diagram of fig. 4.9 is readily done. We obtain

\be
\begin{array}{ll}
\Pi^{(2) \mu\nu}(q)=
& {\displaystyle {\sum_{N,N^{\prime}}}}i
{\displaystyle{\int}}\Frac{d^{4}k}{(2\pi)^{4}}
\Frac{d^{3}p}{(2\pi)^{3}}\Frac{n_{N}(p)[1-n_{N^{\prime}}(p+q-k)]}
{q^{0}-k^{0}+E(p)-E(p+q-k)+i\epsilon} \times
\\
&
\\
& D_{\pi}^{2}(k)\Frac{{f}_{\pi NN}^{2}}{m_{\pi}^{2}}\vec{k}^{2}
{U}_{\lambda}(k)Tr^{Spin}(T^{\mu}{T}^{\dagger \nu})_{NN^{\prime}}
S_{\alpha}F_{\pi}^{4}(k)

\end{array}
\ee

\noindent
where $U_\lambda$ is the Lindhard function for $ph$ by an object of charge
$\lambda$: this is, twice $\bar{U}_{p,n}$ or $\bar{U}_{p,n}$ for the
excitation by a charged pion or $\bar{U}_{p,p}+\bar{U}_{n,n}$ for
the excitation by a neutral pion and $\vec{k}$ is the pion momentum.
The factor $F^4_\pi (k)$, where $F_\pi$ is the pion form factor
appears because now the pions are off shell. Recall that we also
take all form factors equal in order to preserve gauge invariance
(eq. (9)). The factor $S_\alpha$ is the symmetry factor, unity for all 
diagrams and 1/2 for the symmetric one of fig. (4.10).

We can again simplify the expression by taking an average nucleon
momentum of the Fermi sea to evaluate the matrix elements of
$T^\mu \, T^{\dagger\nu}$. This allows us to factorize the Lindhard
function and we get

\be
\begin{array}{ll}

\Pi^{(2)\,\,\mu\nu}(q)=
& {\displaystyle{\sum_{NN^{\prime}}}} i {\displaystyle{\int}}
\Frac{d^{4}k}{(2\pi)^{4}}\bar{{U}}_{NN^{\prime}}(q-k)
 D_{\pi}^{2}(k)\Frac{{f}_{\pi NN}^{2}}{m_{\pi}^{2}}
  \\
   &\\
    & {\vec{k}}^{2}
    {U}_{\lambda}(k)\frac{1}{2}Tr^{Spin}(T^{\mu}T^{\dagger \nu})_{NN^{\prime}}
S_{\alpha}F_{\pi}^{4}(k)
    
    \end{array}
\ee

By applying Cutkosky
rules we find

 \be
   \begin{array}{ll}
     Im\Pi^{(2)\,\,\mu\nu}=&-{\displaystyle{\sum_{NN^{\prime}}}}{\displaystyle{
       \int}}\Frac{d^{4}k}{(2\pi)^{4}}\Theta(q^{0}-k^{0})Im\bar{{U}}_{N
         N^{\prime}}(q-k)\Theta (k^{0})\times\\
           &\\
     & Im {U}_{\lambda}(k)D_{\pi}^{2}(k)\Frac{{f}_{\pi NN}^{2}}{m_{\pi}^{2}}
               {\vec{k}}^{2} F_{\pi}^{4}(k)S_{\alpha}\times\\
                 &\\
    & Tr^{Spin}(T^{\mu}T^{\dagger \nu})_{NN^{\prime}}
    \end{array}
       \ee

The cut which places the two $ph$ on shell in the diagrams of
fig. 4.9 is not the only possible one. In fig. 4.11 we show a different
cut (dotted line) which places one $ph$ and
the pion on shell.

  \vskip 0.2cm
      \centerline{\protect\hbox{\psfig{file=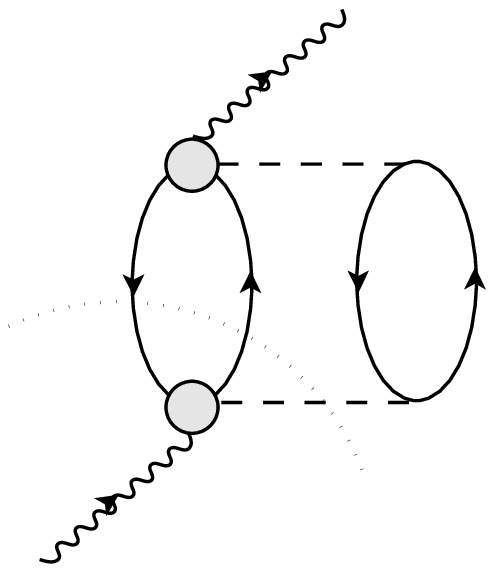,width=5.5cm}}}
            \vskip 0.2cm
\noindent            
    {\small {\bf Fig.4.11} Same as fig. 4.9 and showing the cut which places 
    one $ph$ and the pion on shell.}
\vspace*{0.2cm}
                      
 This contribution is taken into account in the
$\Delta$ excitation term by means of the term $C_Q$. As done for
real photons in \cite{CAR}, we neglect this contribution in the
other terms, because at low energies where the background pieces are
important, the $(\gamma^*, \pi)$ channel is small and at high
energies where the $(\gamma^*, \pi)$ contribution is important, this channel
is dominated by the $\Delta$ excitation and there this correction is
taken into account.

We have also considered two body diagrams where each photon couples
to different bubbles: As found is \cite{CAR} only one of them is
relevant, the one in fig. (4.12), which involves the KR term alone
and which we take into account.

  \vskip 0.2cm
        \centerline{\protect\hbox{\psfig{file=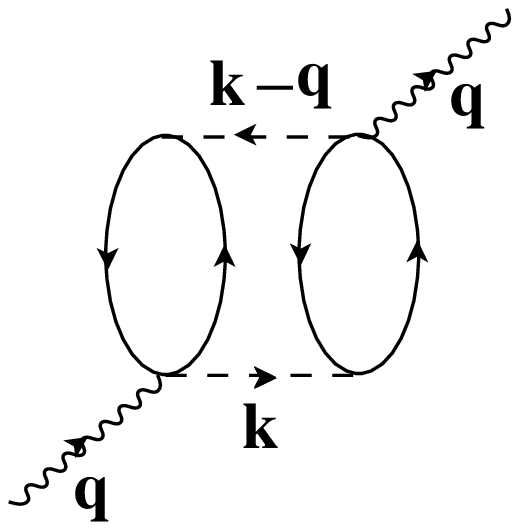,width=4.9cm}}}
                    \vskip 0.1cm
\noindent                    
  {\small {\bf Fig.4.12} Feynman diagram related to the KR term of fig. 4.9
  with outgoing photon from the second nucleon.}
\vspace*{0.3cm}
     
 Following the same rules as above, 
this term is readily evaluated and gives

 \be
 \begin{array}{ll}
 
 Im\Pi^{00}=& -2e^{2}\left(\Frac{f^{2}}{m_{\pi}^{2}}\right)^{2}{2}
 F_{A}^{2}(q^{2})\times\\
 &\\
 & {\displaystyle{\int}}\Frac{d^{4}k}{(2\pi)^{4}}D_{0}(k)D_{0}(k-q)
 \Frac{(\vec{k}\vec{q})(\vec{k}-\vec{q})\vec{q}}{M_{N}^{2}}\times\\
 &\\
 & F_{\pi NN}^{2}(k)F_{\pi NN}^{2}(k-q)\times\\
 &\\
 & \left[Im\bar{{U}}_{pp}(q)Im\bar{{U}}_{pp}(k-q)+
 Im\bar{{U}}_{pn}
 (q)Im\bar{{U}}_{np}(k-q)+\right.\\
 & \\
 & +\left.Im\bar{{U}}_{np}(q)Im\bar{{U}}_{pn}(k-q)\right]
 \Theta(k^{0})\Theta (k^{0}-q^{0})
 \end{array}
 \ee  

%  \be
  $$
  \begin{array}{ll}
  Im(\Pi^{xx}+\Pi^{yy})= & -2e^{2}\left(\Frac{f^{2}}{m_{\pi}^{2}}\right)^{2}{2}
   F_{A}^{2}(q^{2})\times\\
    &\\
     & {\displaystyle{\int}}\Frac{d^{4}k}{(2\pi)^{4}}D_{0}(k)D_{0}(k-q)
     |\vec{k}|^{2}sin^{2}\theta F_{\pi NN}^{2}(k)F_{\pi NN}^{2}(k-q)\times\\
     &\\
     &  \left\{\Frac{1}{2}\left(\Frac{q^{0}}{M_{N}}\right)^{2}
     Im\bar{{U}}_{pp}(q)Im\bar{{U}}_{pp}(k-q)\right.+
%     \\
%     &\\

     \end{array}
      $$
      
\newpage      
     \be
     \begin{array}{ll}

     &\left[-\left(1-\Frac{q^{0}}{2M}\right)\sqrt{2}\right]^{2}
     Im
     \bar{{U}}_{pn}(q)Im\bar{{U}}_{np}(k-q)\\
     &\\
     & +\left.\left[\left(1+\Frac{q^{0}}{2M}\right)\sqrt{2}\right]^{2}
     Im\bar{{U}}_{np}(q)Im\bar{{U}}_{pn}(k-q)\right\}\times\\
          &\\  
         & \Theta(k^{0})\Theta(k^{0}-q^{0})
         \end{array}
         \ee

The contribution of this term is roughly 1/2 of the $KR \times KR$
term in the generic diagram of fig. 4.9.

\subsection{Contributions tied to the ($\gamma^*, 2 \pi)$ channel}

The $\gamma N \rightarrow \pi \pi N$ reaction has been the subject of
recent detailed experimental analyses \cite{BRA,STR} and also of 
recent theoretical analyses, some of 
them spanning a large energy range \cite{GOM1,GOM2} and
others concentrating only very close to threshold in order
to test predictions of chiral perturbation theory \cite{ULF,BEN}.

The model in ref. \cite{GOM1}  for the $\gamma p \rightarrow
\pi^+ \pi^- p$ uses 67 Feynman diagrams, while ref. \cite{GOM2}, 
where the model is extended to the other isospin channels uses
only 20 diagrams which are necessary below $E_\gamma = 800 \, MeV$,
where the new data have been measured.

Although  the model is rather elaborate  and contains many terms,
one can see that the gross features of the reaction 
can be obtained with the two
terms of fig. 4.13,
 accounting
 for about 80$\%$ of the cross section.

    \vskip 0.2cm
      \centerline{\protect\hbox{\psfig{file=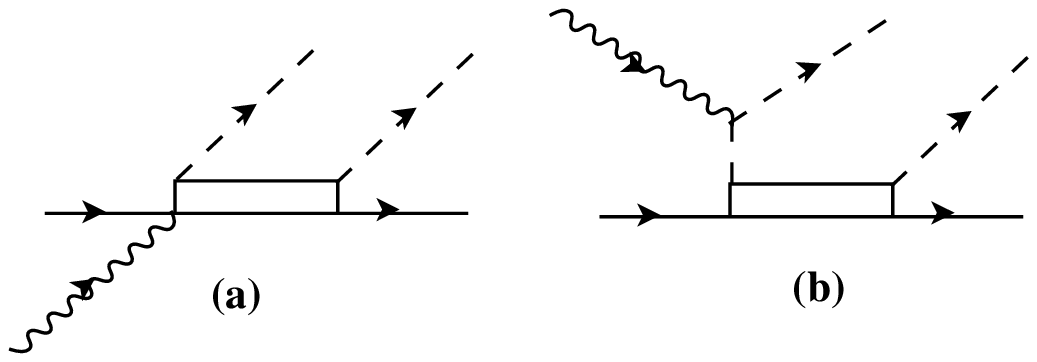,width=8.cm}}}
\noindent      
   {\small {\bf Fig.4.13} Relevant Feynman diagrams that enter in the evaluation
   of the $\gamma^* N \rightarrow N 2 \pi$ cross section. }
\vspace*{0.1cm}   
          
Since here we are only concerned about corrections to the more 
important terms which we have discussed above, it is sensible to just
take these two diagrams. The diagram in fig. (a) is the $\Delta
N \pi \gamma$ Kroll Ruderman (KR) term, while the one in fig. (b) is the 
pion pole term. In both terms the $\Delta$ is excited. The KR term, 
which appears from the $\Delta N \pi$ vertex by minimal coupling,
is given by,

 \be
  \begin{array}{ll}
   \cal{M}^{\mu}= &
    \left\{
     \begin{array}{ccc}
      0 & , & \pi^{0}\\
       1 & , & \pi^{\pm}
        \end{array}\right\}
         \left(e\Frac{f^{*}}{m_{\pi}}\right)\times\\
          & \\
           & \left(1 \frac{1}{2} \frac{3}{2} \left| m_{\pi}
            M_{N} M_{\Delta}\right.\right)
             \left(
              \begin{array}{c}
               \Frac{\vec{S}^{\dagger}\vec{p}_{\Delta}}{\sqrt{s}}\\
                 \\
                   \vec{S}^{\dagger}
                    \end{array}\right)
                     \end{array}
                      \ee

\noindent
corresponding to $\pi^\pm \Delta$ production in the $\gamma^* N
\rightarrow \Delta \pi$ vertex.

By following the same steps as before we obtain the many
body diagrams of fig. (4.14), where the dashed circle indicates any of 
the two terms of fig. 4.13.

  \vskip 0.2cm
    \centerline{\protect\hbox{\psfig{file=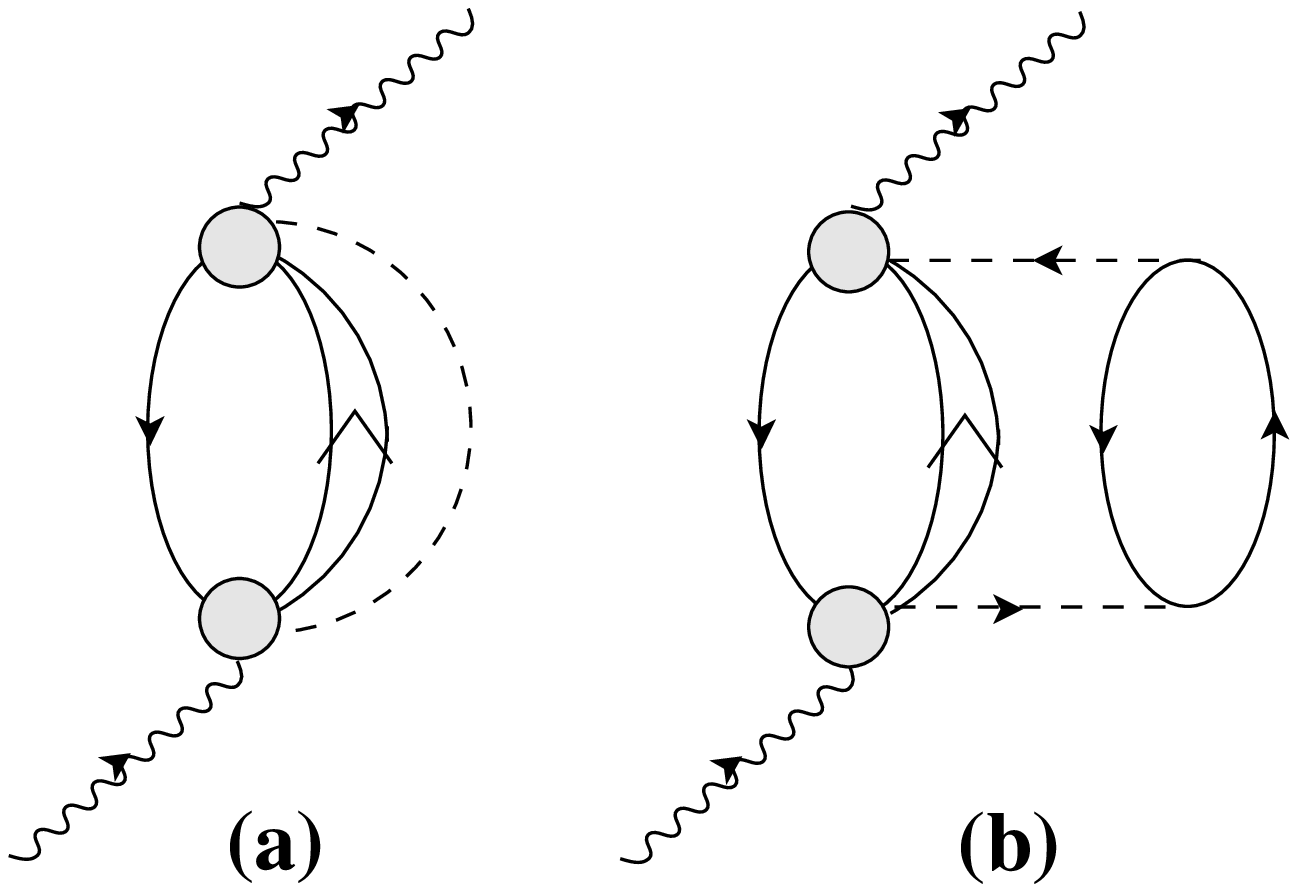,width=6.cm}}}
         \vskip 0.2cm
\noindent                  
       {\small {\bf Fig.4.14} Photon self-energy diagrams obtained by folding
       the terms of fig. 4.13. Diagram (b) is obtained when the pion is allowed
       to produce a $ph$ excitation.}
\vspace*{0.8cm}
                      
Furthermore, as discussed in ref. 
~\cite{CAR}, in the diagrams of fig. 4.14 (b) we keep only
the term with the $KR \times KR$ in the vertices. This is done since
the pion in flight term, can be considered as a two step process of a
$\gamma^* N \rightarrow \pi N$ with $\pi$, a real pion, followed by the
$\pi N \rightarrow \Delta$ excitation. The two step processes 
redistribute strength but do not change  the cross section and 
hence  are not included in our approach.

Since the $\Delta$ in the $\Delta h$ excitation in fig. 4.14 is also
renormalized, we are accounting for the physical channels depicted
in fig. 4.15 when placing on shell the states cut by the dotted
line.

\vskip 0.4cm
       \centerline{\protect\hbox{\psfig{file=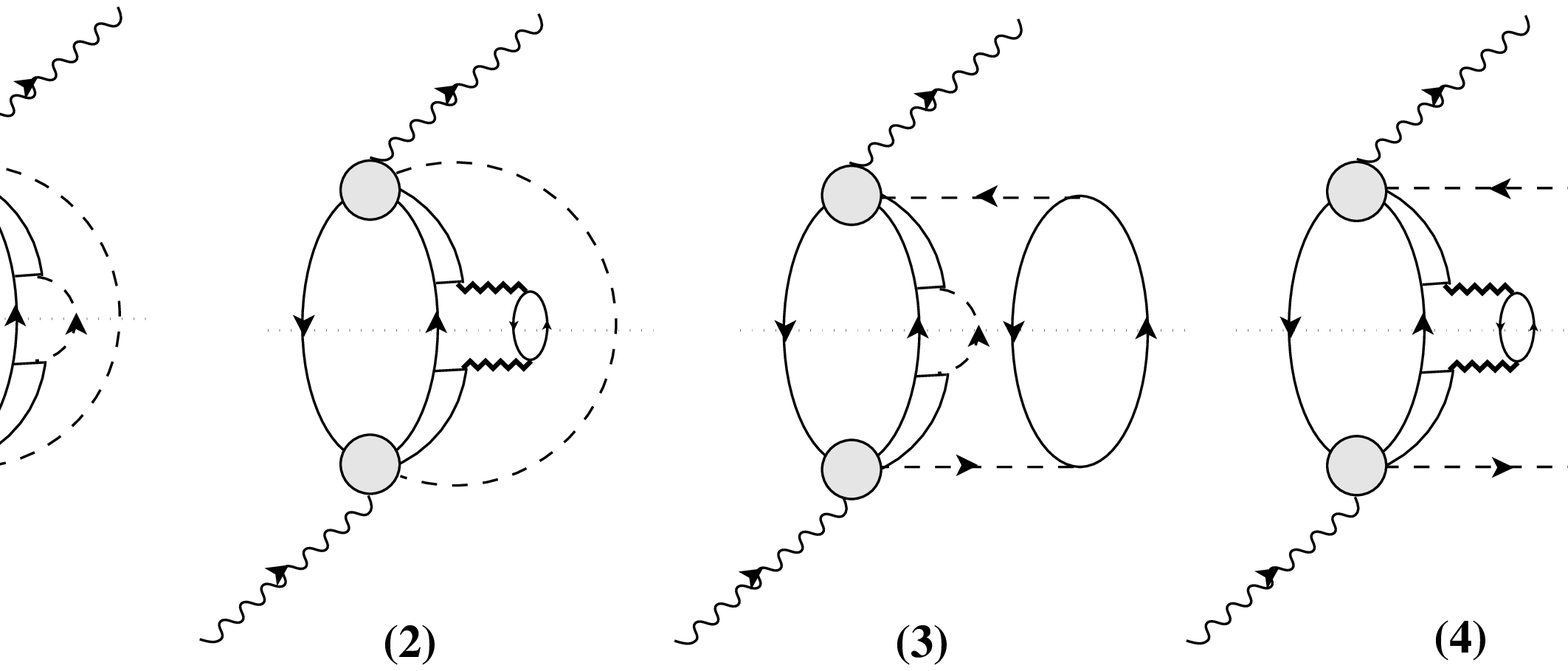,width=13.5cm}}}
            \vspace*{-1.3cm}
            %       \vskip 0.3cm
\noindent            
    {\small {\bf Fig.4.15} Detail of fig. 4.14 indicating the physical channels
    associated to the cuts.}
  \vspace*{0.8cm}  
  
As one can see there, (1) accounts for $1p 1h \, 2 \pi$ excitation,
(2) and (3) for $2 p 2 h \,1 \pi$ excitation and (4) for $3p 3h$
excitation.

The evaluation of these pieces follows exactly the same steps as 
for figs. 4.2 and 4.9, simply replacing the $\gamma^* N 
\rightarrow \pi N$ by the $\gamma^* N \rightarrow \pi \Delta$
amplitudes and one nucleon propagator by the $\Delta$ propagator.
The contribution of these terms below $\omega = 350 \, MeV$ is
very small. Their importance increases with the energy and at 
$\omega = 450 \; MeV$ they account for about 1/5 of the
cross section, as found for real photons.

\subsection{Polarization (RPA) effects}

In the diagrams of fig. 4.9 we can consider the $ph$ as just
the first order of a series of the RPA excitations through
$ph$ and $\Delta h$ excitations. If one replaces the $ph$ by the RPA series,
one is led to the terms implicit in fig. 4.16. A similar
series would appear for the case of the $(\gamma, \pi)$
process depicted in fig. 4.2.

 In practical
terms this is done in a simple way by having a bookkeeping of  both the
spin longitudinal and spin transverse parts and replacing

\be\label{eq:pol}
Im U_N \rightarrow a \frac{Im U_N}{|1 - U_\lambda (q) V_l|^2}
+ b \frac{Im U_N}{|1 - U_\lambda V_t|^2}
\ee

\noindent
where $a,b$ measure the strength of the longitudinal and
transverse parts.

For the transverse part of the photon self-energy $\Pi^{xx},
\Pi^{yy}$ the procedure to follow is identical to the one explained
in section 9 of \cite{CAR} and we refer the reader to this
paper (see also \cite{AMP}). The only novelty here is  $\Pi^{00}$,
but this component is of spin longitudinal character and is 
renormalized by means of eq.(~\ref{eq:pol}) with $a = 1,
b= 0$. In the $\Delta$ term the polarization effects are already
 included in the self-energy of ref. \cite{LOR}, hence, no
further corrections are needed.

\section{Short range correlations}

So far the calculations have been done using implicitly
plane waves for the nucleon states. Short range nuclear correlations
modify the two nucleon  relative wave function and this has a repercusion
in some of the matrix elements which we have calculated. This is
particularly true in those matrix elements which involve
a $p$-wave coupling in the vertex for each of the two nucleons,
because the pion exchange generates a $\delta (\vec{r})$ function,
which is rended inoperative in the presence of short range 
correlations. This is not exactly the case if finite size effects
by means of form factors are taken into account, but the need
to implement the effects of the short range correlations
remains. The correlations can also introduce spin transverse components
in the p-wave-p-wave terms which were originally of the spin
longitudinal nature \cite{WES}. Hence, at the same time that one
introduces the effect of correlations,
one takes advantage of this and introduces the $\rho$ meson exchange 
in this case. In this way, we generated $V_l$ and $V_t$ of eq.(41). 

The method to introduce the effects of correlations is to substitute
a two nucleon amplitude $V (q)$ by 

$$
V (\vec{q}) \rightarrow \frac{1}{(2 \pi)^3} \int d^3k V (\vec{k}) 
\Omega (\vec{q} - \vec{k})
$$

\noindent
where $\Omega (\vec{p})$ is the Fourier transform of a nuclear
correlation function.

Once again the techniques to make these corrections can be seen
in \cite{CAR} (appendix D, see also \cite{AMP}), and we do not
repeat them here.

\newpage

\vskip 0.2cm
\centerline{\protect\hbox{\psfig{file=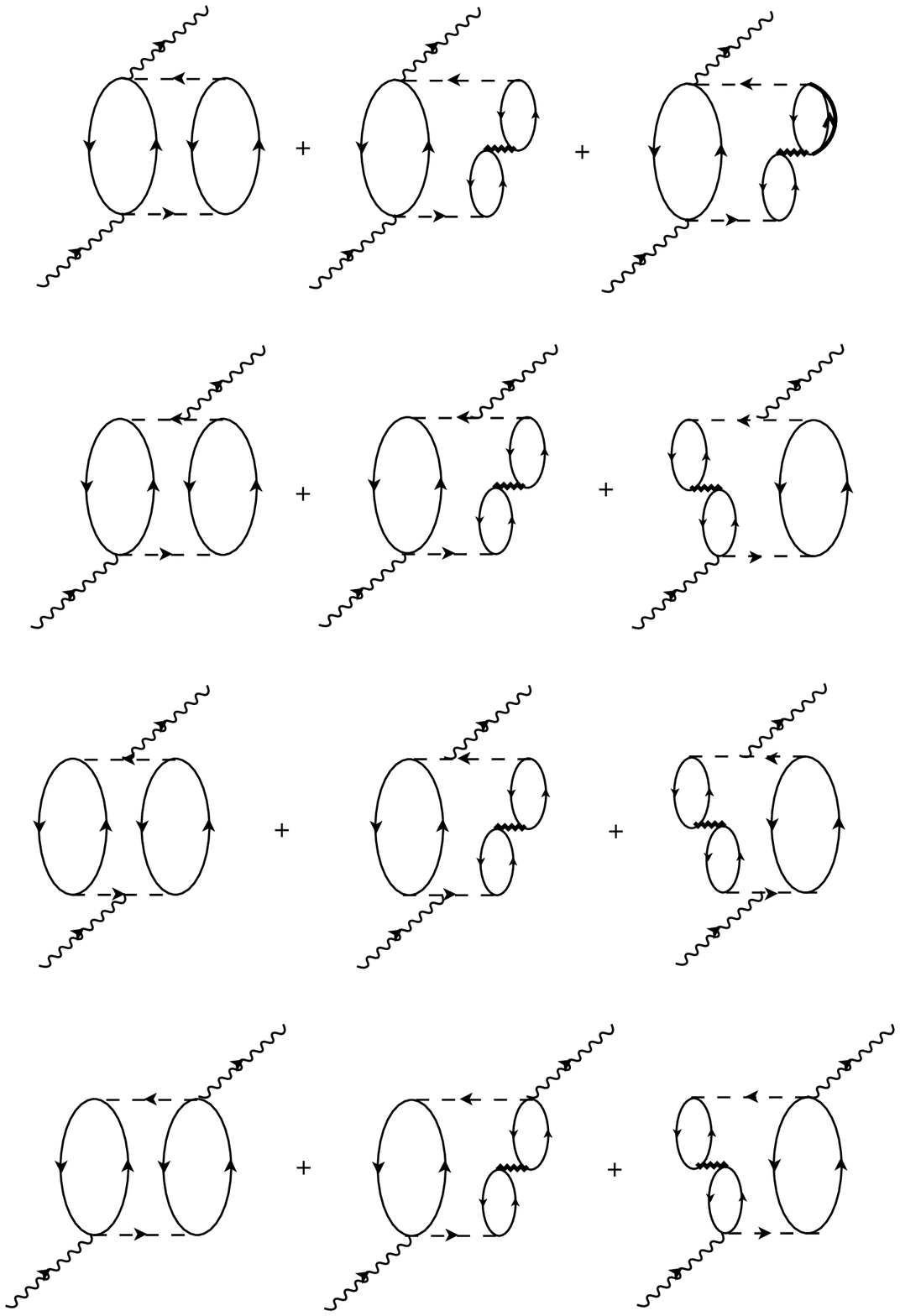,width=14cm}}}
\vskip 0.4cm
\vspace*{0.3cm}
\noindent
{\small {\bf Fig.4.16} Terms of the KR and pion pole
block implicit in fig. 4.9 showing the medium
polarization through RPA $ph$ and $\Delta h$ excitations induced by the pion.}

\newpage

\section{Detailed study of the  quasielastic peak}

\subsection{Formalism}

So far we have studied the $(\gamma^*, \pi)$ process and the
$\gamma^*$ absorption by a pair or trio of particles. This was
done keeping a parallelism to the real photon case. However, unlike
the case with real photons, a virtual photon can be absorbed by
one nucleon leading to the quasielastic peak of the response
function. We have left this problem till the end in order to 
introduce the concepts of many body which proved relevant in the
scattering of real photons with nuclei and in the equivalent 
channels of virtual photons studied before.

Then we can use the same concepts and ideas here in order to
introduce the appropriate many body corrections to the quasielastic
peak.

Thus we begin by evaluating $\Pi^{\mu \nu}$ for the $1ph$ 
excitation driven by the virtual photon, as depicted in fig. 6.1.
\vskip 0.2cm
\centerline{\protect\hbox{\psfig{file=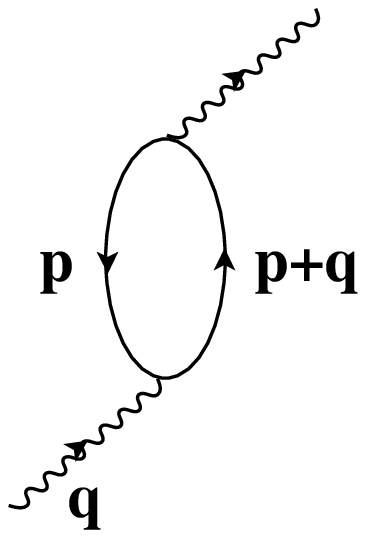,width=4cm}}}
\vskip 0.2cm
%\vspace*{-3.2cm}
{\small {\bf Fig.6.1} Photon self-energy diagram for the 1$ph$ excitation
driven by the virtual photon.}
\vspace*{0.5cm}

 The
photon self-energy associated to this diagram is given by

\be
\begin{array}{ll}
-i\Pi^{\mu\nu}={\displaystyle\int}\Frac{d^{4}p}{(2\pi)^{4}} & \Frac{i
n_{i}(p)}{p^{0}-E(\vec{p}\,)+i\epsilon}\times\\
&\\
& \times\Frac{i(1-n_{j}(p+q))}
{p^{0}+q^{0}-{E}(\vec{p}+\vec{q}\,)-i\epsilon}
Tr(V^{\mu}V^{\dagger\nu})
\end{array}
\ee

\noindent
where $V^\mu$ represents the $\gamma NN$ vertex and $E (\vec{p})$
is the nucleon kinetic energy. The vertex
$V^\mu$ is given by

\be
V^{\mu}=\bar{u}_r(\vec{p}\,)\left\{eF_1\gamma^{\mu}-ie\Frac{G_M}{2M_N}
\mu_n\sigma^{\mu \rho}q_{\rho}\right\}u_{{r}^{\prime}}(\vec{p}\,+\vec{q}\,)
\ee

Once again the application of Cutkosky rules leads to

\be
\begin{array}{ll}
Im\Pi^{00}=-{\displaystyle\int}\Frac{d^{3}p}{(2\pi)^{2}} &
n_{i}(p)(1-n_{j}(p+q))\times\\
&\\
& \times\delta (p^{0}+q^{0}-{E}(\vec{p}+\vec{q}\,)\,)
Tr(V^{0}V^{\dagger 0})
\end{array}
\ee

 \be
  \begin{array}{ll}
   Im\Pi^{xx}=-{\displaystyle\int}\Frac{d^{3}p}{(2\pi)^{2}} &
    n_{i}(p)(1-n_{j}(p+q))\times\\
     &\\
      & \times\delta (p^{0}+q^{0}-{E}(\vec{p}+\vec{q}\,)\,)
       Tr(V^{x}V^{\dagger x})
        \end{array}
         \ee

\noindent
and if we average $Tr (V^\mu V^{\dagger\nu})$ over the nucleon momentum
in the Fermi sea we can write

 \be
  Im\Pi^{00}=\Frac{1}{2}Im\bar{U}(q,\rho)\langle Tr(V^{0}V^{\dagger 0})\rangle
   \ee
   
    \be
      Im\Pi^{xx}=\Frac{1}{2}Im\bar{U}(q,\rho)\langle
Tr(V^{x}V^{\dagger x})\rangle 
 \ee
 
The average over the Fermi momentum can be done keeping terms
up to $q^2 /M_N^2$ and we find in terms of

$$
A^{\mu \nu} = \frac{1}{e^2} < Tr (V^\mu V^{\dagger \nu}) >
$$

\be
 A^{00}=\Frac{1}{{M}_{N}^{2}}\left\{
  \Frac{1}{1-\Frac{q^{2}}{4{M}_{N}^{2}}}
   \left[{G}_{E}^{2}(q)-\Frac{q^{2}}{4{M}_{N}^{2}}{G}_{M}^{2}(q)\right]
    \Frac{1}{2}(2p^{0}+q^{0})^{2}-\Frac{1}{2}{\vec{q}\,}^{2}{G}_{M}^{2}(q)
     \right\}
       \ee

           \be
                A^{xx}=\Frac{1}{{M}_{N}^{2}}\left\{
             \Frac{1}{1-\Frac{q^{2}}{4{M}_{N}^{2}}}
           \left[{G}_{E}^{2}(q)-\Frac{q^{2}}{4{M}_{N}^{2}}{G}_{M}^{2}(q)\right]
                 \Frac{2}{5}{k}_{F}^{2}-\Frac{1}{2}{q}^{2}{G}_{M}^{2}(q)
              \right\}
                      \ee

\noindent
where  $p^0 = M_N + \frac{3}{5} \frac{k_F^2}{2M_N}$ and $G_E, G_M$
are the Sachs form factors \cite{AML,MUL}.
\vspace*{1.3cm}

\subsection{Spectral function description and final state interaction}

One of the corrections to the bare $ph$ excitation studied above is
the one induced by final state interaction, as we indicated in 
section 2, which in our approach can be taken into account by 
dressing up the nucleon propagator of the particle state in the $ph$
excitation, as depicted in fig. 6.2 (there the dashed line would
account for the whole $NN$ interaction not just pion exchange). However,
some caution must be exerted when talking about this diagram.

\vskip 0.2cm
\centerline{\protect\hbox{\psfig{file=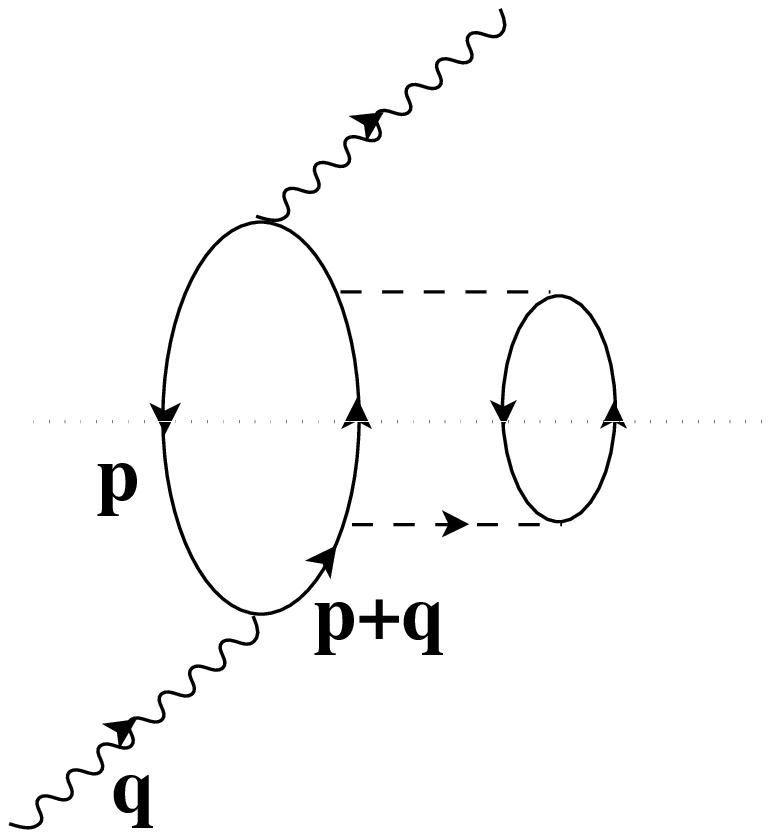,width=5.cm}}}
\vskip 0.2cm
\vspace*{-0.2cm}
\noindent
{\small {\bf Fig.6.2} Photon self-energy diagram obtained from fig. 6.1 
by dressing up the nucleon pro\-pa\-ga\-tor of the particle state in the
$ph$ excitation.
}
\vspace*{0.3cm}

In the first place, this is one of the terms implicit in the generic
diagram of fig. 4.9 when the nucleon pole term is
taken in each of the $\gamma^* N \rightarrow N N$ amplitudes. This term poses
no problem for real photons and leads to a small fraction of the two 
nucleon absorption. However, for virtual photons this diagram is 
divergent. The reason is that when placing the $2p 2h$ excitation on shell
through Cutkosky rules, we still have the square of the nucleon propagator 
with 
momentum $p + q$ in the figure. This propagator can be placed 
on shell for virtual photons (not for real photons) and we get a 
divergence.

\vskip 0.2cm
\centerline{\protect\hbox{\psfig{file=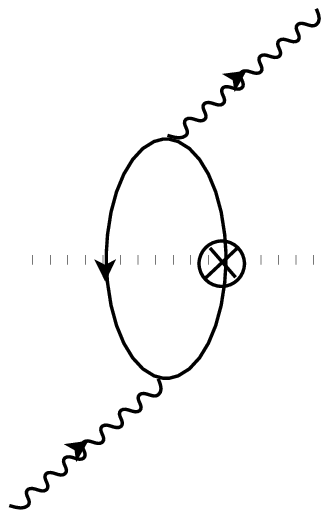,width=3.5cm}}}
\vskip 0.2cm
\vspace*{-0.2cm}
\noindent
{\small {\bf Fig.6.3} Insertion of the nucleon
self-energy on the nucleon line of the particle state.}
\vspace*{0.3cm}

The divergence is physical in the sense that its meaning is the
probability per unit time of absorbing a virtual photon by one
nucleon times the probability of collision of the final nucleon with
other nucleons in the infinite Fermi sea in the lifetime of this
nucleon \cite{STROT}. Since this nucleon is real, its lifetime is infinite
and thus the probability infinite. The problem is physically
solved \cite{STROT} by recalling that the nucleon in the Fermi sea has
a self-energy with an imaginary part which gives it a finite lifetime
(for collisions). This is also immediately taken into account technically
by iterating the nucleon self-energy insertion of fig. 6.3 on the
nucleon line, following the Dyson equation, hence substituting
the particle nucleon propagator by a renormalized nucleon propagator
including the nucleon self-energy in the medium,

 \be
  G(p^{0},\vec{p}\,)=\Frac{1}{p^{0}-\Frac{{\vec{p}\,}^{2}}{2M_N}-\Sigma(p^{0},
   \vec{p}\,)}
    \ee

\noindent
where $\sum (p^0 , \vec{p})$ is the nucleon self-energy. Alternatively
one can use the spectral function representation \cite{WAK}

\be\label{eq:spsh}
G (p^0 \vec{p}) = \int_{- \infty}^\mu d \omega \frac{S_h (\omega, 
\vec{p})}{p^0 - \omega - i \epsilon} + 
\int_{\mu}^\infty \frac{S_p (\omega, \vec{p})}{p^0 - \omega + i
\epsilon} d \omega
\ee

\noindent
where $S_h, S_p$ are the hole and particle spectral functions related to
$\Sigma$ by means of \cite{ANG}

\be
 \begin{array}{lcl}
  \mbox{* If}\,\,\, \omega \geq\mu & , & S_{p}(\omega,p)=-\Frac{1}{\pi}
   Im G(\omega, p)=-\Frac{1}{\pi}\Frac{Im\Sigma(\omega, p)}{A+B}\\
    \\
      \mbox{* If}\,\,\, \omega \leq\mu & , & S_{h}(\omega, p)=\Frac{1}{\pi}
         Im G(\omega, p)=\Frac{1}{\pi}\Frac{Im\Sigma(\omega, p)}{A+B}
          \end{array}
           \ee

\noindent
and $\mu$ is the chemical potential and
 
$$
\begin{array}{l}
A=\left( \omega-\Frac{{\vec{p}\,}^{2}}{2M_N}-Re\Sigma(\omega,\vec{p}\,)\right)^{2}\\
\\
B=\left( Im\Sigma(\omega,{\vec{p}\,})\right)^{2}
\end{array}
$$

By means of eq.(62)
 we can write the $ph$ propagator or new Lindhard
function incorporating the effects of the nucleon self-energy in
the medium and we have for $Im \bar{U}$

 \be
   \begin{array}{ll}
     Im\,\bar{{U}}(q)=-\Frac{1}{2\pi}& \!\!\!
       {\displaystyle \int_{0}^{\infty}}
         dp\, p^{2}
         {\displaystyle \int_{-1}^{1} dx}{\displaystyle \int_{\mu-q^{0}}
           ^{\mu}}d\omega  S_{h}(\omega, p)\times\\
             & \\
               & \times S_{p}(q^{0}+\omega ,{\displaystyle
                  \sqrt{{\vec{p}\,}^{2}+{\vec{q}\,}^{2}+2pqx}})
                    \end{array}
                      \ee

We use the spectral functions calculated in \cite{PED}, but since the
imaginary part of the nucleon self-energy for the hole states is much
smaller than that of the particle states under consideration we make
the approximation of setting to zero $Im \Sigma$ for the hole
states. This was found to be a good approximation in \cite{CIO}. Thus,
we take

 \be
   S_{h}(\omega,p)=\delta(\omega-\tilde{E}(\vec{p}\,))\Theta(\mu-\tilde{E}(p))
     \ee

\noindent
where $\tilde{E} (p)$ is the energy associated to a momentum $\vec{p}$
obtained selfconsistently by means of the equation

\vspace{0.2cm}                           
\be
\tilde{E}(\vec{p}\,)=\Frac{{\vec{p}\,}^{2}}{2M_N}+Re\Sigma (\tilde{E}(\vec{p}\,),
    \vec{p}\,)
\ee

The chemical potential was then taken as

$$
\mu = \frac{k_F^2}{2M_N} + Re \Sigma (\mu, k_F)
$$

\noindent
where $k_F$ is the Fermi momentum.

It must be stressed that it is important to keep the real part of 
$\Sigma$ in the hole states when renormalizing the particle states because
there are pieces in the nucleon  self-energy largely independent of the 
momentum and which cancel in the $ph$ propagator, where the two 
selfenergies subtract.

The effect of the use of the spectral function, accounting for FSI is
a quenching of the quasielastic peak and a spreading of the 
strength at higher energy as can be seen in fig. 6.4.

\vskip 0.3cm
\centerline{\protect\hbox{\psfig{file=spectralnew.ps,width=10cm}}}
\vskip 0.1cm
\noindent
{\small {\bf Fig.6.4} Effect of the use of the spectral function in the
evaluation of the Lindhard function.
The uncorrelated Fermi sea results are obtained from eqs.(57),(58).
Those with the medium spectral function, with the same equations
substituting the bare Lindhard function $\bar{{U}}$ by the medium 
modified one of eq.(64).
}

\subsection{Polarization (RPA) effects in the quasielastic contribution}

We take now into account polarization effects in the $1p 1h$ excitation,
substituting it by an RPA response as shown diagrammatically in fig.
6.5.

\vskip 0.2cm
\centerline{\protect\hbox{\psfig{file=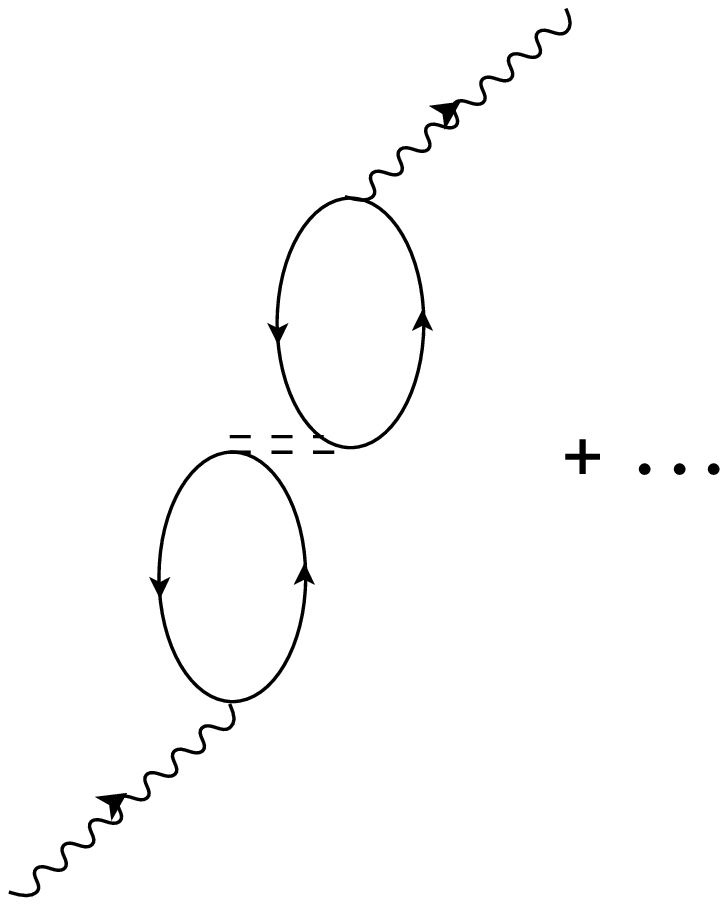,width=4cm}}}
\vskip 0.2cm
\vspace*{-0.5cm}
{\small {\bf Fig.6.5} Diagrammatic representation of the polarization 
effects in the
1$ph$ excitation.}
\vspace*{0.25cm}

For that purpose we use an effective interaction of the Landau-Migdal
type

\be
\begin{array}{ll}
V(\vec{r}_{1},\vec{r}_{2})= & c_{0}\delta(\vec{r}_{1}-\vec{r}_{2})\left\{
f_{0}(\rho)+f_{0}^{\prime}(\rho)\vec{\tau}_{1}\vec{\tau}_{2}\right.+\\
& \\
& +\left.g_{0}(\rho)\vec{\sigma}_{1}\vec{\sigma}_{2}+g_{0}^{\prime}(\rho)
(\vec{\sigma}_{1}\vec{\sigma}_{2})
\vec{\tau}_{1}\vec{\tau}_{2}
\right\}
\end{array}
\ee

and we take the parameterization for the coefficients from ref. 
~\cite{SPH}

\be
f_{i}(\rho (r))=\Frac{\rho (r)}{\rho (0)} f_{i}^{(in)}+
\left[ 1-\Frac{\rho (r)}{\rho (0)}\right] f_{i}^{(ex)}
\ee

\be
\begin{array}{ll}
f_{0}^{(in)}=0.07 & f_{0}^{\prime (ex)}=0.45\\
f_{0}^{(ex)}=-2.15 & c_{0}=380\, MeV fm^{3}\\
f_{0}^{\prime (in)}= 0.33 & \\
g_{0}^{(in)}=g_{0}^{(ex)}=g_{0}=0.575\\
g_{0}^{\prime (in)}=g_{0}^{\prime (ex)}=g_{0}^{\prime}=0.725
\end{array}
\ee

For consistency, in the $S = 1 = T$ channel ($\vec{\sigma} \vec{\sigma}
\vec{\tau} \vec{\tau}$ operator)  we have continued to use the interaction used in
~\cite{CAR} which has been used for the renormalization of the pionic
and pion related channels studied in the former sections. There is only
a minor difference of about 4$\%$ in $g^\prime_0$ between the two 
parametrizations.

Recalling that we had

\be
\Pi^{\mu\nu}_{(1)}=\frac{1}{2} \bar{U}_{N}(q)A_{N}^{\nu\mu}(q)e^{2}
\ee

\noindent
let us take the nonrelativistic reduction of $A^{\nu \mu}$ in
order to see the effects of the RPA renormalizations

\be\label{eq:amunu}
\begin{array}{ll}
A^{\mu\nu}\equiv & {\displaystyle \sum_{r,r^{\prime}}}
\chi_{r}\left[F_{1}^{p}(q)\delta^{\mu 0}-i\Frac{\mu_{p}G_{M}(q)}
{2M_{N}}(\vec{\sigma}\times\vec{q}\,)_{i}\delta^{\mu i}
+F_{1}^{p}\Frac{(2\vec{p}+\vec{q}\,)_{i}}{2M_{N}}\delta^{\mu i}
\right]\chi_{r^{\prime}}\times \\
& \\
&\times\chi_{r^{\prime}}\left[ F_{1}^{p}(q)\delta^{\nu 0}+
F_{1}^{p}\Frac{(2\vec{p}+\vec{q}\,)_{i}}{2M_{N}}\delta^{\nu i}
+i\Frac{\mu_{p}G_{M}(q)}
{2M_{N}}(\vec{\sigma}\times\vec{q}\,)_{i}\delta^{\nu i}
 \right]\chi_{r}+\\
 & \\
 & +(p\leftrightarrow n)
 \end{array}
 \ee

Given the spin-isospin structure,
the electric and magnetic components will be renormalized in the following
way:

{\bf a)} Interaction $\vec{\sigma} \vec{\sigma} \vec{\tau} \vec{\tau}$: is the
one we used to renormalize the pionic related channels in former
sections. It affects only the magnetic components. If we write

  \be
    \begin{array}{ll}
      A_{mag.}^{\mu\nu}= & {\displaystyle \sum_{r, r^{\prime}}}
        \chi_{r}\left[-i\Frac{\mu_{p}G_{M}(q)}
          {2M_{N}}(\vec{\sigma}\times\vec{q})_{i}\delta^{\mu i}
            \right]\chi_{r^{\prime}}\times\\
              & \\
                & \times \chi_{r^{\prime}}\left[i\Frac{\mu_{p}G_{M}(q)}
                    {2M_{N}}(\vec{\sigma}\times\vec{q})_{i}\delta^{\nu i}
                        \right]\chi_{r}
                            \Frac{(1+\tau_{3})}{2}+\\
                              & \\
                                & + \mbox{[neutrons]}\Frac{(1-\tau_{3})}{2}
                                  \end{array}
                                    \ee

\noindent
it is easy to see that the magnetic part of $\Pi^{ij}$  becomes

\be\label{eq:a}
  \begin{array}{ll}
    \Pi^{ij}_{mag.}= & \frac{1}{2}\bar{{U}}_{N}A^{ij}_{mag.}(q)e^{2}+
      \Frac{e^{2}}{4{M}_{N}^{2}}\Frac{{f}_{\pi NN}^{2}}{{m}_{\pi}^{2}}\Frac{
          V_{t}(q)}
                {1-\Frac{{f}_{\pi NN}^{2}}{{m}_{\pi}^{2}}V_{t}(q){U}(q)}\times
                  \\
                    &
    \\
      & \times ({\vec{q}\,}^{2}\delta^{ij}-q^{i}q^{j})G_{M}^{2}(q)
         \left( \mu_{p}\bar{{U}}_{p}-\mu_{n}\bar{{U}}_{n}\right)^{2}
         \end{array}
        \ee

\noindent
where $U = U_N + U_\Delta$.

{\bf b)} Interaction $\vec{\tau} \vec{\tau}$:

        This interaction selects the non magnetic components of $V^\mu$.
Thus $A^{00}$ and the convective terms of $A^{ij}$ (term with $2 \vec{p} + 
\vec{q}$ in eq.(~\ref{eq:amunu})) are renormalized.

However, given the smallness of the convective terms (about 10$\%$
contribution to the transverse response) we shall not consider their 
renormalization.

Thus we consider only the modification to $A^{00}$ from this source.
Since $A^{00}$ is given by

\be
 \begin{array}{ll}
  A^{00} & = \left[ {\displaystyle \sum_{r,r^{\prime}}}\chi_{r}F_{1}^{p}(q)\chi_{
   ^{\prime}}\chi_{r^{\prime}}F_{1}^{p}(q)\chi_{r}\right]
    \Frac{(1+\tau_{3})}{2}+\\
     & \\
      & + \left[ {\displaystyle \sum_{r,r^{\prime}}}\chi_{r}F_{1}^{n}(q)\chi_{r
        ^{\prime}}\chi_{r^{\prime}}F_{1}^{n}(q)\chi_{r}\right]
           \Frac{(1-\tau_{3})}{2}
              \end{array}
               \ee

\noindent
the renormalized expression for $\Pi^{00}$ will be

\be
 \Pi^{00}=e^{2}\left\{(F_{1}^{p})^{2}\bar{{U}}_{p}+
  (F_{1}^{n})^{2}\bar{{U}}_{n}+\Frac{c_{0}f_{0}^{\prime}(
   F_{1}^{p}\bar{{U}}_{p}-F_{1}^{n}\bar{{U}}_{n})^{2}}{
    1-c_{0}f_{0}^{\prime}{U}_{N}(q)}\right\}
     \ee

\noindent
where in the denominator we do not have now $U_\Delta$ since the operator
$\vec{\tau} \vec{\tau}$ cannot excite $\Delta$ components.

{\bf c)} Interaction $\vec{\sigma} \vec{\sigma}$:

Here again, like in case a), only the magnetic components are modified.
We find

  \be\label{eq:c}
     \begin{array}{ll}
       \Pi^{ij}_{mag.}= & \frac{1}{2}\bar{{U}}_{N}A^{ij}_{mag.}(q)e^{2}+
        \Frac{e^{2}}{4M^{2}}c_{0}g_{0}\Frac{
           1}{1-c_{0}g_{0}{U}_{N}(q)}\times  \\
               &
                  \\
             & \times ({\vec{q}\,}^{2}\delta^{ij}-q^{i}q^{j})G_{M}^{2}(q)
           \left( \mu_{p}\bar{{U}}_{p}+\mu_{n}\bar{{U}}_{n}\right)^{2} \\
                            &
                                \\
             &+\,\,\, effect\,\,\,of\,\,\,(a)
                                         \end{array}
                                             \ee

The correction from the RPA sum is taken into account by means of the
second term of the right hand side of eq.(~\ref{eq:c}) to which we
should add the same term in eq.(~\ref{eq:a}) coming from the
renormalization with the $\vec{\sigma} \vec{\sigma} \vec{\tau} \vec{\tau}$
operator.

{\bf d)} Scalar interaction:

        This one affects $A^{00}$ and we find 

\be
  \Pi^{00}=e^{2}\left\{(F_{1}^{p})^{2}\bar{{U}}_{p}+
     (F_{1}^{n})^{2}\bar{{U}}_{n}+\Frac{c_{0}f_{0}(\rho)(
         F_{1}^{p}\bar{{U}}_{p}+F_{1}^{n}\bar{{U}}_{n})^{2}}{
              1-c_{0}f_{0}{U}_{N}(q)}\right\}\,\,+\,\,\, effect\,\, of\,\,(b)
                    \ee

We show in figs. 6.6, 6.7, 6.8 and 6.9 the effects
on $R_L$ and $R_T$ ($R_L=-\Frac{ |\vec{q}\,|^{2} }{q^2}W_L$ and $R_T=2W_T$)
 of the different polarization terms. The solid line corresponds to the 
 calculation including these effects and the dashed line, to the calculation
 without polarization effects (we are using spectral functions in the
  calculation of the Lindhard function): 

 \vspace*{0.2cm}
   \centerline{\protect\hbox{\psfig{file=rlf0esc.ps,width=11.3cm}}}
      \vskip 0.2cm
\noindent
     {\small {\bf Fig.6.6} Polarization (RPA)
      effect (solid line) in the evaluation of
     $R_L$: scalar interaction.}

%\newpage
\vspace*{0.1cm}
   \centerline{\protect\hbox{\psfig{file=rlfprima0.ps,width=11.3cm}}}
         \vskip 0.2cm
\noindent         
   {\small {\bf Fig.6.7}
   Polarization (RPA) effect (solid line) in the evaluation of
        $R_L$: $\vec{\tau}\vec{\tau}$ interaction.
   }

\newpage

       \vspace*{0.4cm}
        \centerline{\protect\hbox{\psfig{file=rtss.ps,width=11.5cm}}}
                         \vskip 0.2cm
\noindent
            {\small {\bf Fig.6.8}
            Polarization (RPA) effect (solid line) in the evaluation of
                 $R_T$: $\vec{\sigma}\vec{\sigma}$ interaction.}
\vspace*{0.7cm}

      \vskip 0.2cm
     \centerline{\protect\hbox{\psfig{file=rtsstt.ps,width=11.5cm}}}
    \vskip 0.2cm
\noindent    
     {\small {\bf Fig.6.9} Polarization (RPA) effect (solid line) in the evaluation of
                      $R_T$: 
                      $\vec{\sigma}\, \vec{\sigma}\,\vec{\tau}\,\vec{\tau}$
                       interaction.}
\vspace*{0.1cm}

\newpage
One possible source of renormalization not yet considered is the one
shown in fig. 6.10, where a $ph$ is attached to one photon and a 
$\Delta h$ to the other one, plus any other possible $ph$ 
or $\Delta h$ excitations in between.

 \vspace*{0.3cm}
  \centerline{\protect\hbox{\psfig{file=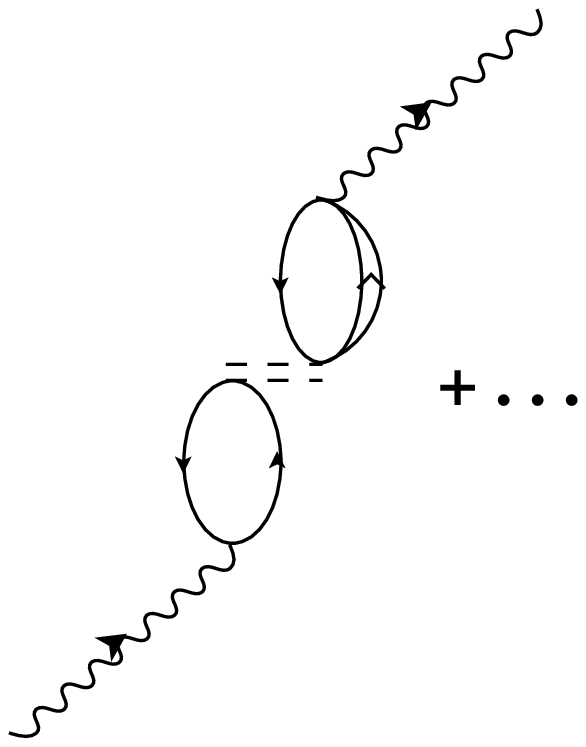,width=5.9cm}}}
   \vskip 0.2cm
    \vspace*{-1.5cm}
\noindent
    {\small {\bf Fig.6.10} Diagrammatic representation of a possible
    source of renormalization: a $ph$ is attached to one photon and a
    $\Delta h$ to the other one, plus any other possible $ph$
    or $\Delta h$ excitations in between.}
\vspace*{0.5cm}
      
This contribution affects the transverse part and both  the quasielastic
as well as the $\Delta$  peak. However, given the small interference
between $ph$ and $\Delta h$ excitations, the contribution of these terms is
not significant. We give, however, the expression here for completeness

\be
 \begin{array}{ll}
  \Pi^{ij}= e\left(\Frac{4}{3}\right)^{2}&\!\! \Frac{f_{\gamma}}{m_{\pi}}
   \Frac{\rho}{\left(\sqrt{s}-M_{\Delta}+i\frac{\bar{\Gamma}}{2}-\Sigma_{\Delta}
    \right)}i\times\\
     &\\
      & \times\Frac{V_{t}}{\left(1-{U}
       \Frac{{f}_{\pi NN}^{2}}{{m}_{\pi}^{2}}
        V_{t}\right)}\Frac{f_{\pi NN}}{m_{\pi}}\Frac{
         f^{*}}{m_{\pi}}[{\vec{q}\,}^{2}\delta^{ij}-q^{i}q^{j}]\times\\
          & \\
           & \times\Frac{G_{M}}{4M_N}[\mu_{p}\bar{{U}}_{p}-\mu_{n}\bar{{U}}_{n}]
            \end{array}
             \ee

The effect of the polarization is moderate, but relevant when aiming
at a precise description of the process.

\vspace*{0.8cm}
We show in figs. 6.11 and 6.12 the effects of the polarization
in the longitudinal and transverse response functions. The net effect 
in the cross section is a quenching in the quasielastic peak and
a spreading of the strength at higher energies.
\vskip 0.1cm
\centerline{\protect\hbox{\psfig{file=sinpolrl.ps,width=11.cm}}}
\vskip 0.1cm
\noindent
{\small {\bf Fig.6.11} 
 Polarization (RPA) effect in the evaluation of
                       $R_L$. }
\vspace*{0.1cm}

\vskip 0.6cm
\centerline{\protect\hbox{\psfig{file=sinpolrt.ps,width=11.cm}}}
\vskip 0.1cm
\noindent
{\small {\bf Fig.6.12}
Polarization (RPA) effect in the evaluation of
                       $R_T$.}
\vspace*{0.1cm}

\subsection{Further considerations}

Some other terms appearing in the generic diagram of fig. 4.11 
require some special thought. These are the terms in which one of the
vertices contains the nucleon pole term of the $\gamma^* N \rightarrow
\pi N$ amplitude, while the other one contains all terms but
that one. This is depicted in fig. 6.13.

 \vskip 0.2cm
   \centerline{\protect\hbox{\psfig{file=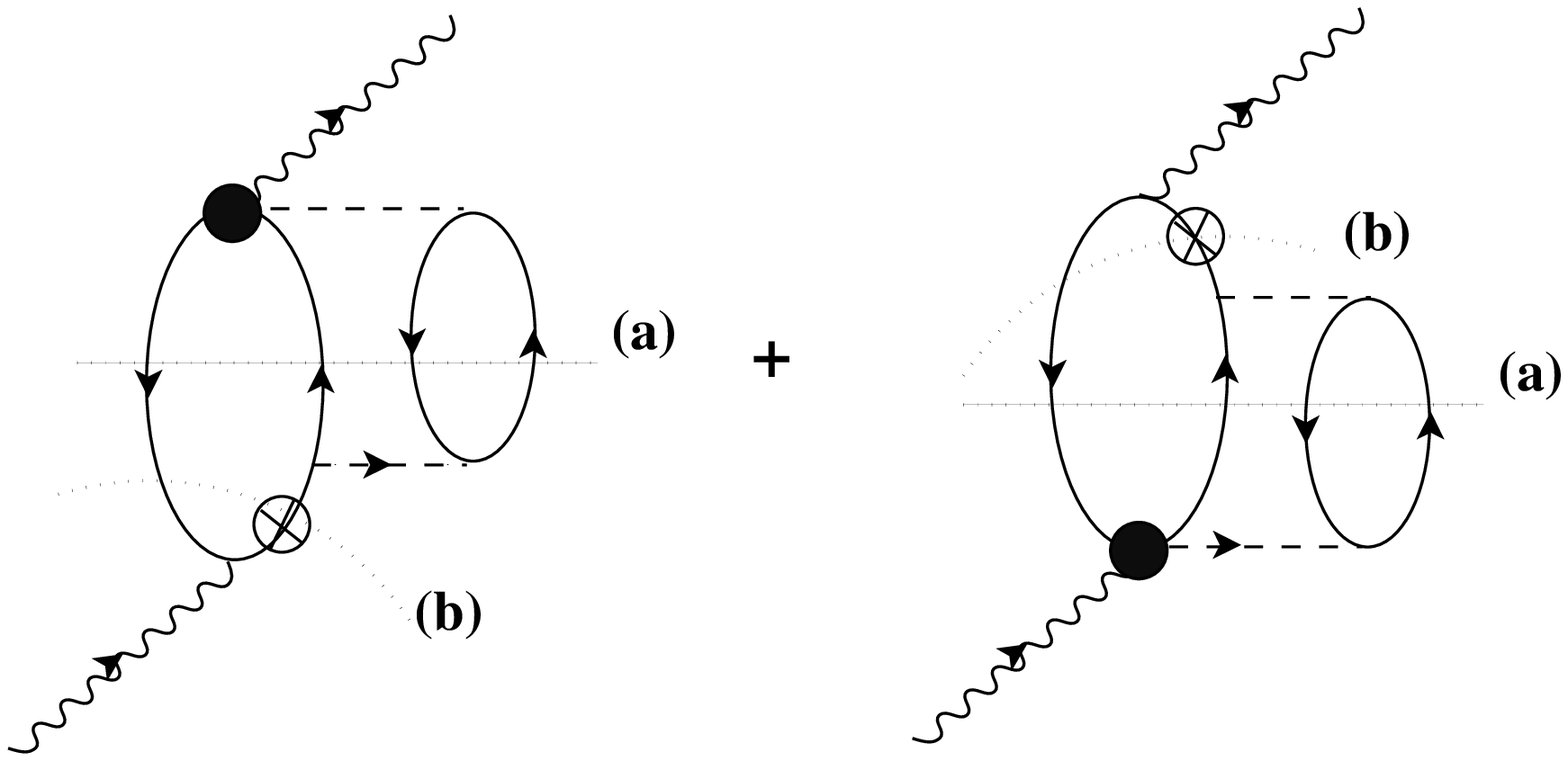,width=9.5cm}}}
      \vskip 0.2cm
          %\vspace*{-3.2cm}
\noindent
{\small {\bf Fig.6.13} Photon self-energy diagrams in which one of
the vertices contains the
nucleon pole term of the $\gamma^* N \rightarrow
\pi N$ amplitude, while the other one contains all terms but that one.}
\vspace*{0.5cm}

Unlike the case of real photons where only the cut (a) exciting
$2p 2h$ gives rise to an imaginary part, now the cut (b) 
placing the $ph$ on shell is a source of imaginary part which produces
strength is the quasielastic peak. The diagrams in fig. (6.13)
for the cut (b) could
be then considered an ordinary $ph$ excitation with a renormalized
vertex. We have evaluated the two sources of
imaginary part in $\Pi^{\mu \nu}$ and find for the cut (a)

\be
\begin{array}{ll}
Im \Pi^{\mu\nu}=&\Frac{{f}_{\pi NN}^{2}}{m_{\pi}^{2}}
{\displaystyle \sum_{ij}}{\displaystyle \int}\Frac{d^{4}p}{(2\pi)^{4}}
n_{i}(\vec{p})(2\pi)\Theta (p^{0})
\delta\left(p^{0}-{E}(\vec{p}\,)-Re\Sigma
\left(\Frac{{\vec{p}\,}^{2}}{2M_{N}},\vec{p}\,\right)\right)\times\\
& \\
& \times {\displaystyle \int} \Frac{d^{4}k}{(2\pi)^{4}}F_{\pi}^{4}(k)
\Frac{{\vec{k}}^{2}_{CM}\Theta (k^{0})}{(k^{2}-m_{\pi}^{2})^{2}}2\pi
\Theta(p^{0}+q^{0}-k^{0})\times \\
& \\
&\delta\left(p^{0}+q^{0}-k^{0}-{E}(\vec{p}+\vec{q}-\vec{k}\,)
 -Re\Sigma\left( \Frac{(\vec{p}+\vec{q}-\vec{k}\,)^{2}}{2M_N} ,
\vec{p}+\vec{q}-\vec{k}\,\right) \right)\times\\
& \\& \\
&\times (1-n_{j}(\vec{p}+\vec{q}-\vec{k}\,))
\Frac{Im {U}_{\lambda}(k)}{1-\frac{{f}_{\pi NN}^{2}}{m_{\pi}^{2}}V_{l}(k)
{U}(k)}(1-n_{i}(\vec{p}+\vec{q}\,))\times\\
& \\
& \times 2 Re\left\{\Frac{1}{p^{0}+q^{0}-{E}(\vec{p}+\vec{q}\,)
-\Sigma_{N}\left(\frac{(\vec{p}+\vec{q}\,)^{2}}{2M_N}, \vec{p}+\vec{q}\,\right)}
\right.\times \\
& \\
& \left.Tr\left(\bar{\cal{M}}^{\mu}_{NP}(i\rightarrow j)
\bar{\cal{M}}^{\dagger \nu}_{{}_
{KR,PP,NPC,\Delta,\Delta C}}
(i\rightarrow j)\right)\right\}

\end{array}
\ee

\noindent
where $\bar{\cal{M}}_{NP}$ is the nucleon pole amplitude of $\gamma^* N
\rightarrow \pi N$ omitting the nucleon propagator. On the other
hand  the contribution of the cut (b) is given by

\be
\begin{array}{ll}
Im\Pi^{\mu\nu}=& -{\displaystyle \sum_{ij}}{\displaystyle \int}
\Frac{d^{4}p}{(2\pi)^{4}}n_{i}(p) \delta\left(p^{0}
-E(\vec{p}\,)-Re\Sigma\left(\Frac{{\vec{p}\,}^{2}}{2M_N}, \vec{p}\,\right)
\right)
\times\\
& \\
&\times\Theta (p^{0}) (2\pi)^{2}(1-n_{i}(\vec{p}+\vec{q}\,))
\Theta(p^{0}+q^{0})\times\\
&\\
&\times\delta\left(p^{0}+q^{0}-{E}(\vec{p}+\vec{q}\,)
-Re\Sigma\left( \Frac{(\vec{p}+\vec{q}\,)^{2}}{2M_N} ,
\vec{p}+\vec{q}\,\right)\right)\times\\
&\\
&{\displaystyle \int }
\Frac{d^{4}k}{(2\pi)^{4}}F_{\pi}^{2}(k) Im\left\{\Frac{(1-n_{j}(p+q-k))}{
p^{0}+q^{0}-k^{0}-{E}(\vec{p}+\vec{q}-\vec{k}\,)+i\epsilon}\right.\times\\
& \\
& \times\left(\Frac{1}{k^{2}-m_{\pi}^{2}-\Pi}-\Frac{1}{
{k}^{2}-m_{\pi}^{2}+i\epsilon}\right)\times\\
& \\
& \times Tr\left.\left(\bar{\cal{M}}^{\nu}_{{}_
{KR,PP,NPC,\Delta,\Delta C}}
(i\rightarrow j)\bar{\cal{M}}^{\dagger\mu}_{NP}(i\rightarrow j)
\right)\right\}
\end{array}
\ee

\noindent
where $\Pi$ is the pion self-energy in the nuclear medium

\be
\Pi={\vec{k}}^2_{CM}\left(\Frac{f_{\pi NN}}{m_{\pi}}\right)^2
{F}^{2}_{\pi}(k^2)\Frac{U_{\lambda}(k)}{1-g^{\prime}
\left(\Frac{f_{\pi NN}}{m_{\pi}}\right)^2
U(k)}
\ee

\noindent
the subtraction of the free pion propagator 
 appearing in eq.(80) guarantees that in the limit
$\rho \rightarrow 0$ the correction to the $\gamma NN$ vertex
vanishes as it should be.

\subsection{Considerations on gauge invariance}

Most of the theoretical models found in the literature 
 on inclusive $(e,e^\prime)$ scattering from nuclei do not preserve 
gauge invariance. The requirement of invariance under gauge transformations
leads to relations between the components (charge and 
spatial current) of the hadronic current which determines the nuclear
response. Actually, the longitudinal (charge) multipoles are related
to the two transverse (spatial current) multipoles of the electric
type~\cite{Hei}. Several prescriptions have been used to restore
gauge invariance~\cite{Hei},~\cite{Friar}. However, 
in a recent work~\cite{AAL} the arbitrariness
of the most common prescriptions 
is discussed in detail
with the conclusion that
the standard procedures to impose  gauge invariance in
calculations, based on models which do not verify it, are misleading
and for very low nuclear excitation energies 
do not ensure at all that a better, or a more reasonable,
description of the data will be obtained.

Our model, as we shall see, is gauge invariant  at the lowest order
(impulse approximation) of 
the density expansion  and this symmetry is
only partially broken when some non-leading density corrections are
included. In what follows, we will study the consequences of the
partial breaking
of the gauge symmetry for the kinematics studied in this paper (from the
quasielastic peak to the $\Delta$ excitation region), which involves
larger nuclear excitation energies than those studied
in~\cite{AAL}. We will also study the feasibility of non-gauge
invariant models to  disentangle between longitudinal and transverse channels.

The unpolarized  cross section for inclusive $(e,e^\prime)$ scattering from
nuclei is given by (eq.(29)):

 \begin{equation}\label{eq:j1}
  \frac{\displaystyle{d^{2}\sigma}}{\displaystyle{d\Omega^{\prime}_e 
dE^{\prime}_e}}
   =\frac{\displaystyle{\alpha^{2}}}{\displaystyle{q^{4}}}\frac{\displaystyle{
    |\vec{k}^{\,\prime}|}}{\displaystyle{|\vec{k}|}}L^{\mu\nu}W_{\mu\nu}
     \end{equation}

Lorentz, space-inversion, time-reversal and gauge invariance 
constraint the form of the hadronic 
tensor $W_{\mu\nu}$, which determines the nuclear
response. Indeed, the most general expression for this tensor assuming 
the latter symmetries is given by~\cite{Muta}:

\begin{eqnarray}\label{eq:j2}
W^{\mu\nu} & = & \{\frac{q^\mu q^\nu}{q^2}-g^{\mu\nu}\}W_1 \nonumber\\ 
& + & \left \{ \left ( P^\mu -\frac{P.q}{q^2}q^\mu\right ) 
\left ( P^\nu -\frac{P.q}{q^2}q^\nu\right ) \frac{W_2}{M^2_A}     \right \}
\end{eqnarray}
with $q$ and $P$ the virtual photon and initial hadronic system four-momenta
respectively and $M_A^2 = P^2$. 
The structure functions $W_{1,2}$ are unknown  scalar functions
of the virtual photon variables which determine the nuclear response.

Using the expression for the hadronic tensor of eq.~(\ref{eq:j2})  
and taking $\vec{q}$ in the $z$ direction the cross
section of eq.~(\ref{eq:j1}) in the lab system becomes (eq. (31)):

 \begin{equation}\label{eq:j3}
  \frac{\displaystyle{d^{2}\sigma}}{\displaystyle{d\Omega^{\prime}_e
    dE^{\prime}_e}}=
     \left(\frac{\displaystyle{d\sigma}}{\displaystyle{d\Omega}}\right)_{Mott}
      \left(-\frac{\displaystyle{q^{2}}}{\displaystyle{|\vec{q}\,|^{2}}}\right)
       \left\{W_{L}(\omega ,|\vec{q}\,|)+
        \frac{\displaystyle{W_{T}(\omega,|\vec{q}\,|)}}
         {\displaystyle{\epsilon}}\right\}
          \end{equation}
\noindent
where the structure functions $W_L$ and $W_T$ are given in terms of
$W_1$ and $W_2$ by:

\begin{eqnarray}\label{eq:j4}
W_L &\equiv& -W_1 -\frac{|\vec{q}\,|^2}{q^2}W_2 =
-\frac{q^2}{\omega^2}W^{zz}= -\frac{q^2}{|\vec{q}\,|^2}W^{00}= 
-\frac{q^2}{\omega|\vec{q}\,|}W^{0z}\\
W_T  &\equiv& W_1 = W^{xx}= W^{yy}
\end{eqnarray}

Let us now suppose  that the hadronic model does not preserve gauge
invariance. In these circumstances the hadronic tensor (we will call
it ${\cal W}^{\mu\nu}$ to differentiate it from the one defined in
eq.~(\ref{eq:j2}))  is not conserved (ie, $q_\mu {\cal W}^{\mu\nu} \ne
0$, $q_\nu {\cal W}^{\mu\nu} \ne
0$ ), and  is now given in terms of four independent
functions: 
 
\begin{eqnarray}\label{eq:j5}
{\cal W}^{\mu\nu} & = & \{\frac{q^\mu q^\nu}{q^2}-g^{\mu\nu}\}W_1 \nonumber\\ 
& + & \left \{ \left ( P^\mu -\frac{P.q}{q^2}q^\mu\right ) 
\left ( P^\nu -\frac{P.q}{q^2}q^\nu\right ) \frac{W_2}{M^2_A}     \right \}
\nonumber\\ 
& + & W_3 \frac{q^\mu q^\nu}{q^2} + W_4 \frac{P^\mu P^\nu}{M^2_A}
\end{eqnarray}

Because of the loss of gauge invariance now ${\cal W}^{00}$,
${\cal W}^{0z}$ and ${\cal W}^{zz}$ are no longer related and become
independent. Thus, we have now:

\begin{eqnarray}\label{eq:j6}
W_L &\equiv& -W_1 -\frac{|\vec{q}\,|^2}{q^2}W_2 = 
\frac{|\vec{q}\,|}{\omega}{\cal W}^{0z}-{\cal W}^{zz}\\
W_T  &\equiv& W_1 = {\cal W}^{xx}= {\cal W}^{yy}\\
W_3 & = & \frac{\omega}{|\vec{q}\,|}{\cal W}^{0z}-{\cal W}^{zz}\\
W_4 & = & {\cal W}^{00} + {\cal W}^{zz} - \left
(\frac{|\vec{q}\,|}{\omega} + \frac{\omega}{|\vec{q}\,|}\right ){\cal W}^{0z} 
\end{eqnarray}

If gauge invariance is restored and therefore the hadronic tensor is
conserved, the response functions $W_3$ and $W_4$ vanish and $W_L$
reduces to any of the expressions of eq.~(\ref{eq:j4}). With this new 
hadronic tensor the differential cross section is now given by:

 \begin{equation}\label{eq:j7}
  \frac{\displaystyle{d^{2}\sigma}}{\displaystyle{d\Omega^{\prime}_e
    dE^{\prime}_e}}=
     \left(\frac{\displaystyle{d\sigma}}{\displaystyle{d\Omega}}\right)_{Mott}
      \left(-\frac{\displaystyle{q^{2}}}{\displaystyle{|\vec{q}\,|^{2}}}\right)
       \left\{ {\cal W}_{L}(\omega ,|\vec{q}\,|) +
        \frac{\displaystyle{W_{T}(\omega,|\vec{q}\,|)}}
         {\displaystyle{\epsilon}}\right\}
          \end{equation}

with 
\begin{eqnarray}\label{eq:j8}
{\cal W}_L (\omega ,|\vec{q}\,|) &=&  \left ( W_{L}(\omega ,|\vec{q}\,|) -
\frac{|\vec{q}\,|^2 }{q^2}W_{4}(\omega ,|\vec{q}\,|)\right
)\nonumber\\
&=& -\frac{\omega^2}{q^2}{\cal W}^{zz}-\frac{|\vec{q}\,|^2}{q^2}{\cal
W}^{00} + 2 \frac{|\vec{q}\,|\omega}{q^2} {\cal
W}^{0z}
\end{eqnarray}

The function $W_3$ does not appear in the expression for the cross
section because the leptonic tensor is conserved. Note that, one can
still factor out the differential cross section in the form $A+B/\epsilon$
and therefore, despite the breaking of gauge invariance, one can still
compare the results to the experimental response functions obtained via the
Rosenbluth plot. Then, the breaking of gauge invariance in the theoretical
model for the nuclear response leads to a redefinition of the response
function which has to be compared to the experimental one, in the
longitudinal channel, and thus one should compute ${\cal W}_L$ given 
in eq.~(\ref{eq:j8}) instead of $W_L$ of eq.~(\ref{eq:j4}). In fact,
the latter one is not well defined and there is an arbitrariness in
its definition because the $00$, $0z$ and $zz$ components of 
${\cal W}^{\mu\nu}$ are no longer related.

Traditionally, the longitudinal response function is calculated from
the ``charge-charge'' component of the hadronic tensor 
($ -q^2/|\vec{q}\,|^2{\cal W}^{00}$). In general, the response
function calculated in this way will differ from that calculated by
means of eq.~(\ref{eq:j8}). Now we will examine 
the difference between both approaches 
as a function of  the energy and momentum 
transferred to the nucleus. In order to do that, we define
\begin{eqnarray}\label{eq:j9}
{\cal W}^{zz} & = & \frac{\omega^2}{|\vec{q}\,|^2}{\cal W}^{00} +
\Delta {\cal W}^{zz}\\
{\cal W}^{0z} & = & \frac{\omega}{|\vec{q}\,|}{\cal W}^{00} +
\Delta {\cal W}^{0z} 
\end{eqnarray}
where $\Delta {\cal W}^{zz}$ and $\Delta {\cal W}^{0z}$ account for
the breaking of the gauge symmetry. By construction one expects:  
\begin{eqnarray}\label{eq:j10}
\frac{\Delta {\cal W}^{zz}}{{\cal W}^{00}} &\approx & 
\delta_1\frac{\omega^2}{|\vec{q}\,|^2} \\
\frac{\Delta {\cal W}^{0z}}{{\cal W}^{00}} &\approx & 
\delta_2\frac{\omega}{|\vec{q}\,|}
\end{eqnarray}  
\noindent
where $\delta_1$ and  $\delta_2$ would be 1 in the case that 
$\Delta {\cal W}^{zz}={\cal W}^{zz}$ and $\Delta {\cal W}^{0z}={\cal W}^{0z}$.
We certainly expect that in our case $\delta_1$ and $\delta_2$ are smaller
than 1 because at order $\rho$ in the density expansion
we exactly fulfill gauge invariance (see comments below).
Taking a conservative point of view one sees that 
the ratios 
in eqs. (96), (97) are at most of order 1. Using the definitions of
eq.~(\ref{eq:j9}), we can now write
\begin{equation}\label{eq:j11}
{\cal W}_L = -\frac{q^2}{|\vec{q}\,|^2}{\cal W}^{00} \left\{
1 + \frac{\omega^2|\vec{q}\,|^2}{q^4}\frac{\Delta 
{\cal W}^{zz}}{{\cal W}^{00}} - 2\frac{\omega|\vec{q}\,|^3}{q^4}\frac{\Delta 
{\cal W}^{0z}}{{\cal W}^{00}}  \right\}
\end{equation}
The size of the corrections to the ``charge-charge'' prescription (the
term proportional to 1 in the formula)
traditionally used in the literature depends on the kinematics 
under study. Here we will pay a special attention to three different
regions (keeping always the momentum transferred 
to the nucleus smaller than about 500 MeV) : $\Delta$-resonance and quasielastic peaks and 
the region
({\it dip}) between both peaks. 

\begin{itemize}
\item {\underbar{Quasielastic peak}}: In this region we
have $\omega \approx |\vec{q}\,|^2/2M_N$ and then the coefficients
of the ratios $\frac{\Delta {\cal W}^{zz}}{{\cal W}^{00}}$ and 
$\frac{\Delta {\cal W}^{0z}}{{\cal W}^{00}}$ in eq.~(\ref{eq:j10}) 
turn out to be
\begin{eqnarray}
\frac{\omega^2|\vec{q}\,|^2}{q^4}&\approx&
\frac{|\vec{q}\,|^2}{4M_N^2} \\
2\frac{\omega|\vec{q}\,|^3}{q^4}&\approx&\frac{|\vec{q}\,|}{M_N}
\end{eqnarray}
Thus, taking also into account the estimates of eq.~(\ref{eq:j10}),
one finds  corrections to the ``charge-charge'' prescription of the
order of $\delta_2|\vec{q}\,|^2/2M_N^2$ which at most could be of the order of
$5-10\%$ for the momenta and energies transfers studied in this
paper,
assuming $\delta_2\approx 1$,
which is certainly an overestimate for the
reasons pointed above. From this discussion,  in this  region 
we have decided to use the traditional
prescription  ``charge-charge'' to compute the longitudinal response 
function.

\item {\underbar{{\it Dip} area}}: In this region the conclusions are
similar to those drawn in the previous
point. However, we would like to point out that gauge symmetry
breaking corrections to the longitudinal response function are not now
as small as before.

\item {\underbar{$\Delta$-resonance peak}}: In this region 
$\omega/|\vec{q}\,| \approx 1 $ and the situation is
radically different. The corrections due to gauge
symmetry breaking are significantly more important than in the quasi-free 
scattering region. For instance, taking the incoming electron energy
equal to 620 MeV and the outgoing electron scattering angle equal to
$60^0$, one finds values for the coefficients
of the ratios $\frac{\Delta {\cal W}^{zz}}{{\cal W}^{00}}$ and 
$\frac{\Delta {\cal W}^{0z}}{{\cal W}^{00}}$ in eq.~(\ref{eq:j11}) of the
order of two. 
Thus the corrections to unity
in the bracket of eq.~(\ref{eq:j11}), are of the order 
of $2(\delta_1-\delta_2)$, much larger than in the quasielastic peak.
We have evaluated $\delta_1$ and $\delta_2$ in this region
and we find $\delta_1 \approx 0.01-0.06$, $\delta_2 \approx 0.01\,-\,0.02$
and $2(\delta_1-\delta_2)\approx 0.0\,-\,0.08$. Hence a 10$\%$ error in
the longitudinal response due to the breaking of gauge invariance
of our results seems realistic in this region.
In any case, we should mention that the contribution of the 
longitudinal response to the cross section is very small 
here and hence theoretical cross sections are largely free
of uncertainties due to the small breaking of gauge invariance.
Since there is no experimental separation of $R_L$ and $R_T$
in this region, we do not give these results either.

\end{itemize}

We finish this section discussing the origin of the breaking of the
gauge invariance within our model. Let us consider the
diagram of fig.~6.1 whose contribution to the virtual photon
self-energy in the medium is
given in eqs.(53) and (54). One can easily check that the imaginary
part (when the intermediate nucleons are put on shell) 
of this self-energy is gauge invariant 
($q_\mu\Pi^{\mu\nu}\propto q_\mu V^\mu=0$). The
first medium correction is given by the diagram
depicted in fig.6.2. When the two $ph$
excitations are put on shell, this diagram contributes to the
imaginary part of the photon self-energy. This new
contribution is not gauge invariant because the intermediate nucleon
with four momentum $p+q$ is not on shell and then the contraction 
 $q_\mu V^\mu$ does  not  vanish now.
However, to this level we restore gauge invariance
because we consider not only the term of fig. 6.2, but also  
all terms implicit in fig. 4.9. Though the $NP$ amplitude in not gauge
invariance by itself, the thick dots of fig. 4.9 account for the six
amplitudes (NP+NPC+KR+PP+DP+DPC) of our model for the $eN \to e^\prime
N\pi$ reaction. In section 3, we fixed the different form-factors
entering in the amplitudes to end up with 
a gauge invariant model (Eqs.(8-9)). 
Thus, the leading terms in our density expansion (fig.6.1, fig. 4.2 and
fig. 4.9) lead to a gauge invariant photon self-energy.

However, we break again the gauge invariance in section 4.7 
when we include the polarization corrections to the 36 diagrams of 
fig. 4.9: we do not
renormalize in the same way for instance the $KR \times KR $ term
(which is purely longitudinal and therefore gets renormalized only
with $V_l$) than the $NP \times NP $ term (which
contributes to both longitudinal and transverse channels and thus
gets renormalized not only
with $V_l$ but also with $V_t$). As a consequence,  the cancellations
in eq.(8) which ensured the gauge invariance of the model are altered.

\section{Results}

We have already shown results on the different effects in previous
sections. Here we will show results with emphasis in comparison with
experiment.

Let us first show results in the quasielastic peak in figs. 7.1, 7.2
we show results for $R_L$ and $R_T$ 
for $^{12}$C and compare them to the
data of \cite{DLA}. The lower line shows the results obtained
with the medium spectral function, while the upper one includes also the rest of the
effects discussed in the former section.

\vskip 0.3cm 
\centerline{\protect\hbox{\psfig{file=rlc12300.ps,width=11.5cm}}}
\vskip 0.1cm
\noindent
{\small {\bf Fig.7.1} Calculation of $R_L$ for $^{12}$C. The lower line
in the high energy region
corresponds to the result obtained with 
the contribution of the $1p1h$ excitation (fig. 6.3) using the
medium spectral function of eq.(64). The upper line is the result 
when one adds the rest of contributions: vertex corrections
(fig. 6.13), two body absorption diagrams (fig. 4.9), $(\gamma ^*, \pi)$
 terms (fig. 4.2), $(\gamma ^*, 2 \pi)$ related terms (fig. 4.15), etc.
  Experimental data from \cite{DLA}.}
\vspace*{0.1cm}

%\vspace*{0.5cm}
\centerline{\protect\hbox{\psfig{file=rtcarbon300.ps,width=11.5cm}}}
\vskip 0.1cm
\noindent
{\small {\bf Fig.7.2} 
Calculation of $R_T$ for $^{12}$C. Same meaning of the lines
as in fig. 7.1. Experimental data from \cite{DLA}.
}
\vspace*{0.6cm}

In fig. 7.3, 7.4, we show results for $^{40}$Ca compared to the data
of~\cite{JOU} (fig. (7.3) and lower points in fig. (7.4)) 
 and those of~\cite{MEZ} (upper points in fig. (7.4)).

\centerline{\protect\hbox{\psfig{file=rlca410.ps,width=11.5cm}}}
\vskip 0.2cm
\noindent
{\small {\bf Fig.7.3} 
Calculation of $R_L$ for $^{40}$Ca. 
 Same meaning of the lines
 as in fig. 7.1.
 Experimental data from \cite{JOU}.
}

%\newpage

\vspace*{1cm}
\centerline{\protect\hbox{\psfig{file=rtcalcio410.ps,width=11.5cm}}}
\vskip 0.2cm
\noindent
{\small {\bf Fig.7.4}  Calculation of $R_T$ for $^{40}$Ca.
 Same meaning of the lines
  as in fig. 7.1.
Experimental data from \cite{JOU} (lower points) and \cite{MEZ} (upper points). }

\vspace*{0.1cm}

As one can see, we find a good agreement with the recent 
reanalysis of \cite{JOU}.

On the other hand much of the work done here has gone into the 
evaluation of two body mechanisms. In fig. 7.5 we show the results
for two body photon absorption (solid line) and compare them to pion
production (dotted line). 
Similarly, in fig. 7.6 we show the
contribution of three body photon absorption (solid line) versus the
two body one (dotted line). We can see that at low energies,
the contribution of three body absorption is negligible while at
energies around 450 MeV the three body contribution becomes sizeable.
These results agree qualitatively with the findings of \cite{AO}
for real photons. Let us recall that this classification corresponds
to the primary step in the collision. The particles produced still
undergo secondary collisions in their way out of the nucleus. This
does not change the inclusive cross section but redistributes the
strength. The treatment of this FSI and the evaluation of the exclusive
channels will be treated  in a forthcoming paper \cite{JUA}.

%\vskip 0.2cm
\centerline{\protect\hbox{\psfig{file=1pion2body.ps,width=11.3cm}}}
\vskip 0.2cm
\noindent
{\small {\bf Fig.7.5}
Two body photon absorption (solid line) versus pion
production (dotted line) for $^{12}$C.
}
\vspace*{0.1cm} 
 
 \vskip 0.2cm
 \centerline{\protect\hbox{\psfig{file=3bodysec.ps,width=11.3cm}}}
 \vskip 0.2cm
 \noindent
 \centerline{\small {\bf Fig.7.6}
 Three body photon absorption (solid line) versus the
 two body one (dotted line) for $^{12}$C.
 }
 \vspace*{0.1cm}
 
Finally let us see the global results including the quasielastic
peak, the dip region and the delta region. They can be
seen in figs. 7.7, 7.8 and 7.9 for the nuclei of $^{12}C$ and $^{208}Pb$.

The global agreement is good and the three regions are well reproduced (a bit
overestimated for $^{208}Pb$).

%\vspace*{0.5cm}
\centerline{\protect\hbox{\psfig{file=newcarbonsec.ps,width=11.5cm}}}
\vskip 0.2cm
\noindent
{\small{\bf Fig.7.7} Inclusive $(e,e^{\prime})$ cross section for $^{12}C$.
$E_e=620$ MeV and $\theta_e=60^0$. The dotted line
corresponds to the pion production contribution. 
Experimental data from \cite{BAR}. 
 }

\centerline{\protect\hbox{\psfig{file=carbon36inclusivo.ps,width=11.5cm}}}
\vskip 0.2cm
\noindent
{\small{\bf Fig.7.8}
Inclusive $(e,e^{\prime})$ cross section for $^{12}C$.
$E_e=680$ MeV and $\theta_e=36^0$. Experimental data from \cite{BAR}.
}
\vspace*{0.1cm}

\vskip 0.2cm
\centerline{\protect\hbox{\psfig{file=plomoinclusivo.ps,width=12cm}}}
\vskip 0.2cm
\noindent
{\small{\bf Fig.7.9}
Inclusive $(e,e^{\prime})$ cross section for $^{208}Pb$.
$E_e=645$ MeV and $\theta_e=60^0$. The dotted line
corresponds to the $1p 1h$ excitation 
contribution. Experimental data from \cite{ZGH}.
}
\vspace*{0.1cm}

In fig. 7.7 we  also show with a dotted line the results for pion
production.

In fig. 7.9 instead we show with a dotted line the results for
the $1p 1h$ excitation alone.

\section{Conclusions}

   We have undertaken the task of constructing a microscopic many body 
   model of the $(e,e^{\prime})$ reaction including all the reaction channels
   which appear below $\omega=500-600$ MeV, and which is suited to study
   the inclusive $(e,e^{\prime})$ reaction from the quasielastic peak 
   up to the $\Delta$ peak, passing through the dip region. Although
   many studies have been devoted to particular
    energy regions of the spectrum, this is the first work, to our knowledge,
    which ranges this wide energy spectrum.
    
    Our model has no free parameters. All the input consists of basic couplings
    of photons to nucleons and isobars, and some phenomenological inputs,
    as correlations, which has been tested in former pionic reactions.
    
    We include explicitly the $1N$ knockout channel, the virtual photon 
    absorption by pairs or trios of particles, the pion production
    plus exchange currents mechanisms tied to the $(\gamma^*,2\pi)$
    channel and which contribute to $(\gamma^*,NN\pi)$ or 
    $(\gamma^*,NNN)$ channels.
    
       We include effects which have been found important in earlier works,
       like polarization, renormalization of $\Delta$ properties
       in a nuclear medium, FSI effects through the use of spectral
       functions and meson exchange currents.
       
       The meson exchange currents are generated in 
       a systematic way from a model 
       for the elementary pion electroproduction on the nucleon, which
        reproduces accurately the experimental data.
        
               We have payed some attention to the question  of gauge invariance, showing
               that it is preserved in our approach in leading order
               of the density expansion. We also show that the appropriate
               prescription to evaluate the longitudinal response
               is from the $W^{00}$ component of the hadronic tensor, which minimizes the breaking 
               of gauge invariance at higher orders in $\rho$.
               
               We evaluate cross sections in the energy range from the quasielastic peak
               to the $\Delta$ peak and find good agreement with experimental 
               data. The three traditional regions: quasielastic peak, dip
               region  and delta peak, are well reproduced in our scheme.
               
               We also separate the longitudinal and transverse response functions
               in the quasielastic peak and find good agreement with the latest
               results of the analysis of Jourdan from the world set of data.
             
        We have used the technique of the local density
        approximation, which has been shown before to be particularly suited
        to deal with inclusive cross sections and which makes
        unnecessary the use of sophisticated finite nuclei wave functions.
        
               Finally, the method used here allows the separation of the
               contribution of different channels to the inclusive 
               cross section. This information is the seed to produce
               exclusive cross sections like $(e,e^{\prime}N)$, 
               $(e,e^{\prime}NN)$, $(e,e^{\prime}\pi)$, $(e,e^{\prime}\pi N)$
               etc. However, this still requires to follow the fate of all the particles
               produced from their production point in the nucleus,
               which is usually done using Monte Carlo simulation techniques,
               and this will be the subject of some future work.
               
\vspace*{0.5cm}

       We would like to acknowledge useful discussions with 
       R.C. Carrasco, C. Garc\'{\i}a-Recio and
       A. Lallena. This paper is partially
       supported by CICYT contract no. AEN 96-1719. 
       One of us (J. Nieves) thanks to DGES contract PB95-1204.

\newpage
{\Large {\bf Appendix}}

\vspace*{0.6cm}
The Galilean invariant vertices 
which appear in the model for
 $e N \rightarrow e \pi N$, are:

\begin{description}

\item{{\bf (a)}} $\gamma NN$ vertex (fig. 3.1(a)):

\be               
\begin{array}{rcl}
{V}^{\mu}_{\gamma N N} & = &-ie
\left\{
\begin{array}{c}
F_{1}^{N}(q^{2})\\
\\
F_{1}^{N}(q^{2})\left[ \frac{\displaystyle{
\vec{p}+{\vec{p}\,}^{\prime}}}{\displaystyle{2M_{N}}}\right]
+i\frac{\displaystyle{\vec{\sigma}\times\vec{q}}}{\displaystyle{2M_{N}}}
G^{N}_{M}(q^{2})
\end{array}
\right\}
\end{array}
\ee

\item{{\bf (b)}}  $\gamma N\Delta$ vertex 
(fig. 3.1(c)):

\be
\begin{array}{rcl}
{V}^{\mu}_{\gamma N \Delta} & = &
\displaystyle{\sqrt{\Frac{2}{3} }}
\frac{\displaystyle{f_{\gamma}(q^{2})}}{\displaystyle{m_{\pi}}}
\frac{\displaystyle{\sqrt{s}}}{\displaystyle{M_{\Delta}}} 
\left\{
\begin{array}{c}
\frac{\displaystyle{\vec{p}_{\Delta}}}{\displaystyle{\sqrt{s}}}
(\vec{S}^{\dagger}\times \vec{q}\,) \\
\\
\frac{\displaystyle{{p}^{0}_{\Delta}}}{\displaystyle{\sqrt{s}}}
\left\{
\vec{S}^{\dagger}\times \left(\vec{q}\,-\,
\frac{\displaystyle{{q}^{0}}}{\displaystyle{{p}^{0}_{\Delta}}}
\vec{p}_{\Delta}
\right)  \right\}
\end{array}
\right\}
\end{array}
\ee

\item{{\bf (c)}}  $\pi N\Delta$ vertex 
(fig. 3.1(d)):

\be\label{eq:pind}
   V_{\pi N \Delta}
   =I
   \frac{\displaystyle{{f}^{*}}}{\displaystyle{{m}_{\pi}}}
\vec{S}^{\dagger}. \left(\vec{k}\,-\,
\frac{\displaystyle{{k}^{0}}}{\displaystyle{\sqrt{s}}}
\vec{p}_{\Delta}
\right) 
\ee

\item{{\bf (d)}} $\pi N N$ vertex 
(fig. 3.1(b)):

\be\label{eq:piNN}
   V_{\pi N N} 
   =\frac{\displaystyle{{f}_{\pi NN} }}{\displaystyle{{m}_{\pi}}}
   B(N,N^{\prime}\pi)
   \left\{\vec{\sigma}\,\vec{k}\,-\,
   \frac{\displaystyle{{k}^{0}}}{\displaystyle{2M_{N}}}
   \vec{\sigma}\,(\vec{p}\,+\,\vec{p}\,^{\prime})
   \right\} 
\ee

\end{description}

\noindent
where $\vec{q}$, $\vec{p}$, ${\vec{p}\,}^{\prime}$, 
 $\vec{p}_{\Delta}$ y $\vec{k}$ are
the photon, incoming nucleon, outgoing nucleon and pion
momenta, respectively;  
  $\sqrt{s}$, the invariant energy in the
 $\gamma^{*}\,N$ system
 and $M_N$, $m_{\pi}$ and $M_{\Delta}$ 
 are the nucleon, pion and delta resonance masses.
 In equations (\ref{eq:pind}) 
  and (\ref{eq:piNN})  we are including
 the corresponding isospin factors
 $I$ y $B(N,N^{\prime}\pi)$, respectively.

 Besides the vertices shown in fig. 3.1,
in our elementary model for electroproduction two more vertices appear: 
  
\vskip 0.2cm
\centerline{\protect\hbox{\psfig{file=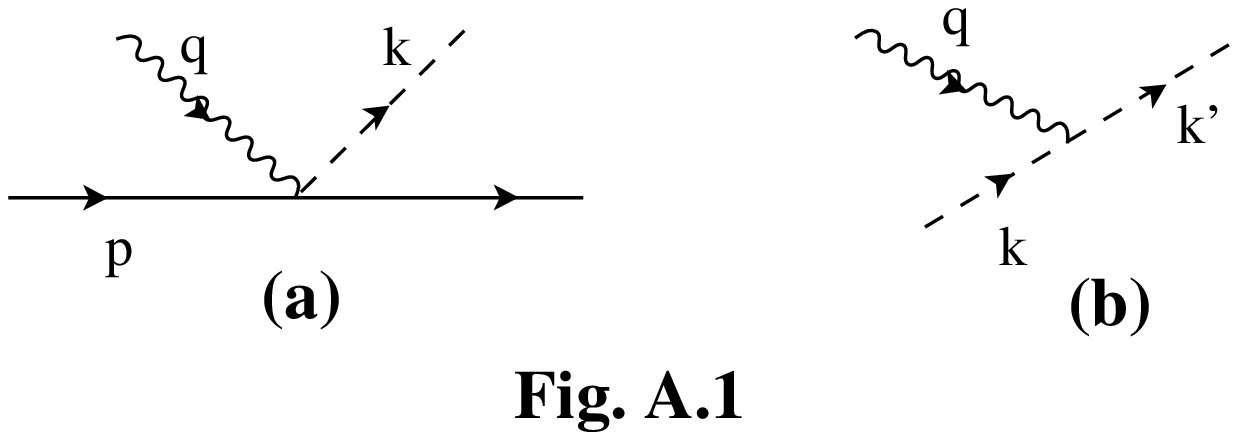,width=6.7cm}}}

 In fig. A.1(a) one can see  the seagull  vertex. It appears from the
  ${\cal{L}}_{\pi NN}$ lagrangian via minimal coupling.
This vertex is exactly zero for the  
 ${\pi}^{0} n$ and
  ${\pi}^{0} p$ channels and  has the following expression:
  
  \be
  {V}^{\mu}_{seagull}=e
  \frac{\displaystyle{{f}_{\pi NN} }}{\displaystyle{{m}_{\pi}}}
     B(N,N^{\prime}\pi)F_AC^{\mu}
  \ee
  
\noindent
where

$$
\begin{array}{lll}
\!\!\!\!\!\!\!\!\!\!\!\!\!\!\!\!\!\!\,\,\,\,\,\,\,\,\,
\,\,\,C^{\mu}(\pi^{-}p)=
\left(
\begin{array}{c}
\frac{\displaystyle{\vec{\sigma}\,(2\vec{p}\,+\,\vec{q}\,-\,\vec{k})}}
{\displaystyle{2M_{N}}}\\
\\
\vec{\sigma}
\end{array}
\right)
&\,\,\,\, ; &
C^{\mu}(\pi^{0} n)=0
\end{array}
$$
$$
\begin{array}{lll}
C^{\mu}(\pi^{+}n)=
\left(
\begin{array}{c}
\frac{\displaystyle{\vec{\sigma}(\vec{k}\,-\,\vec{q}\,-\,2\vec{p})}}
{\displaystyle{2M_{N}}}\\
\\
\vec{\sigma}
\end{array}
\right)
& ; &
C^{\mu}(\pi^{0}p)=0
\end{array}
$$
   
In figure A.1, the vertex (b) corresponds to the 
 $\pi \pi \gamma^{*}$ coupling and it is defined as:

\be
   {V}^{\mu}_{\pi \pi \gamma^{*}}=ie(k^{\mu}\,+\,{k^{\prime}}^{\mu})
\ee

With respect to form factors and coupling constants,
their expressions are the following:

\vspace*{0.3cm}
We use Sachs Form Factors:

\be
\begin{array}{ll}
{G}^{N}_{M}(q^2) = 
  \frac{\displaystyle{{\mu}_{N}}}{\displaystyle{
  \left(1-                     
  \frac{\displaystyle{{q}^{2}}}{\displaystyle{{\Lambda}^{2}}}
  \right)^2
 }} & ; \,\,\,
 {G}^{N}_{E}(q^2) = 
   \frac{\displaystyle{1}}{\displaystyle{
     \left(1-                     
       \frac{\displaystyle{{q}^{2}}}{\displaystyle{{\Lambda}^{2}}}
         \right)^2
          }}
 
 \end{array}
\ee

\noindent
 with $\Lambda^{2}=0.71\,\,GeV^2$; $\mu_{p}=2.793$; $\mu_n=-1.913$.
 The relationship between ${F}_{1}^{p}(q^2)$ (Dirac form factor)
 and ${G}_{E}^{p}$ is:
 
 \be\label{eq:f1p}
  {F}^{p}_{1}(q^2) ={G}_{E}^{p} 
  \frac{\displaystyle{
   \left(1-                     
  \frac{\displaystyle{{q}^{2}}}{\displaystyle{4{M_N}^{2}}}
  \mu_p
  \right)
   }}{\displaystyle{
  \left(1-                     
  \frac{\displaystyle{{q}^{2}}}{\displaystyle{4{M_N}^{2}}}
  \right)
 }}
 \ee

\noindent 
and  ${F}^{n}_{1}=0$.

\vspace*{0.2cm} 
 For the rest of form factors and coupling constants we take:

 \be
 \begin{array}{lll}
 \frac{\displaystyle{{f}^{2}_{\pi NN}}}{\displaystyle{4\pi}}\, =\, 0.08\,\,;&
 \frac{\displaystyle{{{f}^{*}}^{2}}}{\displaystyle{4\pi}}\, =\, 0.36\,\,;&
{F}_{A}(q^2) = 
  \frac{\displaystyle{1}}{\displaystyle{
  \left(1-                     
  \frac{\displaystyle{{q}^{2}}}{\displaystyle{{M_{A}}^{2}}}
  \right)^2
 }}
 \end{array}
\ee

\noindent
 where ${M}_{A}=1.08\,GeV$.

 \be
{F}_{\gamma \pi \pi}(q^2) = 
  \frac{\displaystyle{1}}{\displaystyle{
  \left(1-                     
  \frac{\displaystyle{{q}^{2}}}{\displaystyle{p^{2}_{\pi}}}
  \right)
 }}
\ee

\noindent
 with ${p}^{2}_{\pi}=0.47\,GeV^2$.

\be
{F}_{\pi}(q^2) =
  \Frac{{\Lambda}_{\pi}^2-{m}_{\pi}^2}{{\Lambda}_{\pi}^2-q^2}
  \,\,\,;\,\,\,\,{\Lambda}_{\pi}\sim 1250 \,MeV
\ee

\be 
{f}_{\gamma}(q^2) = 
f_{\gamma}(0)
\frac{\displaystyle{
  \left(1-                     
  \frac{\displaystyle{{q}^{2}}}{\displaystyle{
  \left(M_{\Delta}+M_{N}\right)^{2}}}
  \right)                      
 }}
{\displaystyle{
  \left(1-                     
  \frac{\displaystyle{{q}^{2}}}{\displaystyle{4{M_{N}}^{2}}}
  \right)
 }}\,
 \frac{\displaystyle{{G}_{M}^{p}(q^2)}}{\displaystyle{\mu_{p}}}
 \frac{\displaystyle{ \left(M_{\Delta}+M_{N}\right)^{2} }}
 {\displaystyle{({M_{\Delta}+M_{N}})^{2}-q^2}}
 \ee
 
\noindent
 where $f_{\gamma}(0)=0.122$. It is 
 the  $\gamma N \Delta$ coupling constant 
 for real photons.

  With respect to the $\vec{S}$ and $\vec{T}$ operators
   (transition operator between $\Frac{3}{2}$ spin states
   to
  $\Frac{1}{2}$ spin states  and respectively
   between 
  $\Frac{3}{2}$ isospin states to
 $\Frac{1}{2}$ isospin states),
  their normalization is:

  \be
  <\frac{3}{2},M|S^{\dagger}_{\lambda}|\frac{1}{2},m>=
  (1\frac{1}{2}\frac{3}{2}|\lambda m M)
   \ee   

 \be
   <\frac{3}{2},M|T^{\dagger}_{\lambda}|\frac{1}{2},m>=
     (1\frac{1}{2}\frac{3}{2}|\lambda m M)
        \ee
        
\noindent
where $\lambda$  is a spherical basis index.


\newpage
\begin{thebibliography}{99}
\bibitem{MEZ}Z.E. Meziani et al. , Phys. Rev. Lett. 52 (1984) 2130;
ibid 54(1985)1233

\bibitem{ALT} R. Altemus et al., Phys. Rev. Lett. 44 (1980) 965

\bibitem{BAR} P. Barreau et al., Nucl. Phys. A402 (1983) 515

\bibitem{GIU} G. Orlandini and M. Traini, Rep. Prog. Phys.54 (1991) 257

\bibitem{BLA} C. C. Blatchley et al. , Phys Rev. C34 (1986) 1243

\bibitem{KAR} T. C. Yates et al. , Phys. Lett. B312 (1993) 382

\bibitem{JOU} J. Jourdan, Phys. Lett. B353 (1995) 189; J. Jourdan, Nucl.
Phys. A603 (1996) 117

\bibitem{DEN} D. Wilkinson, talk at the PANIC 96 conference,
Williamsburg, June 1996, C.E. Carlson and J.J. 
Domingo Edts., World Scientific, pag. 217

\bibitem{NOB} J. Noble et al., Phys. Rev. Lett. 46(1981) 412

\bibitem{SHA} C. M. Shakin, Nucl. Phys. A446 (1985) 323

\bibitem{GOE} M. Bergmann, K. Goeke and S. Krewald, Phys. Lett. B243 (1990)
185

\bibitem{WAL} J. D. Walecka, Ann. of Phys. 83 (1974) 491
B.D. Serot and J. D. Walecka, Adv. in Nucl. Phys. 16, ed. J.W. Negele and
E. Vogt (plenum, N.J. 1986)

\bibitem{HOR} C. J. Horowitz and J. Piekarewicz, Nucl. Phys. A511 (1990)
461

\bibitem{PRI} R. J. Furnstahl and C. E. Price, Phys. Rev.C40 (1989) 1398

\bibitem{GIM} A. Gil, M. Kleinmann, H. M\"uther and E. Oset, Nucl. Phys.
A584 (1995) 621

\bibitem{RIS} M. Kirschbach, D.O. Riska and K.Tsushima, Nucl. Phys. A542 (1992) 616

\bibitem{GRA} E.D. Izquierdo, G. Barenboim and A.O. Gattone , 
Nucl. Phys. A609 (1996) 437

\bibitem{HOD} C. J. Horowitz, Phys. Lett. B208 (1988) 8; Ibid, Phys. Rev.
Lett. 62 (1989) 391

\bibitem{SUZ} H. Kurasawa and J. Suzuki, Nucl.Phys. A 490 (1988) 571

\bibitem{MAH} C. Mahaux, P.F. Bortignon, R.A. Broglia  and C.H. Dasso, 
Phys. Reports 120 (1985) 1

\bibitem{PAL} E. Oset and A. Palanques, Nucl. Phys A 359 (1981) 289

\bibitem{CARL} J. Carlson and R. Schiavilla, Phys. Rev. C491 (1994) 2880

\bibitem{WAN} W. M. Alberico, T.W. Donnelly and A. Molinari, Nucl. Phys.
A 512 (1990) 541

\bibitem{MAG} W. M. Alberico, M. Ericson and A.Molinari, Ann. of Phys. 154 
(1984) 356

\bibitem{ORD} J. W. Van Orden and T.W. Donnelly, Ann. of Phys 131 (1981)
451

\bibitem{MIS} M. Kohno and N. Ohtsuka, Phys. Lett. B98 (1981) 335

\bibitem{RIN} W. M. Alberico, R. Cenni, A. Molinari and P. Saracco, Phys
 Rev. Lett. 65 (1990) 1845

\bibitem{GCO} J. E. Amaro, G. Co', E. M. V. Fasanelli and A. M. Lallena,
Phys. Lett. B 277 (1992) 249

\bibitem{AMA} J. E. Amaro, G. Co' and A. M. Lallena, Ann. of Phys. 221
(1993)306

\bibitem{LAN} J. E. Amaro, A.M. Lallena and G. Co', Nucl. Phys. A578 (1994) 365

\bibitem{TAI} K. Takayanagi, Phys. Lett. B233 (1989) 271,ibid, Phys. 
Lett. B230 (1989) 12; ibid, Nucl. Phys. A 516 (1990) 276

\bibitem{CAR} R. C. Carrasco and  E. Oset, Nucl. Phys. A536 (1992) 445

\bibitem{CRO} G.E. Cross et.al., Nucl. Phys. A593 (1995) 463

\bibitem{GRO} P.D. Harty et.al., Phys. Lett. B380 (1996) 247

\bibitem{helh} T. Helh, Prog. Part. Nucl. Phys. 34 (1995) 385
 

\bibitem{DRE} M. Cavinato, D. Drechsel, E. Fein, M. Marangoni and A.M.
Saruis, Nucl. Phys. A423 (1989)376

\bibitem{WAB} S. Drozdz, G. Co', J. Wambach and J. Speth, Phys. Lett.
B 185 (1987) 287

\bibitem{WAG} W. M. Alberico, M. Ericson, A. Molinari and Zi Xing Wang,
Phys. Lett. B233 (1989) 37

\bibitem{CHI} C. R. Chinn, A. Picklesimer and J. W. Van Orden, Phys.Rev.
C40 (1989) 790

\bibitem{NIMAI} Y. Horikawa, F. Lenz and N. Mukhopadhyay, Phys. Rev. C 22
 (1980) 1680
 
\bibitem{FAN} A. Fabrocini and S. Fantoni, Nucl. Phys. A 503 (1989)375

\bibitem{PED} P. Fern\'{a}ndez de C\'{o}rdoba and E. Oset, Phys. Rev. C 46 
(1992) 1697

\bibitem{ANG} A. Ramos, A. Polls, and W, H. Dickhoff, Nucl. Phys. A503
(1989) 1

\bibitem{ART} H. M\"uther, G. Knehr and A. Polls, Phys. Rev. C52 (1995) 2955

\bibitem{CIO} C. Ciofi degli Atti, S. Liuti and S. Simula, Phys. Rev. 
C41 (1990) 2474

\bibitem{TJON} M. J. Dekker et al., Phys. Lett. B 266 (1991) 249
\bibitem{ANGH} M. Anghinolfi et al.,  Nucl. Phys. A602 (1996) 405

\bibitem{NOZ} S. Nozawa and J. S. H. Lee, Nucl. Phys. A513 (1990) 511

\bibitem{WOL} E. Oset, H. Toki and W. Weise, Phys. Reports 83 (1982) 281
         

\bibitem{THI} Y. Horikawa, M. Thies and F. Lenz, Nucl. Phys. A 345
(1980) 386

\bibitem{LOR} E. Oset and L.L. Salcedo, Nucl. Phys. A 468 (1987) 631

\bibitem{CARMEN} C. Garcia Recio, E.Oset, L.L. Salcedo, D. Strottman 
and M.J. Lopez,  Nucl. Phys. A 526 (1991) 685

\bibitem{MAN} L. L. Salcedo, E. Oset, M.J. Vicente Vacas and C. Garcia
Recio Nucl. Phys. A 484 (1988) 557

\bibitem{EUG} P. Fern\'{a}ndez de C\'{o}rdoba, E. Marco, H. M\"uther, E. Oset
and A. Faessler, Nucl. Phys. A611 (1996) 514

\bibitem{ANT} J. E. Amaro, A. M. Lallena and G.Co', Int. Jour. Mod. Phys.
E, 3 (1994) 735

\bibitem{ALB} W. M. Alberico, A. Drago and C. Villavecchia, Nucl. Phys
A 505 (1989) 309

\bibitem{JUA} A. Gil, J. Nieves and E. Oset, University of Valencia preprint,
 FTUV/97-40

\bibitem{AMP} A. Gil, PhD Thesis, University of Valencia, December 1996.


\bibitem{OLS} M. G. Olsson, Nucl. Phys. B78 (1974) 55

\bibitem{AML} E. Amaldi, S. Fubini and G. Furlan, {\it Pion electroproduction},
 Springer Tracts in Modern Physics, Vol. 83 (Springer, Berlin, 1979)

\bibitem{BAE} K. Baetzner et al., Phys. Lett. { B39} (1972) 575

\bibitem{BRE} H. Breuker et al., Nucl. Phys. { B146} (1978) 285

\bibitem{MIST}  C. Mistretta et al., Phys. Rev.  Vol.184 (1969) 5

\bibitem{GUE} L. Ghedira, PhD Thesis, University of Orsay (1986)

\bibitem{WES} E. Oset and W. Weise, Nucl. Phys. A319 (1979) 477

\bibitem{STROT} L.L. Salcedo et al., Phys. Lett. B208 (1988) 339

\bibitem{WAK} A.L. Fetter and J.D. Walecka,
{\it Quantum Theory of many-particle systems}, McGraw-Hill, 1971

\bibitem{SPH} J. Speth et al., Nucl. Phys. { A343} (1980) 382
\bibitem{RAFAPI} R.C. Carrasco, E. Oset and L.L. Salcedo, Nucl. Phys.
A 541 (1992) 585

\bibitem{DLA} A. Dellafiore et al., Phys. Rev C {31} (1985) 1088

\bibitem{MUL} P. J. Mulders, Phys. Reports 185 (1990) 83

\bibitem{GIO} A. Gil and E. Oset, Nucl. Phys. A 580 (1994) 517

\bibitem{ZGH} A. Zghiche et al. Nucl. Phys. A 572 (1994) 513
 
\bibitem{BRA} A. Braghieri et al., Phys. Lett. B363 (1995) 46

 \bibitem{STR} A. Str\"oher et al., Workshop on $N^*$ resonances, INT, Seattle,
 
  \bibitem{GOM1} J.A. G\'omez-Tejedor and E. Oset,
   Nucl. Phys. A571 (1994) 667
   
    \bibitem{GOM2} J.A. G\'omez-Tejedor and E. Oset, Nucl. Phys. A600 (1996) 413.
    
     \bibitem{ULF} V. Bernard et al., Nucl. Phys. A580 (1994) 475.
  \bibitem{BEN} M. Benmerrouche and E. Tomusiak, Phys. Rev. Lett. 73 (1994) 400.
      

\bibitem{Hei} J. Heisenberg, Adv. Nucl. Phys. 12 (1981) 61;
J. Heisenberg and H.P. Blok, Annu. Rev. Nucl. Part. Sci. 33 (1983) 569.

\bibitem{Friar} J.L. Friar and S. Fallieros, Phys. Lett. B 114
(1982) 403; Phys. Rev. C 29 (1984) 1645.

\bibitem{AAL} J.E. Amaro, B. Ameziane and A.M. Lallena, Phys. Rev. C53
(1996) 1430.

\bibitem{Muta} T. Muta, Foundations of Quantum Chromodynamics, World
Scientific 1987.

\bibitem{AO} E. Oset and A. Ramos, Phys. Rev. C53 (1996) 305
\end{thebibliography}
\end{document}